\documentclass[aps,showpacs,twocolumn,prd,superscriptaddress,nofootinbib]{revtex4-1}

\usepackage{amsmath}
\usepackage{amsfonts}
\usepackage{amssymb}
\usepackage{latexsym}
\usepackage{graphicx}
\usepackage{bm}
\usepackage{enumerate}
\usepackage{tabularx}
\usepackage{braket}

\usepackage{color}
\usepackage[usenames,dvipsnames,svgnames,table]{xcolor}
\usepackage[colorlinks=true,
            linkcolor=YellowOrange,
            urlcolor=RoyalBlue,
            citecolor=RedViolet]{hyperref}

\newcommand{\be}{\begin{equation}}
\newcommand{\ee}{\end{equation}}
\newcommand{\bsub}{\begin{subequations}}
\newcommand{\esub}{\end{subequations}}

\newcommand\calT{{\mathcal{T}}}

\newcommand\calL{{\mathcal{L}}}

\newcommand{\nn}{\nonumber}

\newcommand{\SNR}{\,\mathrm{SNR}}
\newcommand{\Hz}{\,\mathrm{Hz}}
\newcommand{\mHz}{\,\mathrm{mHz}}
\newcommand{\au}{\,\mathrm{au}}
\newcommand{\sinc}{\,\mathrm{sinc}}
\newcommand{\Msol}{M_{\odot}}

\newcommand\betaL{{\beta_{L}}}
\newcommand\lambdaL{{\lambda_{L}}}

\newcommand\psiL{{\psi_{L}}}
\newcommand{\tf}{t_{f}}

\newcommand{\sYlm}{{}_{-2}Y_{\ell m}}
\newcommand{\sYlmstar}{{}_{-2}Y_{\ell m}^{*}}
\newcommand{\sYlminusmstar}{{}_{-2}Y_{\ell, -m}^{*}}

\newcolumntype{C}[1]{>{\centering\arraybackslash}p{#1}}
\newcolumntype{L}[1]{>{\raggedright\arraybackslash}p{#1}}

\begin{document}

\title{Exploring the Bayesian parameter estimation of binary black holes with LISA}

\author{Sylvain Marsat}
\affiliation{APC, AstroParticule et Cosmologie, Universit\'{e} Paris Diderot, CNRS/IN2P3, CEA/Irfu, Observatoire de Paris, Sorbonne Paris Cit\'{e}, 10, rue Alice Domon et L\'{e}onie Duquet 75205 PARIS Cedex 13, France}
\author{John G. Baker}
\affiliation{Gravitational Astrophysics Laboratory, NASA Goddard Space Flight Center, 8800 Greenbelt Rd., Greenbelt, MD 20771, USA}
\author{Tito Dal Canton}
\thanks{NASA Postdoctoral Program fellow}
\affiliation{Gravitational Astrophysics Laboratory, NASA Goddard Space Flight Center, 8800 Greenbelt Rd., Greenbelt, MD 20771, USA}
\affiliation{Université Paris-Saclay, CNRS/IN2P3, IJCLab, 91405 Orsay, France}

\date{\today}

\begin{abstract}

The space-based gravitational wave detector LISA will observe mergers of massive black hole binary systems (MBHBs) to cosmological distances, as well as inspiralling stellar-origin (or stellar-mass) binaries (SBHBs) years before they enter the LIGO/Virgo band. Much remains to be explored for the parameter recovery of both classes of systems. Previous MBHB analyses relied on inspiral-only signals and/or a simplified Fisher matrix analysis, while SBHBs have not yet been extensively analyzed with Bayesian methods. We accelerate likelihood computations by (i) using a Fourier-domain response of the LISA instrument, (ii) using a reduced-order model for non-spinning waveforms that include a merger-ringdown and higher harmonics, (iii) setting the noise realization to zero and computing overlaps in the amplitude/phase representation. We present the first simulations of Bayesian inference for the parameters of massive black hole systems including consistently the merger and ringdown of the signal, as well as higher harmonics. We clarify the roles of LISA response time and frequency dependencies in breaking degeneracies and illustrate how degeneracy breaking unfolds over time.  We also find that restricting the merger-dominated signal to its dominant harmonic can make the extrinsic likelihood very degenerate. Including higher harmonics proves to be crucial to break degeneracies and considerably improves the localization of the source, with a surviving bimodality in the sky position. We also present simulations of Bayesian inference for the extrinsic parameters of SBHBs, and show that although unimodal, their posterior distributions can have non-Gaussian features.

\end{abstract}


\maketitle


\section{Introduction}
\label{sec:intro}

%
%
%

Gravitational waves from the coalescence of black hole binaries are now regularly observed \cite{GWTC-1} by the ground-based interferometers Advanced LIGO~\cite{LIGO}, Advanced Virgo~\cite{Virgo} and, soon, KAGRA~\cite{KAGRA}.

Ground-based interferometers are fundamentally limited at low frequency by terrestrial noise, and cannot be used to study mergers of compact objects much heavier than a few $ 10^2 \Msol$. The spacebourne detector LISA~\cite{LISA2017} will overcome this limitation, enabling the observation and precise characterization of gravitational waves from binary black holes, from coalescences of systems with millions of $\Msol$ (massive black hole binaries, MBHB~\cite{Klein+15}) to the inspiral of systems with tens of $\Msol$ (stellar-mass black hole binaries, SBHB~\cite{Sesana16}). The observation of black hole mergers and inspirals in the LISA frequency band will yield major scientific rewards, and they form the main target of LISA among other classes of sources.

In order to realize the full scientific objectives of LISA with respect to black hole mergers, adequate data analysis tools must be prepared in advance. Similarly to what happens in LIGO and Virgo, one has to first identify the presence of a merger waveform in the LISA data, possibly in the presence of other superposed signals (notably galactic white dwarfs binaries~\cite{Nelemans+01}), and then infer the distribution of the physical parameters of its source from the data, enabling the construction of a catalog of black hole binaries. Inferring the parameters of binaries, in particular their distance and sky location, and refining the analysis as the signal accumulates with observation time, will also be necessary to organize multimessenger observations of their late inspirals and mergers~\cite{ArmitageNatarajan02, DalCanton+19}. Cosmological applications using LISA observations as standard sirens~\cite{Schutz86, Tamanini+16} also depend on the ability to localize individual sources. Understanding parameter estimation of SBHBs will also be important to understand the outcomes and challenges of possible multiband gravitational-wave observations~\cite{Sesana16}. In this paper we focus on the latter part of the analysis problem, i.e.~the inference of the black hole parameters from the LISA data, once we have reasons to believe the presence of the signal in the data, leaving aside the identification problem. We also limit our analysis to extrinsic parameters and ignore the effect of superposed gravitational wave signals from various sources.

The inference problem amounts to producing samples from the posterior distribution for source parameters, related by Bayes theorem to the likelihood function, i.e.~the probability of observing the measured data given the source parameters and a model of the source and detector. In LISA, this problem is more complex than in kilometer-scale ground-based interferometers, as additional challenges arises from the much larger expected signal-to-noise ratios (SNRs) for high-mass systems, or the much longer duration of the waveform for low-mass systems. In particular, for MBHBs, the large SNR will require a very accurate modeling of the waveform, including the merger and ringdown regimes, the effects of spin and corrections of higher harmonics in the signal beyond the quadrupole.

The response of the LISA instrument to a given waveform shows a time- and frequency-dependence~\cite{Cutler97, Larson+99, CR02, MB18}.
Contrary to LIGO and Virgo, which are approximately fixed in an inertial frame during the observation of any merger, the LISA constellation will move around the Sun and change its orientation appreciably if the signal is observable for a significant fraction of a year, leading to additional modulations of the signals. Furthermore, due to the LISA arm length, for many sources the long-wavelength approximation will not hold, which introduces a frequency-dependence in the response. As we will see in this study, a proper treatment of these effects is necessary in order to understand degeneracies between the source parameters.

Previous parameter recovery studies with LISA used mainly simplified signal and response models, and often relied on a Fisher matrix approximation to parameter recovery instead of full Bayesian simulations. Numerous works used the combination of inspiral-only post-Newtonian waveforms (sometimes including precession), the low-frequency approximate response and Fisher matrix estimates~\cite{Cutler97, Vecchio03, Arun06, Berti+04, LangHughes06}. The Mock LISA Data Challenges~\cite{MLDC09} used such a setup for the signals and the response, albeit moving towards MCMC tools with a focus on detection. Bayesian methods going beyond Fisher, albeit still restricted to inspiral signals, were developed in~\cite{Brown+07, CornishPorter06a, Crowder+06, Wickham+06, Roever+07, Feroz+09, GairPorter09, Petiteau+09, PorterCarre13, PorterCornish15}. The importance of the higher harmonics for LISA was stressed already in several studies~\cite{Arun+07a, TriasSintes07, PorterCornish08, McWilliams+09}. To explore the importance of the merger-ringdown using Numerical relativity waveforms, the studies~\cite{Thorpe+08, McWilliams+09, McWilliams+10, McWilliams+11} used a Fisher approach, while~\cite{Babak+08} used Bayesian analyses limited to extrinsic parameters. More recently, in the context of the redesign of the LISA mission~\cite{elisa13}, the study~\cite{Klein+15} explored the performance of various LISA instrumental designs using Fisher matrix estimates and inspiral signals, but used a reweighting procedure to represent the role of the merger-ringdown signal. We also note a recent work~\cite{Baibhav+20} investigating parameter recovery for ringdown-dominated signals. To this date, no full Bayesian parameter estimation studies with IMR signals have been performed.

On the other hand, important advances have been registered in recent years in providing fast and accurate waveforms including the merger and ringdown for the LIGO/Virgo data analysis, through phenomenological waveforms~\cite{Khan+15, London+17}, Reduced Order Models (ROM) of Effective One Body Waveforms~\cite{Puerrer14, Bohe+16}, and numerical relativity surrogates~\cite{Blackman+17b, Varma+18}. This progress has yet to be transposed to LISA applications.

Here we demonstrate that standard Bayesian inference can be performed for MBHBs using a self-consistent waveform model that includes inspiral, merger, ringdown and higher-order modes (albeit no spins) and a full model of the LISA response. We investigate and explain degeneracies in the posterior distribution of the source's parameters, and highlight the crucial role played by the higher harmonics in the signal and the frequency-dependency in the instrument response.

We start by introducing a fast method to calculate the likelihood for a black hole merger. This method includes: a fast computation of a Fourier-domain inspiral-merger-ringdown waveform with higher-order modes based on a reduced order model; a fast Fourier-domain computation of the LISA response based on~\cite{MB18}; and a fast method for computing the noise-weighted product between waveforms when setting the noise realization to zero. We couple our likelihood with two codes for Bayesian inference based on different sampling techniques, constructing an inference engine which enables us to perform a variety of investigations with simulated black hole mergers.

Focusing on two examples of moderately massive black holes binaries, we perform Bayesian parameter estimation simulation for full, inspiral-merger-signals, and we show the crucial role of higher-order modes. We highlight the challenges we encounter in the sampling, and compare our two samplers with each other and with the Fisher matrix approximation. We explore degeneracies appearing between some of the parameters when ignoring these higher harmonics, and derive an analytic explanation of their origin via a simplified approximations to the instrument response. We show for our examples how the posterior distributions evolve over time as more and more data becomes available for the inference, and explain how degeneracies between different positions of the source in the sky are broken by features in the instrument response.

Finally, we repeat some of our investigations for stellar-origin black hole mergers. Parameter estimation studies for SBHB systems have relied so far on the Fisher matrix approach~\cite{Sesana16, Vitale16, Nishizawa+16a, Nishizawa+16b}, with Bayesian inference tools yet to be developed. Focusing on the posterior distribution of extrinsic parameters (fixing the masses and time to coalescence), we demonstrate that Bayesian analyses can be run with a speed for the likelihood computation that is comparable to the MBHB case, and discuss the features of the posteriors.

Section~\ref{sec:response} introduces comprehensive notations for the frequency-domain signal and response, and Section~\ref{sec:method} explains our methods for likelihood computations and Bayesian sampling. Section~\ref{sec:SMBH} presents the MBHB example signals that we analyze and their accumulation with time, together with their instrumental response in various approximations. In Section~\ref{sec:SMBHPE}, we present our main results for the parameter estimation of MBHBs; we contrast results obtained with and without higher harmonics, contrast Fisher matrix with Bayesian results, describe and explain degeneracies in parameter space. In Section~\ref{sec:MBHBPEacctime}, we turn to the refinement of parameter estimation as the signal accumulates with ongoing observations, describing how multiple inferred sky positions can coexist before features of the full instrument response break the degeneracies. In Section~\ref{sec:SBHB}, we turn to the application of our fast likelihood computation to investigate extrinsic parameter inference of SBHBs. We summarize and discuss our findings in Section~\ref{sec:summarydiscussion}.


\section{LISA instrument response}
\label{sec:response}

In this section, we introduce a complete set of notations for the LISA instrument response that will be useful in the later discussion of degeneracies in parameter space. We will use units with $G=c=1$ and assume the proposed armlength for LISA $L = 2.5\mathrm{Gm}$~\cite{LISA2017}.


\subsection{The GW signal}
\label{sec:gwsignal}

To describe the gravitational wave signal, we first need to define a conventional source frame associated to the binary system, as detailed in App.~\ref{app:conventions}. The gravitational wave propagation vector $k$, pointing from the source to the observer, has polar angles $(\iota, \varphi)$ in this source frame. Next, one introduces polarization vectors $p$, $q$ so that $(p, q, k)$ is a direct triad. The precise choice of $(p,q)$ is a matter of convention, for which we refer to App.~\ref{app:conventions}.

The gravitational waveform in the transverse-traceless gauge is described by the two polarizations $h_{+}$, $h_{\times}$. If $H_{ij} = h_{ij}^{\rm TT}$ represents the gravitational wave signal in matrix form,
\be
	H = h_{+} P_{+} + h_{\times} P_{\times}
\ee
with the polarization tensors
\bsub
\begin{align}
	P_{+} &= p \otimes p - q \otimes q \,,\\
	P_{\times} &= p \otimes q + q \otimes p \,.
\end{align}
\esub
Conversely, the polarizations are
\bsub
\begin{align}
	h_{+} &= \frac{1}{2} \left(p \otimes p - q \otimes q \right) : H \,,\\
	h_{\times} &= \frac{1}{2} \left( p \otimes q + q \otimes p \right) : H \,,
\end{align}
\esub
with the notation $A : B = A_{ij}B_{ij}$.

One can further decompose the gravitational wave signal, seen as a function of the direction of emission $(\iota, \varphi)$ in the source frame, in spin-weighted spherical harmonics~\cite{Goldberg+67} as
\be\label{eq:hpcmodes}
	h_{+} - i h_{\times} = \sum_{\ell \geq 2} \sum_{m = -\ell}^{\ell} {}_{-2}Y_{\ell m} (\iota, \varphi) h_{\ell m} \,.
\ee
Explicit expression of the ${}_{-2}Y_{\ell m}$ can be found in e.g. Sec.~3 of~\cite{BlanchetLiving}. In the following, we will use exclusively this parametrization of the waveform as a set of modes $h_{\ell m}$. The dominant harmonic is $h_{22}$, while the others are called higher modes (HM) or higher harmonics.

We will both generate the waveforms and apply the response directly in the Fourier domain. Our convention for the Fourier transform of a function $F$ is\footnote{Note that this differs by a change $f\rightarrow -f$ from the more usual convention used e.g. in \texttt{LAL}~\cite{lal}.}
\be\label{eq:defFourier}
	\tilde{F}(f) = \int dt \; e^{2 i \pi f t} F(t) \,.
\ee


\subsection{Mode decomposition and polarization angle}
\label{sec:modespol}

We now translate the mode decomposition~\eqref{eq:hpcmodes} in the Fourier domain. We have
\bsub
\begin{align}
	h_{+} = \frac{1}{2} \sum_{\ell, m} \left( \sYlm h_{\ell m} + \sYlmstar h_{\ell m}^{*} \right) \,,\\
	h_{\times} = \frac{i}{2} \sum_{\ell, m} \left( \sYlm h_{\ell m} - \sYlmstar h_{\ell m}^{*} \right) \,,
\end{align}
\esub
which is valid in general (we dropped the $(\iota, \varphi)$ arguments of the $\sYlm$). Now, for non-precessing binary systems, an exact symmetry relation between modes reads
\be\label{eq:nonprecsymmetry}
	h_{\ell, -m} = (-1)^{\ell} h_{\ell, m}^{*} \,.
\ee
Using this symmetry, we can write
\be
	h_{+,\times} = \sum_{\ell, m} K_{\ell m}^{+, \times} h_{\ell m} \,,
\ee
with
\bsub
\begin{align}
	K_{\ell m}^{+} =\frac{1}{2} \left( \sYlm + (-1)^{\ell} \sYlminusmstar \right) \,,\\
	K_{\ell m}^{\times} = \frac{i}{2} \left( \sYlm - (-1)^{\ell} \sYlminusmstar \right) \,.
\end{align}
\esub

Going to the Fourier domain, an approximation often used is to neglect support for negative/positive frequencies according to
\be\label{eq:approxnegf}
	\tilde{h}_{\ell m} (f) \simeq 0 \;\; \text{for} \;\; m<0, \; f>0 \;\; ( m>0, \; f<0 )\,,
\ee
and neglecting modes $h_{\ell 0}$. We will use this approximation throughout this paper. Note that we picked our Fourier convention~\eqref{eq:defFourier} to ensure that modes $\tilde{h}_{\ell m} (f)$ with $m>0$ have support for $f>0$. Using~\eqref{eq:approxnegf}, for $f>0$ we have
\be
	\tilde{h}_{+,\times} = \sum_{\ell} \sum_{m>0} K_{\ell m}^{+, \times} \tilde{h}_{\ell m} \,.
\ee

Next, it is convenient to introduce mode-by-mode polarization matrices
\be
	P_{\ell m} = P_{+} K_{\ell m}^{+} + P_{\times} K_{\ell m}^{\times} \,,
\ee
so that
\be\label{eq:Hsummodes}
	H = \sum_{\ell, m} P_{\ell m} h_{\ell m} \,.
\ee

The polarization angle $\psi$ can be seen (see App~\ref{app:conventions}) as a degree of freedom in the relation between the source frame and the detector frame, parametrizing a rotation around the wave vector $k$. To define this angle, one introduces reference vectors $(u,v)$ orthogonal to $k$, that define the zero of the polarization angle as $(p,q)(\psi=0) = (u,v)$. Our convention for $(u,v)$ is detailed in App~\ref{app:conventions}.

To make explicit the dependence in polarization, we can define polarization tensors for 0 polarization angle as
\bsub
\begin{align}
	P_{+}^{0} &= P_{+}(\psi = 0) = u \otimes u - v \otimes v \,,\\
	P_{\times}^{0} &= P_{\times}(\psi = 0) = u \otimes v + v \otimes u \,.
\end{align}
\esub
This allows to write the dependence in polarization as
\be
	P_{+} + i P_{\times} = e^{-2 i \psi} \left( P_{+}^{0} + i P_{\times}^{0} \right) \,,
\ee
or explicitly in $P_{\ell m}$ as
\begin{align}
	P_{\ell m} (\iota, \varphi, \psi) &= \frac{1}{2} \sYlm(\iota, \varphi) e^{-2 i \psi} \left( P_{+}^{0} + i P_{\times}^{0} \right) \nn\\
	& + \frac{1}{2} (-1)^{\ell} \sYlminusmstar (\iota, \varphi) e^{+2 i \psi} \left( P_{+}^{0} - i P_{\times}^{0} \right) \,.
\end{align}
Writing the $P_{\ell m}$ matrices in this way allows us to factor out explicitly all dependencies in the extrinsic parameters $(\iota, \phi, \psi)$. Together with the luminosity distance $D$ scaling the overall amplitude of the signal, these parameters enter as constant (time and frequency independent) prefactors in the response for each mode $h_{\ell m}$.

Since $\psi$ always appears with a factor of 2, it has an exact $\pi$-degeneracy, and we choose the convention of restricting $\psi \in [0,\pi]$.


\subsection{Frequency domain LISA response}
\label{subsec:FDresponse}

The LISA response~\cite{EW75, Cutler97, Larson+99, CR02, RCP04} can be built from single-link observables $y_{slr} = (\nu_{r} - \nu_{s})/\nu$ representing a laser frequency shift between the transmitting spacecraft $s$ and the receiving spacecraft $r$ along the link $l$. We use the expression~\cite{Vallisneri04, Krolak+04}
\be\label{eq:defyslr}
	y_{slr} = \frac{1}{2} \frac{n_{l} \otimes n_{l}}{1 - k\cdot n_{l}} : \left[ H(t - L - k\cdot p_{s}) - H(t - k\cdot p_{r}) \right] \,,
\ee
with $H$ the transverse-traceless matrix representing the gravitational wave, $k$ the wave propagation vector, $L$ the delay along one arm, taken to be fixed, $n_{l}$ the link unit vectors (from $s$ to $r$) and $p_{s}$, $p_{r}$ the positions of the spacecrafts. Here $n_{l}$, $p_{s}$ and $p_{r}$ are evaluated at the same time $t$. In the following, we will use interchangeably the notation $y_{sr}$ instead of $y_{slr}$, since $n_{l}$ can be deduced from the sending and receiving indices $s,r$.

Our response formalism applies to waveforms which can be represented as a combination of harmonics with
slowly varying amplitude and phase as\footnote{Note our Fourier convention~\eqref{eq:defFourier}.}
\be\label{eq:hlmampphase}
	\tilde{h}_{\ell m} (f) = A_{\ell m} (f) e^{-i\Psi_{\ell m} (f)}\,.
\ee
We will use the analysis of~\cite{MB18} and write the response in individual observables $y_{slr}$ with a transfer function for each spherical harmonic mode as
\be
	\tilde{y}_{slr} = \sum_{\ell, m}\calT_{slr}^{\ell m}(f) \tilde{h}_{\ell m} \,.
\ee
Applying the perturbative formalism of~\cite{MB18} to leading order in the separation of timescales, we have simply
\be\label{eq:Tlmslr}
	\calT_{slr}^{\ell m}(f) = G_{slr}^{\ell m}(f, t_{f}^{\ell m}) \,,
\ee
where
\begin{align}\label{eq:Gslr}
	G_{slr}^{\ell m}(f,t) &= \frac{i \pi f L}{2} \sinc \left[ \pi f L\left(1-k\cdot n_{l} \right) \right] \nn\\
	& \quad \cdot \exp\left[ i \pi f \left( L + k\cdot \left( p_{r} + p_{s} \right) \right) \right] \; n_{l} \cdot P_{\ell m} \cdot n_{l} \,,
\end{align}
is the kernel from~\cite{MB18} with $n_{l}$, $p_{s}$, $p_{r}$ evaluated at $t$, with $P_{\ell m}$ defined by the decomposition~\eqref{eq:Hsummodes}. In~\eqref{eq:Tlmslr},
\be\label{eq:deftflm}
	t_{f}^{\ell m} = -\frac{1}{2\pi} \frac{d\Psi_{\ell m}}{df}
\ee
is the effective time-frequency correspondence, defined across the whole frequency band and including the merger and ringdown. This definition generalizes the Stationary Phase Approximation (SPA).

The analysis of~\cite{MB18} has shown that higher-order corrections in the separation of timescales in the LISA response are small in general for MBHB systems, and are also small for SBHB systems provided they are not too far from the coalescence. The fact that we use the same Fourier-domain treatment of the transfer functions for both the signal and the templates should also mitigate the importance of modelling errors. We will therefore limit ourselves to the leading order in the treatment of~\cite{MB18} in the rest of this paper.

In~\eqref{eq:Gslr} it is convenient to decompose the spacecraft positions as
\be\label{eq:defp0}
	p_{r,s} = p_{0} + p_{r,s}^{L} \,,
\ee
with $p_{0}$ the position of the center of the LISA constellation. This allows us to note an important feature of~\eqref{eq:Gslr}: apart from a global phase delay factor $\exp[i \pi f k\cdot p_{0}]$, frequency-dependent terms only feature the projections
\be
	k_{\parallel} \cdot n_{l}, \; k_{\parallel} \cdot p_{s,r}^{L} \,,
\ee
with $k_{\parallel}$ the projection of the wave vector $k$ in the (instantaneous) LISA plane, thus the frequency-dependent factors are invariant when we reflect $k$ across this plane. The delay $k\cdot p_{0}$ is the only one with a baseline outside the LISA plane (see~\eqref{eq:defPhiR} below). This will remain true when constructing TDI variables below\footnote{Strictly speaking it will not be true in the most general response formalism of~\cite{MB18}, where corrections are either slightly non-local in time or involve velocities, that are out of the LISA plane.}. We will investigate sky degeneracies in detail in Sections~\ref{subsec:MBHBPEdegen} and~\ref{subsec:MBHBacctimebreakdegen}.

In~\eqref{eq:defyslr} and~\eqref{eq:Gslr}, one can distinguish two types of delays: on one hand the delay associated to the position $p_{0}$ of the center of the constellation on its orbit around the Sun, with baseline $R=1 \au$, and on the other hand the delays associated to the individual spacecraft positions in the constellation, with baseline the armlength $L$. This defines the transfer frequencies
\bsub\label{eq:transferfrequencies}
\begin{align}
	f_{R} &= 1/R = 2.0\times10^{-3}\Hz \,,\\
	f_{L} &= 1/L = 0.12\Hz \,.
\end{align}
\esub
The transfer frequency $f_{L}$ correspond to fitting a full wavelength in the armlength; as we will see below in Sec.~\ref{sec:MBHBPEacctime}, departures from the long-wavelength approximation starts to be important at significantly lower frequencies.


\subsection{Time Delay Interferometry}
\label{subsec:TDIresponse}

The basic one-arm observables of Eq.~\eqref{eq:defyslr} are affected by laser noise whose amplitude is orders of magnitude larger than the astrophysical signal.
However, time delay interferometry (TDI) allows one to construct a new set of observables from delayed combinations of $y_{slr}$, where laser noise is suppressed by orders of magnitude~\cite{TintoArmstrong99, Armstrong+99, Estabrook+00, Dhurandhar+02, Tintoliving}.
Various generations of TDI schemes have been proposed in order to deal with a non-rigid and rotating LISA constellation~\cite{Shaddock03, CornishHellings03, Shaddock+03, Tinto+04}.
However, these refinements affect only marginally the response to the gravitational waves.
Hence, in this work we only consider first-generation TDI and adopt a rigid approximation for the constellation, where delays are all constant and equal to $L$.
Using the notation $y_{slr,nL} = y_{slr}(t - nL)$, the first-generation TDI Michelson observable $X$ reads~\cite{Vallisneri04}
\begin{align}
	X &= y_{31} + y_{13,L} + \left( y_{21} + y_{12,L} \right)_{,2L} \nn\\
	& - \left( y_{21} + y_{12,L} \right) - \left( y_{31} + y_{13,L} \right)_{,2L} \,,
\end{align}
with the other Michelson observables $Y$, $Z$ being obtained by cyclic permutation. Uncorrelated combinations $A$, $E$ and $T$~\cite{Prince+02} are then expressed as
\bsub\label{eq:defAET}
\begin{align}
	A &= \frac{1}{\sqrt{2}} \left( Z - X \right) \,,\\
	E &= \frac{1}{\sqrt{6}} \left( X - 2Y + Z \right) \,,\\
	T &= \frac{1}{\sqrt{3}} \left( X + Y + Z \right) \,.
\end{align}
\esub
These channels are independent under the assumption of an identical and uncorrrelated noise in the detector arms. Note that various conventions coexist in the literature.
With constant delays in the rigid approximation, and using the notation $z\equiv \exp[2i\pi fL]$, the TDI combinations in the frequency domain take the form
\bsub\label{eq:aet}
\begin{align}
	\tilde{a} &= (1+z) \left( \tilde{y}_{31} + \tilde{y}_{13} \right) - \tilde{y}_{23} - z \tilde{y}_{32} - \tilde{y}_{21} - z \tilde{y}_{12} \,,\\
	\tilde{e} &= \frac{1}{\sqrt{3}} \left[ (1-z)\left( \tilde{y}_{13} - \tilde{y}_{31} \right) + (2+z) \left( \tilde{y}_{12} - \tilde{y}_{32} \right) \right. \nn\\
	&\qquad\quad \left. + (1+2z) \left( \tilde{y}_{21} - \tilde{y}_{23} \right) \right] \,,\\
	\tilde{t} &= \frac{\sqrt{2}}{\sqrt{3}} \left[ \tilde{y}_{21} - \tilde{y}_{12} + \tilde{y}_{32} - \tilde{y}_{23} + \tilde{y}_{13} - \tilde{y}_{31} \right] \,,
\end{align}
\esub
where we have eliminated frequency-dependent prefactors that are common to the signal and to the noise by introducing the rescalings
\bsub\label{eq:scalingAET}
\begin{align}
	\tilde{a}, \tilde{e} &= \frac{e^{-2i\pi fL}}{i \sqrt{2} \sin (2\pi f L)}\times \tilde{A}, \tilde{E} \,,\\
	\tilde{t} &= \frac{e^{-3i\pi fL}}{2\sqrt{2}  \sin (\pi f L) \sin (2\pi f L)} \times \tilde{T} \,.
\end{align}
\esub
We will use mode-by-mode transfer functions for these reduced channels as
\be\label{eq:defTlmaet}
	\tilde{a}, \tilde{e}, \tilde{t} = \sum_{\ell, m} \calT_{a,e,t}^{\ell m} \tilde{h}_{\ell m}\,.
\ee
To make the connection between these reduced TDI observables and the gravitational strain, more familiar in the context of ground-based intruments, we also introduce the notations
\bsub\label{eq:defhaet}
\begin{align}
	\tilde{h}_{a,e,t} \equiv \frac{1}{(-6i \pi f L)} \times \tilde{a}, \tilde{e}, \tilde{t} \,, \\
	\calT_{h_{a}, h_{e}, h_{t}}^{\ell m} = \frac{1}{(-6i \pi f L)} \calT_{a,e,t}^{\ell m}\,.
\end{align}
\esub

Scaling out the same square factors from the noise power spectral density (PSD) as
\bsub
\begin{align}
	S_{n}^{A}, S_{n}^{E} &= 2 \sin^{2} (2\pi f L) \times S_{n}^{a}, S_{n}^{e} \,,\\
	S_{n}^{T} &= 8 \sin^{2} (\pi f L) \sin^{2} (2\pi f L) S_{n}^{t} \,,
\end{align}
\esub
the reduced PSD for the three channels take the form
\bsub\label{eq:Snaet}
\begin{align}
	S_{n}^{a} = S_{n}^{e} &= 2 \left( 3 + 2\cos (2\pi f L) + \cos (4 \pi f L) \right) S^{\rm pm}(f) \nn\\
	& \quad + \left( 2 + \cos (2\pi f L) \right) S^{\rm op}(f) \,,\\
	S_{n}^{t} &= 4 \sin^{2} (2\pi f L) S^{\rm pm}(f) + S^{\rm op}(f) \,,
\end{align}
\esub
with $S^{\rm pm}$ the test-mass noise PSD and $S^{\rm op}$ the optical noise PSD. We also include a confusion noise coming from the background of galactic binaries in the LISA band that is added to the instrumental noise. For the instrument performance defining these noise levels, we take values from~\cite{LISANoiseBudget16} (see App.~\ref{app:LISAnoise}). We can define a strain-like noise PSD associated to the strain-like TDI observables~\eqref{eq:defhaet} as
\be\label{eq:defShaet}
	S_{h}^{a,e,t} (f) = \frac{S_{n}^{a,e,t}(f)}{(6 \pi f L)^{2}}\,.
\ee

The prefactors~\eqref{eq:scalingAET} are oscillatory and have zero-crossings at high frequencies, with the first one occuring at $f_{L}/2 = 0.06 \mathrm{Hz}$, which is why it is convenient to factor them out to avoid $0/0$ numerical instabilities. This treatment would not apply directly with a more realitic model for LISA with varying arm-lengths and residual laser noise. In that case imperfect cancellations in the vicinity of the zero-crossings would likely result in localized loss of sensitivity.


\subsection{The low-frequency limit}
\label{subsec:lowfresponse}

Though our parameter estimation calculation primarily applies the full LISA response, it will be useful to consider some some simplifying asymptotic limits to understand parameter degeneracies.

As is well known~\cite{Cutler97}, in the low-frequency limit (also called the long-wavelength approximation), the finite-armlength effects vanish, and the response of LISA is analogous to the response of of two LIGO-type detectors rotated from each other by $\pi/4$ and set in motion.

For $f \ll f_{L}$, we have $2\pi f L \ll 1$ and the kernel~\eqref{eq:Gslr} reduces to
\be\label{eq:Gslrlowf}
	G^{\ell m}_{slr} \simeq \frac{i \pi f L}{2} \exp\left[ 2 i \pi f k\cdot p_{0} \right] n_{l} \otimes n_{l} : P_{\ell m}\,.
\ee

The exponential factor is a delay phase, for which we introduce the notation:
\be\label{eq:defPhiR}
	\Phi_{R} \equiv 2 \pi f k\cdot p_{0} \,, \quad \Delta \Phi_{R} = \Phi_{R} - \Phi_{R}(t=t_{f}^{\rm peak}) \,.
\ee
The quantity $\Phi_{R}$ is often called the Doppler phase in the literature; it may appear to be large for $f \gg f_{R}$, but we should remember that it corresponds in part simply to the fixed delay between the time of arrival at the SSB and the time of arrival at the LISA constellation. In the limit of short-lived coalescence signals, only the Doppler phase variation $\Delta \Phi_{R}$ carries useful information about the sky position.

For $2\pi f L \ll 1$, we have $z\simeq 1$, and the link reversal symmetry $\tilde{y}_{slr} \simeq \tilde{y}_{r-ls}$, so that~\eqref{eq:aet} becomes (dropping $l$ indices and symmetrizing $r,s$)
\bsub\label{eq:aetlowf}
\begin{align}
	\tilde{a} &\simeq 4\tilde{y}_{31} - 2 \tilde{y}_{23} - 2 \tilde{y}_{12} \,,\\
	\tilde{e} &\simeq 2\sqrt{3} \left[ \tilde{y}_{12}  - \tilde{y}_{23} \right] \,,\\
	\tilde{t} &\simeq 0 \,,
\end{align}
\esub
with the $T$-channel becoming negligible in this limit. Using~\eqref{eq:Gslrlowf}, we can write
\be\label{eq:aelowfmodes}
	\tilde{a}, \tilde{e} = (-2i\pi f) \exp\left[ 2 i \pi f k\cdot p_{0} \right] \sum_{\ell, m>0} \tilde{h}_{\ell m} D_{a,e} : P_{\ell m} \,,
\ee
where we introduced the detector tensors
\bsub\label{eq:DaDe}
\begin{align}
	D_{a} &= \frac{L}{2} \left( n_{1}\otimes n_{1} + n_{3} \otimes n_{3} - 2 n_{2} \otimes n_{2} \right) \,,\\
	D_{e} &= \frac{L\sqrt{3}}{2} \left( n_{1} \otimes n_{1} - n_{3} \otimes n_{3} \right) \,.
\end{align}
\esub
Here, we have made apparent factors $(-2i\pi f)$, which correspond to a time derivative in the Fourier-domain.
The observables $\tilde{a}$, $\tilde{e}$ are therefore analogous to time derivatives of the strain commonly used for ground-based detectors.

We can now map these two channels to two fictitious LIGO-type detectors as follows. For an orthogonal detector of the same setup as the ground-based LIGO and Virgo with armlength $L_{D}$, rotated by an angle $\epsilon_{D}$ from the basis vectors $x,y$, the detector tensor is
\begin{align}
	D &= \frac{L_{D}}{2} \left[ \cos 2\epsilon_{D} \left( x\otimes x - y\otimes y \right) \right. \nn\\
	& \qquad\quad + \left. \sin 2\epsilon_{D} \left( x\otimes y + y \otimes x\right) \right] \,.
\end{align}
By comparison with~\eqref{eq:DaDe}, using the LISA frame as detector frame (see App.~\ref{app:conventions}) we obtain
\be
	L_{a} = L_{e} = 3L \,, \; \epsilon_{a} = \frac{2\pi}{3} \,, \; \epsilon_{e} = \frac{5\pi}{12} \,.
\ee
Note a degeneracy $\pm \pi$ in $\epsilon_{a,e}$, reflecting a freedom in the choice of the orientation of these effective detectors. The effective armlength $3L$ is a matter of convention, since it depends on an arbitrary overall scaling in~\eqref{eq:defAET}.

Next, we factor out the effective length, defining
\be\label{eq:defFapcFepc}
	F_{a,e}^{+,\times} \equiv \frac{1}{3L} D_{a,e} : P_{+,\times}^{0} \,,
\ee
which gives in terms of the LISA-frame sky position angles $\lambdaL$, $\betaL$ (see App.~\ref{app:LISAframe}):
\bsub\label{eq:FapcFepc}
\begin{align}
	F_{a}^{+} &= \frac{1}{2} \left( 1 + \sin^{2}\beta_{L} \right) \cos \left(2\lambda_{L} - \frac{\pi}{3} \right) \,,\\
	F_{a}^{\times} &= \sin\beta_{L} \sin \left(2\lambda_{L} - \frac{\pi}{3} \right) \,,\\
	F_{e}^{+} &= \frac{1}{2} \left( 1 + \sin^{2}\beta_{L} \right) \cos \left(2\lambda_{L} + \frac{\pi}{6} \right) \,,\\
	F_{e}^{\times} &= \sin\beta_{L} \sin \left(2\lambda_{L} + \frac{\pi}{6} \right)  \,.
\end{align}
\esub
These functions are the familiar pattern functions of ground-based detectors (for $\psi_{L}=0$) (see e.g.~\cite{FlanaganHughes05}).
One can check that the expressions for the channel $e$ are obtained from the expressions for $a$ with the replacement $\lambdaL \rightarrow \lambdaL + \pi/4$.

If we introduce
\begin{align}\label{eq:Flmae}
	F_{a,e}^{\ell m} &= \frac{1}{2} \sYlm e^{-2 i \psi_{L}} \left( F_{a,e}^{+} + i F_{a,e}^{\times}\right) \nn\\
	& + \frac{1}{2} (-1)^{\ell} \sYlminusmstar e^{+2 i \psi_{L}} \left( F_{a,e}^{+} - i F_{a,e}^{\times} \right) \,,
\end{align}
the mode transfer functions~\eqref{eq:defTlmaet} become
\begin{align}\label{eq:transferlowfae}
	\tilde{a}, \tilde{e} &= \sum_{\ell,m} \calT^{\ell m}_{a,e} \tilde{h}_{\ell m} \nn\\
	&= \left(-6 i \pi f L\right) \exp[2i \pi f k\cdot p_{0}] \sum_{\ell,m} F_{a,e}^{\ell m} \tilde{h}_{\ell m}\,.
\end{align}
We see that scaling out $(-6 i \pi fL)$ as in~\eqref{eq:defhaet} brings us back to strain-like observables. The only difference with the response of a ground-based observatory is then the time dependency entering $p_{0}(t)$ and the LISA frame angles $\lambda_{L}(t)$, $\beta_{L}(t)$, $\psi_{L}(t)$ (see~\eqref{eq:Lframeangles}), with the time evaluated at $t_{f}^{\ell m}$.

Finally, we note that we could include the polarization angle in the pattern functions as follows. Above we have chosen to write $\psi$ as an outside prefactor, but we could write $\psi$-dependent pattern functions as
\bsub
\begin{align}
	F^{+}_{a,e}(\lambdaL, \betaL, \psiL) &= \cos 2\psiL F^{+}_{a,e} + \sin 2\psiL F^{\times}_{a,e} \,,\\
	F^{\times}_{a,e}(\lambdaL, \betaL, \psiL) &= -\sin 2\psiL F^{+}_{a,e} + \cos 2\psiL F^{\times}_{a,e} \,.
\end{align}
\esub
so that in the low-frequency approximation the response for the strain-like observables~\eqref{eq:defhaet} is
\be
	\tilde{h}_{a,e} = F^{+}_{a,e}(\lambdaL, \betaL, \psiL) \tilde{h}_{+} + F^{\times}_{a,e}(\lambdaL, \betaL, \psiL) \tilde{h}_{\times} \,,
\ee
if we ignore the delay $k \cdot p_{0}$ and treat the angles as constants. This is the more familiar form of the instrument response used for ground-based detectors.



\section{Methodology}
\label{sec:method}


\subsection{Bayesian setting}
\label{subsec:bayes}

Introducing the standard matched-filter inner product~\cite{FlanaganHughes05} (also called overlap) as
\be\label{eq:definnerproduct}
	( \tilde{a} | \tilde{b} ) = 4 \mathrm{Re} \int_{0}^{+\infty} df \; \frac{\tilde{a}(f) \tilde{b}^{*}(f)}{S_{n}(f)} \,,
\ee
for stationary Gaussian noise with PSD $S_{n}$, the likelihood $\calL = p(d | \bm{\theta})$ for a given gravitational wave channel takes the form
\be\label{eq:deflnL}
	\ln \calL = -\frac{1}{2} \left( h(\bm{\theta}) - d | h(\bm{\theta}) - d \right) \,,
\ee
with the data stream being a superposition of the gravitational wave signal for the true parameters $\bm{\theta}_{0}$ and the noise realization in the experiment, $d = h(\bm{\theta}_{0}) + n$.
The posterior distribution for the physical parameters $\bm{\theta}$ given the observed data $d$ is then \be\label{eq:defBayes}
	p\left( \bm{\theta} | d \right) = \frac{p(d | \bm{\theta})p(\bm{\theta})}{p(d)}\,,
\ee
with $p(\bm{\theta})$ the priors on the parameters, and $p(d)$ the evidence.
The focus of this work is not selecting between different models, so we will treat $p(d)$ as a normalization constant and do not consider it further.

In general, the templates $h(\bm{\theta})$ used for the analysis will only be an approximation of the physical signals present in the data stream $d$.
In this work we will ignore entirely this distinction and assume that our template waveforms exactly match the astrophysical waveforms.
In other words, we will not explore the issue of systematic errors arising from incomplete modeling of compact binary merger waveforms.

In our analysis, we will simulate signals and perform Bayesian analyses of them with Eq.~\eqref{eq:deflnL} and ~\eqref{eq:defBayes} to obtain the parameter posteriors.
In doing so, we will use the so-called zero-noise approximation, i.e.~setting $n=0$.
This allows us to accelerate the likelihood computation as explained in Sec.~\ref{subsec:likelihood}.
The approximation is sufficient to explore the structure of the likelihood and the parameter degeneracies, greatly improving with respect to the Fisher matrix approach.
The extension of our analysis to include nonzero noise realizations will be discussed in future work.

The log-likelihood function that we use, with a zero noise realization, is a sum over the three independent channels $A$, $E$, $T$~\eqref{eq:defAET} rescaled according to~\eqref{eq:scalingAET}:
\begin{align}\label{eq:deflnLaet}
	\ln\calL &= -\frac{1}{2} \left[ \left( \tilde{a} - \tilde{a}_{\rm inj} | \tilde{a} - \tilde{a}_{\rm inj}  \right) + \left( \tilde{e} - \tilde{e}_{\rm inj} | \tilde{e} - \tilde{e}_{\rm inj}  \right) \right. \nn\\
	& \qquad\qquad + \left. \left( \tilde{t} - \tilde{t}_{\rm inj} | \tilde{t} - \tilde{t}_{\rm inj} \right) \right] \,,
\end{align}
with the ``$\rm inj$'' subscript indicating the simulated signals, and where the inner products $(\cdot | \cdot)$ are given by~\eqref{eq:definnerproduct} with the noise PSDs~\eqref{eq:Snaet}.


\subsection{Reduced order model for EOBNRv2HM waveforms.}
\label{subsec:ROM}

As opposed to previous studies of Bayesian parameter recovery for LISA, we wish to use full inspiral-merger-ringdown signals. Since higher harmonics will play a crucial role in our analysis, in this study we use the Effective-One-Body (EOB) waveform model \texttt{EOBNRv2HM}~\cite{Pan+11}. These waveforms are based on the EOB formalism~\cite{BD99, BD00}, and are calibrated to numerical relativity waveforms. The model is limited to non-spinning systems on quasicircular orbits, but includes a set of higher harmonics in the signal, among the most important quantitatively:
\be\label{eq:listmodes}
	(\ell, m) = (2,2) ,\, (2,1) ,\, (3,3) ,\, (4,4) ,\, (5,5) \,.
\ee
The waveforms are generated in the time domain by integrating a system of ordinary differential equations, an operation requiring up to seconds of computing time for long signals. For this reason, we developed a reduced order model (ROM) for these waveforms, \texttt{EOBNRv2HMROM}, following the methods of~\cite{Puerrer14}, enabling a much faster generation of the Fourier-domain amplitudes and phases of the modes in Eq.~\eqref{eq:hlmampphase}. As the total mass $M$ scales out of the problem, the parameter space reduces to the mass ratio $q=m_{1}/m_{2}$ only. The parameter space interpolation in this ROM is therefore significantly simpler than the models~\cite{Puerrer14, Puerrer15, Bohe+16}, which include aligned spins. The inclusion of higher harmonics, however, mandates modeling the relative dephasing of different modes. The output of the code consists of a Fourier domain amplitude and phase for each mode, sparsely sampled on $\sim 300$ frequencies. The unfaithfulness with the original waveforms, assuming Advanced LIGO noise curves~\cite{UnofficialNoiseCurves18}, is $\lesssim 10^{-4}$. The computational cost is submillisecond and smaller than other stages of our likelihood computation.

We note that the assumption of non-spinning black holes is an important limitation to a realistic assessment of intrinsic parameter uncertainties. In particular, it means that our results ignore the well-known degeneracy between mass ratio and spin~\cite{CF94, PW95, Baird+2013}, thus underestimating the uncertainty in the recovered mass ratio. By ignoring misaligned spins, we also neglect the possible effects of orbital precession on parameter degeneracies. Nevertheless, consistently including merger, ringdown and higher harmonics already represents a significant improvement over previous studies, as these features in the signal carry a lot of SNR by themselves~\cite{Babak+08, Thorpe+08, McWilliams+09, McWilliams+10, McWilliams+11, Klein+15}.
In the absence of precession, we expect intrinsic and extrinsic parameters to be weakly correlated, so that  our investigations of the extrinsic part of the likelihood should be valid for aligned spins.


\subsection{Likelihood computation with zero noise}
\label{subsec:likelihood}

Since typical Bayesian analyses will need to evaluate millions of likelihood values, the computational performance of the likelihood function is crucial. In general, both the template and the injection in~\eqref{eq:deflnL} are given as frequency series sampled with the Nyquist criterion $\Delta f = 1/(2T)$ with $T$ the duration of the signal, and the overlaps~\eqref{eq:definnerproduct} are simply computed as discrete sums over those frequency samples. Although transforming~\eqref{eq:definnerproduct} in a discrete sum is a straightforward operation, the size of the frequency series can lead to a significant cost. In the context of LIGO/Virgo observations, this has prompted the development of Reduced Order Quadratures (ROQ)~\cite{Smith+14, Canizares+14, Smith+16} to accelerate the likelihood computation. In the LISA case, the computational cost of the standard likelihood implementation depends primarily on the mass of the system and the duration of the signal. This cost can range from tens of milliseconds for MBHBs and short-lived signals (a few days), to impracticable values for SBHBs signals lasting for years and sampled at high frequency.
By setting the noise realization to zero, however, we can represent the amplitude and phase of the signals over a sparse frequency grid and obtain a large speedup.

Applying the response as in~\ref{subsec:FDresponse} mode-by-mode on the sparsely sampled amplitude and phase generated as in~\ref{subsec:ROM}, we obtain a complete representation of both the waveform and of the instrument transfer functions~\eqref{eq:defTlmaet} on a few hundred points in the Fourier domain, with the full signal being implicitly reconstructed with a standard cubic spline interpolation over frequencies. Decomposing~\eqref{eq:deflnL} as
\be\label{eq:lnLoverlapstructure}
	\ln \calL = \left( h(\bm{\theta}) | h(\bm{\theta}_{0}) \right)- \frac{1}{2} \left( h(\bm{\theta}) | h(\bm{\theta}) \right) - \frac{1}{2} \left( h(\bm{\theta}_{0}) | h(\bm{\theta}_{0}) \right) \,,
\ee
we have to compute inner products of the form $\left( h(\bm{\theta}) | h(\bm{\theta'}) \right)$. Note that individual terms in~\eqref{eq:lnLoverlapstructure} can be individually large (of the order of $\mathrm{SNR}^{2}/2$), and computing $\ln \calL$ relies on accurate cancellations between terms.

Taking the likelihood~\eqref{eq:deflnLaet} decomposed in TDI channels, decomposing further in harmonics as in~\eqref{eq:defTlmaet}, we have a sum of terms with the structure (symbolically)
\be
	\sum_{\rm chan.}\sum_{\ell m, \ell'm'}\int \frac{df}{S_{n}} \; \calT_{1, \mathrm{chan}}^{\ell m} \calT_{2, \mathrm{chan}}^{\ell' m'} A_{1}^{\ell m} A_{2}^{\ell' m'} e^{-i(\Psi_{1}^{\ell m} - \Psi_{2}^{\ell' m'})} \,,
\ee
with transfer functions $\calT$ for each TDI channel, and Fourier-domain amplitudes $A$ and phases $\Psi$ (see~\eqref{eq:hlmampphase}).

The difficulty of numerically computing such an overlap depends on the phase difference. A large phase difference causes the integrand to be very oscillatory, requiring more frequency resolution than non-oscillatory integrands. On the other hand, oscillatory integrands result in a small integral due to cancellation effects, while non-oscillatory integrands contribute the most. For a single mode, $\mathrm{SNR}$ terms of the type $(h|h)$ have a zero phase difference by construction, but including different modes generates cross-terms $(\ell, m) \neq (\ell', m')$ with a large phase difference.

Note that in a Bayesian analysis, a large fraction of the time (after burn-in) will be spent exploring signals that are rather close to the injection; with a single mode phase differences will be mostly small. In this work, we wish to include oscillatory cross-terms between modes, and we use a generic numerical treatment for the integrals~\eqref{eq:lnLoverlapstructure} and keep all terms with no further approximation. Other methods to accelerate likelihoods, applicable in the presence of noise, include heterodyning with a reference signal~\cite{Cornish10}, using a variable frequency resolution (multibanding)~\cite{Porter14, Vinciguerra+17}, and, as discussed before, ROQs~\cite{Smith+14, Canizares+14, Smith+16}. We leave for future work the generalization of our likelihood computations to accomodate for noise.

First, we built a joint sparse frequency grid suitable to represent both signals and the transfer functions. We then resample the integrand on this grid, separating prefactors (noise PSD, transfer function without the Doppler phase~\eqref{eq:defPhiR}, signal amplitude) from phases (Doppler phases, signal phase difference). We build a cubic spline for the prefactor, and a quadratic spline for the phase. We obtain the structure
\be\label{eq:fastlikestructure}
	\sum_{j \in \mathrm{grid}} \int_{f_{j}}^{f_{j+1}} df \; P_{j} (f) e^{i \phi_{j} (f)} \,,
\ee
where in each grid interval $P_{j}$ is a cubic polynomial and $\phi_{j}$ a quadratic polynomial. We can then compute the elementary integrals~\eqref{eq:fastlikestructure} using a combination of numerical methods. Integrations by parts can reduce the polynomial order in~\eqref{eq:fastlikestructure}, leaving only Fresnel integral functions to be evaluated. Numerical instabilities can occur in these integration by parts when the coefficients of the phase polynomial are very small or very large, in which case we resort to asymptotic expansions instead.

We combine this implementation of overlap computations with our fast waveform generation~\ref{subsec:ROM} and fast treatment of the Fourier-domain response~\ref{subsec:FDresponse}. Since the number of overlaps to compute in~\eqref{eq:defTlmaet} is quadratic in the number of modes, likelihoods with higher harmonics are significantly more expensive. The final likelihood cost depends also on the number of grid intervals, that is chosen to limit spline interpolation errors (Table~\ref{tab:likecost}).
\begin{table}
	\begin{tabular}{|c|c|c|c|}
		\hline
		Source type & Modes & Grid & Cost \\
		\hline
		SBHB & $(2,2)$ & $500$ & $2$ ms \\
		MBHB & $(2,2)$ & $300$ & $1.2$ ms \\
		MBHB & $(2,2),(2,1),(3,3),(4,4),(5,5)$ & $300$ & $15$ ms \\
		\hline
	\end{tabular}
	\label{tab:likecost}
	\caption{Approximate frequency grid resolution and likelihood cost for different choices of source type and number of modes in the waveform.}
\end{table}
The accuracy of our likelihood computations is typically $|\Delta \ln \calL| \lesssim 0.2$, due to the numerical errors being magnified by the required cancellations between overlaps terms in~\eqref{eq:lnLoverlapstructure}.


\subsection{Bayesian sampling}
\label{subsec:samplers}

For this exploratory work we take a ``brute force'' approach to inferring the posterior probability distribution of the parameters, without applying specific knowledge or expectations about distributions and degeneracies. Such assumptions might in fact lead to an incomplete exploration of the posterior probability in ways that would be difficult to recognize and diagnose. In order to demonstrate that the problem is tractable without critically depending on the details of the methodology, we also use two independent approaches to sample the posterior distributions: parallel tempering Markov-chain Monte Carlo (MCMC) and nested sampling.

Our MCMC code \texttt{ptmcmc}\footnote{https://github.com/JohnGBaker/ptmcmc} has been developed through several astrophysics projects, tested on several sampling problems, and modularly designed with the aim of minizing opportunities for errors with new applications. The code performs parallel tempered MCMC \cite{SwendsenWang86, LittenbergCornish09} with temperatures slowly adjusted to achieve nearly equal exchange rates among all pairs of adjacent temperatures.  Most runs here performed comparably with either 80 or 240 parallel temperature chains.
We applied a general-purpose proposal distribution composed of a weighted set of sub-proposals, including differential-evolution steps~\cite{Braak06, BraakVrugt08} (although the proposal in this case is based on each single chain, not an ensemble) and a collection of several Gaussian step draws of varying sizes scaled off the prior domain at about 0.01--1\% of its scale and weighted to favor the smaller scale proposal draws.
Effective sample sizes (ESS) were estimated from the number of post-burn-in samples, reduced by the autocorrelation length, with the burn-in size chosen to maximize the ESS. In most cases, the runs were continued until achieving ESS~$>2000$.

For comparison and verification we also computed posterior samples using the nested sampling code \texttt{bambi} \cite{Graff+11}, a variant\footnote{\texttt{bambi} comes with the additional option to train a neural network to learn the likelihood, but this was not used in the present study.} of \texttt{multinest} \cite{Feroz+09}. The nested sampling algorithm~\cite{Skilling06} evolves a set of 'live points' toward regions of high probability by iteratively resampling the lowest probability point from within an ever-narrowing region covering the set of live-point samples; in \texttt{multinest}, this region is built as a collection of overlapping ellipsoids. The computation proceeds until the estimated Bayesian evidence, or marginal probability, within the region covered by the live points is smaller than some threshold. Posterior-distributed samples are then drawn based on the set of sub-regions and sample posterior values thereby generated. In our runs we used 4000 live points.


\subsection{Fisher matrix parameter estimation}
\label{subsec:Fisher}

For comparison with our Bayesian inference results we also compute estimates of parameter uncertainties using the Fisher information matrix, which despite its limitations has been the common workhorse of LISA science studies to date \cite{Vallisneri08}. For measurements with additive Gaussian noise, as assumed here, the Fisher information matrix can be computed by
\be
	F_{ij} = \left( \partial_{i} h | \partial_{j} h \right),
\ee
where $\partial_i$ is the derivative with respect to component $i$ of the parameter vector $\bm{\theta}$. In the Fisher matrix approach, the likelihood is approximated as
\be
	\ln \calL \simeq -\frac{1}{2} F_{ij} \Delta \theta^{i} \Delta \theta^{j} \,,
\ee
with $\Delta \theta^{i}$ the parameter deviation from the true signal. The inverse of the Fisher matrix, $\Sigma = F^{-1}$, is the Gaussian covariance matrix that can be used as an estimate of the true parameter uncertainties.

We compute the derivatives by second-order finite differences, with the signal inner-products expanded and computed before differencing. For these calculations, where numerical smoothness is more of a concern than speed, we compute the inner products~\eqref{eq:definnerproduct} explicitly on a fine grid. Generally, finite difference Fisher matrices can be problematic, with small numerical defects potentially having outsized impact, so the finite-difference step size $\epsilon_i$ used in each parameter derivative $\partial_{i}h$ must be chosen carefully. We target the step-size to be scaled off the diagonal Fisher matrix elements by ${\epsilon_i}\approx\delta/\sqrt{F_{ii}}$ using $\delta=0.001$ for the results shown here.
To achieve this, we begin with an initial choice of $\epsilon_i$ scaled off the parameter values or prior widths, then we estimate the Fisher diagonals and iterate until convergence.
In this process we also impose the constrain that the step is never larger than $10^{-10}$ times the initial scaling.


\section{Massive black holes signals}
\label{sec:SMBH}


\subsection{Signals and transfer functions}
\label{sec:signaltransfer}

We focus our analysis on a single choice of mass parameters, representative of canonical MBHB sources expected for LISA~\cite{Barausse12, Klein+15, LISA2017}: we pick a total redshifted mass $M=m_{1}+m_{2} = 2\times10^{6} M_{\odot}$ and a mass ratio $q=m_{1}/m_{2} = 3$, and place the source at a redshift $z=4$\footnote{We convert between luminosity distances and redshifts using the cosmological parameters of~\cite{Planck15param}}. The corresponding source-frame total mass is $M/(1+z) = 4\times10^{5} M_{\odot}$. The inclination plays an important role in deciding the importance of higher harmonics; we set it to the value $\iota = \pi/3$. We will study two such systems in detail, named System I and System II, stemming from a series of twelve runs with randomized orientation parameters, and chosen to exemplify qualitative differences in the parameter estimation. The parameters are summarized in Table~\ref{tab:MBHBparams}, along with the SNR of these signals with and without higher harmonics. The randomized orientations and the inclusion of higher harmonics both have a sizeable impact on the SNR.

We note that, by contrast with current ground-based observations that are horizon-limited and that have a strong selection bias towards systems with favorable orientations (face-on or face-off), the LISA horizon for MBHB systems extends to essentially the entirety of the observable universe~\cite{LISA2017}; it is therefore natural to consider randomized orientations. Here, although we focus on just two systems, we take $\iota = \pi/3$, the median value when inclination is randomized.

\begin{table}
	\begin{tabular}{|c||c|c|}
		\hline
		Identifier 		& I 	& II   \\
		\hline
		\hline
		Mass 1 ($\Msol$) 	& \multicolumn{2}{|c|}{$1.5\times10^{6}$} \\
		\hline
		Mass 2 ($\Msol$) 	&  \multicolumn{2}{|c|}{$0.5\times10^{6}$}\\
		\hline
		Source-frame Mass 1 ($\Msol$) 	& \multicolumn{2}{|c|}{$3\times10^{5}$} \\
		\hline
		Source-frame Mass 2 ($\Msol$) 	&  \multicolumn{2}{|c|}{$1\times10^{5}$}\\
		\hline
		Redshift 		&  \multicolumn{2}{|c|}{$4.$}  \\
		\hline
		Lum. Distance (Mpc) 		&  \multicolumn{2}{|c|}{$36594.3$} \\
		\hline
		Inclination (rad) 		& \multicolumn{2}{|c|}{$\pi/3$}  \\
		\hline
		Phase (rad) 			& $2.140$	& $-1.249$ \\
		\hline
		Ecliptic longitude (rad) 			& $3.335$	& $2.275$ \\
		\hline
		Ecliptic latitude (rad) 			& $1.468$	& $-1.376$ \\
		\hline
		Polarization (rad) 			& $2.237$	& $1.635$ \\
		\hline
		\hline
		LISA SNR $h_{22}$			& 857.4 & 645.7 \\
		\hline
		LISA SNR $h_{\ell m}$			& 944.8 & 666.0 \\
		\hline
	\end{tabular}
	\caption{Parameters of the simulated MBHB mergers. Angles are given in the SSB-frame. The $\mathrm{SNR}$ labelled $h_{\ell m}$ is obtained including all harmonics listed in~\eqref{eq:listmodes}.}
	\label{tab:MBHBparams}
\end{table}

\begin{figure}
  \centering
  \includegraphics[width=.99\linewidth]{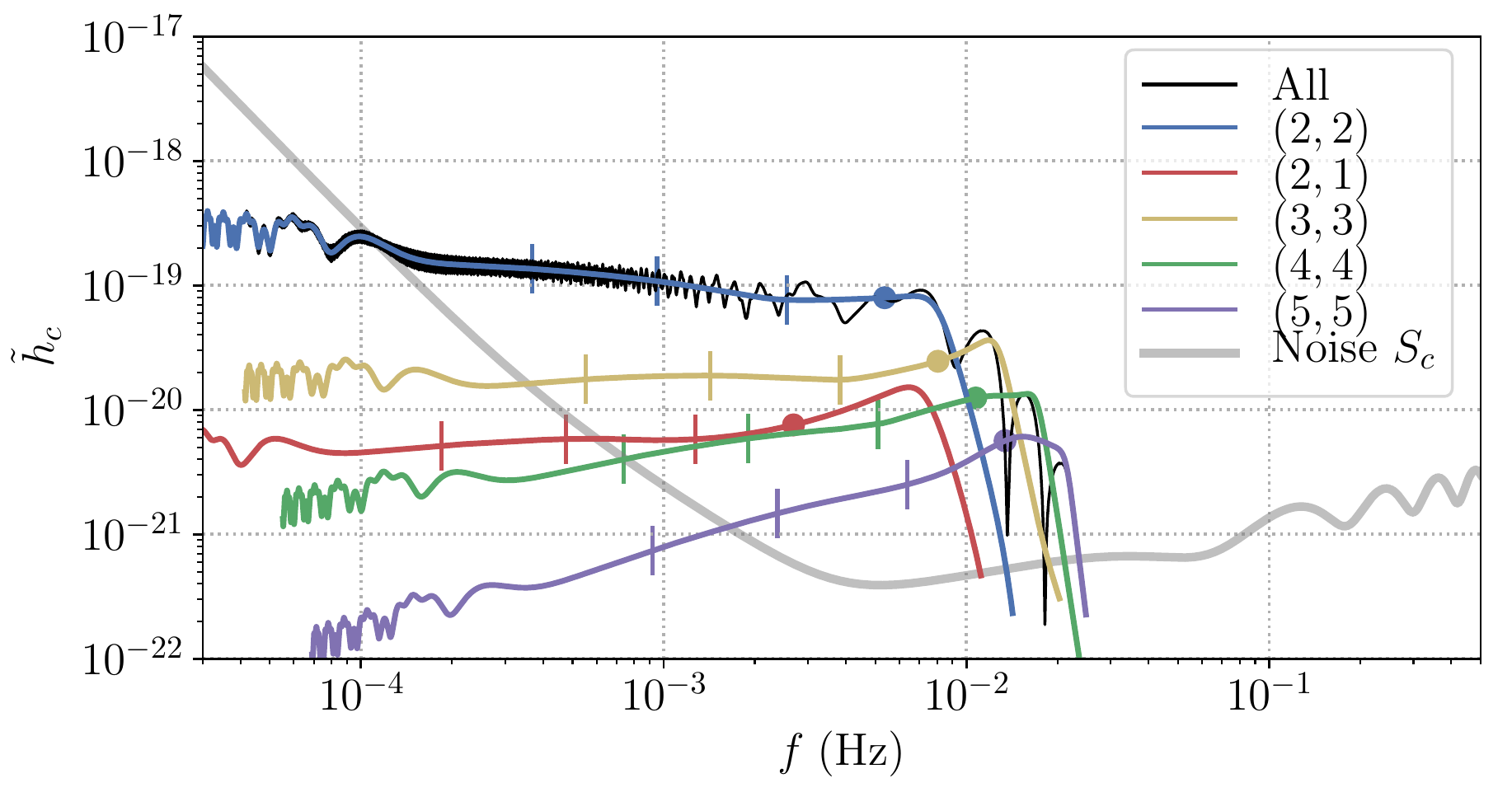}
  \caption{Characteristic strain signal~\eqref{eq:defhc}, shown here for the TDI channel $A$ of System II, decomposed as a set of $(\ell,m)$ contributions from the modes available in the waveform model. The grey curve shows the characteristic noise PSD~\eqref{eq:defhc}. Vertical bars indicate for each mode the frequencies corresponding to the time-to-merger cuts marking the points where (1/64,1/16,1/4) of the final SNR has accumulated as in our pre-merger analysis as described in~\ref{subsec:MBHBacctime}, while the dots mark the merger frequency.}
  \label{fig:hctdiahm}
\end{figure}

\begin{figure*}
  \centering
  \includegraphics[width=.98\linewidth]{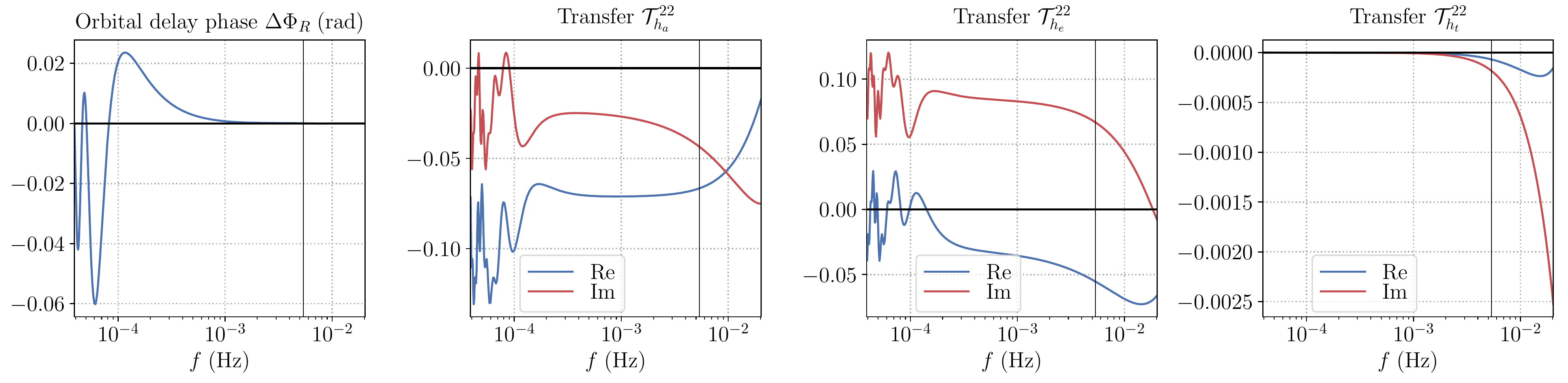}
  \caption{Example of instrument response for System II in Table~\ref{tab:MBHBparams}. The first panel isolates the Doppler phase $\Delta\Phi_{R}$ as defined in~\eqref{eq:defPhiR}, while the three other panels show the TDI transfer functions for $A,E,T$. We show here the rescaled transfer functions for the mode 22 as defined below~\eqref{eq:defTlmaet}. The vertical lines correspond to the frequency at merger.}
  \label{fig:tranferSMBHCase9}
\end{figure*}

We illustrate the Fourier-domain signal in Fig.~\ref{fig:hctdiahm}, with mode-by-mode contributions to the characteristic strain, as well as the characteristic noise PSD, defined to be
\bsub\label{eq:defhc}
\begin{align}
	\tilde{h}_{a,e,t}^{c} (f) &= 2 f \tilde{h}_{a,e,t} (f) \,,\\
	S^{a,e,t}_{c} (f) &= f S^{a,e,t}_{h} (f) \,,
\end{align}
\esub
with the strain-like observables and noise PSD defined in~\eqref{eq:defhaet} and~\eqref{eq:defShaet}.
We only show the $A$ TDI channel, as the $E$ channel is qualitatively similar and the $T$ channel is negligible at low frequencies.
Although the individual harmonics have fairly smooth amplitudes as a function of frequency, the full signal shows strong oscillations caused by the beating between the harmonics.
The effect of the LISA response can be seen in the amplitude oscillations at low frequency, which are caused by the modulation resulting from the LISA motion.

Fig.~\ref{fig:tranferSMBHCase9} focuses on the details of the transfer functions.
The Doppler phase variation $\Delta \Phi_{R}$~\eqref{eq:defPhiR} is a small effect at low frequencies, being suppressed by a factor $\sim 2\pi f R$; it is also small at higher frequencies, because the chirp happens over a short time interval during which $\Phi_{R}$ does not have time to vary. We also show transfer functions for the mode $h_{22}$ in the three TDI channels $A,E,T$, ($T$ is suppressed at low frequencies), after factoring out the Doppler phase. The low-frequency features show the time-dependency of the response created by the LISA motion, while frequency-dependency in the response appears at high frequencies.
Sec.~\ref{subsec:MBHBresponseapprox} will further investigate these different physical effects in the response.


\subsection{Accumulation of signal with time}
\label{subsec:MBHBacctime}

\begin{figure}
   \includegraphics[width=.99\linewidth]{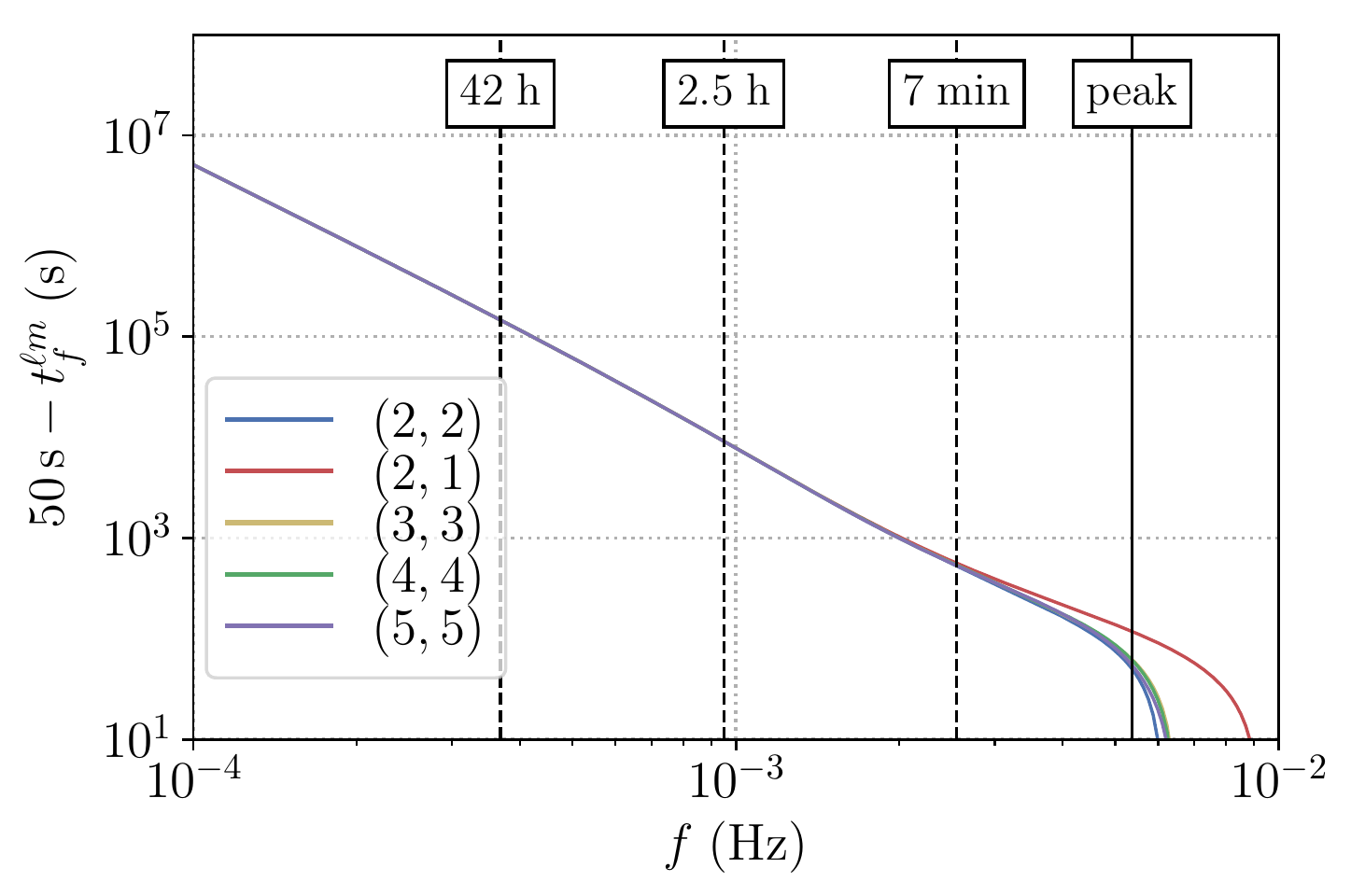}
  \caption{Relation between time and frequency for MBHB systems I and II with $M=2\times 10^6 \; M_{\odot}$, $q=3$ and $z=4$. The different lines correspond to the $(\ell, m)$ harmonics available in the waveform model. The time-to-frequency correspondence is computed according to $t_{f}^{\ell m}$ defined in~\eqref{eq:deftflm}, evaluated at $m f / 2$ to map to the same time according to~\eqref{eq:fscalinglm}. The vertical lines represent the time cuts explained in~\ref{subsec:MBHBacctime} and the merger frequency.}
  \label{fig:t-f-relation-lm}
\end{figure}

\begin{figure*}
  \centering
  \includegraphics[width=.98\linewidth]{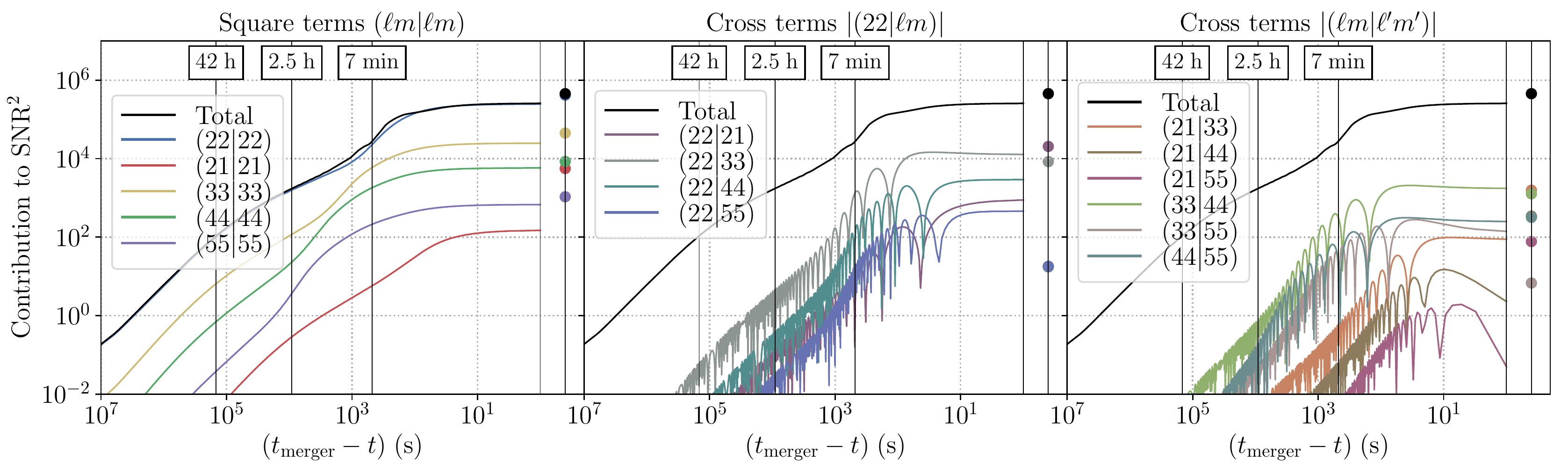}
  \caption{Cumulative contribution to $\mathrm{SNR}^{2}$ as defined in~\eqref{eq:defcontribSNR2} for system II. The left panel shows inner products of modes with themselves, the middle panel shows cross-terms involving $h_{22}$, while the right panel shows other cross-terms. Note that oscillatory cross-terms change sign, we show their absolute value. In all panels the black line represents the total. The horizontal axis gives the time to merger, with vertical lines for the cuts of~\ref{subsec:MBHBacctime}, and the dots on the right of each panel show the result for the full signal including the ringdown.}
  \label{fig:modeoverlapsoftime}
\end{figure*}

We now illustrate the accumulation of the gravitational wave signal with time, which is crucial to understand the ability of the instrument to indentify and localize MBHB signals in advance of their coalescence, as well as the requirements put on the instrumental configuration (data gaps and downlink cadence). As will be shown in more details in a separate publication~\cite{MarsatBabak20} (see also~\cite{KatzLarson18}), within the uncertainties of astrophysical models we can expect the bulk of MBHB signals to be detectable only a few days prior to merger, while a tail of more favorable events could be observable for significantly longer times, up to months.

We will focus on system II of Table~\ref{tab:MBHBparams}, and we will highlight four different epochs, corresponding to the time before merger where a certain fraction of the total $\SNR = 666$ has been accumulated. We include higher harmonics here. These epochs are:
\begin{itemize}
	\item $\SNR/64$: $\sim 42$ hr before merger, which corresponds roughly, with $\SNR \simeq 10$, to the first time we could confidently claim a detection;
	\item $\SNR/16$: $\sim 2.5$ hr before merger;
	\item $\SNR/4$: $\sim 7$ min before merger.
\end{itemize}

We approximate the abrupt interruption of the observation as an upper frequency cutoff.
The cutoff frequency is derived from the end time of the observation via the time-to-frequency correspondence~\eqref{eq:deftflm}.
In reality, an abrupt interruption of the signal would produce non-local features in the Fourier domain (unless tapering is applied) which we do not consider here.

The time-to-frequency correspondence~\eqref{eq:deftflm} gives time as a function of frequency for each $(\ell, m)$ mode. Due to the scaling $h_{\ell m} \propto e^{-i m \phi_{\rm orb}}$ with $\phi_{\rm orb}$ the orbital phase, a given time corresponds to different frequencies according to
\bsub\label{eq:fscalinglm}
\begin{align}
	\omega_{\ell m}(t) &\simeq \frac{m}{2} \times \omega_{22}(t) \quad (\mathrm{Inspiral})\,, \\
	t_{f}^{\ell m} \left( \frac{m}{2} f\right) &\simeq t_{f}^{22} (f) \quad (\mathrm{Inspiral}) \,,
\end{align}
\esub
with $\omega_{\ell m} = \dot{\phi}_{\ell m}$ the instantaneous mode frequency in the time-domain. Fig.~\ref{fig:t-f-relation-lm} illustrates these relations and we use them to mark in Fig.~\ref{fig:hctdiahm} the frequencies corresponding to our time cuts, that differ for each mode, together with the merger frequency (also called the peak frequency).

The relations~\eqref{eq:fscalinglm} are only accurate for the inspiral regime, as such a correspondence is at the heart of the SPA.
Although $t_{f}^{\ell m}$~\eqref{eq:deftflm} can be formally extended even past the merger time, close to merger it starts to lose accuracy and physical interpretation, eventually becoming non-monotonic with frequency~\cite{MB18}.
The departure from the scaling~\eqref{eq:fscalinglm} between modes can be seen in Fig.~\ref{fig:t-f-relation-lm}: the scaling holds up to a few minutes before merger, and the mode $h_{21}$ shows the earliest signs of a deviation.

Fig.~\ref{fig:modeoverlapsoftime} shows the cumulative contributions to the total SNR of individual mode combinations. For a generic signal $s = \sum_{\ell m} s_{\ell m}$ with mode contributions $s_{\ell m}$ ($s$ is any among the channels $\tilde{a}$, $\tilde{e}$, $\tilde{t}$)
\be\label{eq:defcontribSNR2}
	\SNR^{2} = (s|s) = \sum_{\ell m} \sum_{\ell' m'} (s_{\ell m} | s_{\ell' m'}) \,.
\ee
We choose to show contributions to $\SNR^{2}$ because this is the scale on which the contributions of different modes and channels are additive, and because it is the relevant scale for the log-likelihood~\eqref{eq:deflnL}, since $\ln \calL \sim \SNR^{2}$. We also sum over the three channels. Together with the total contribution to $\SNR^{2}$, the three panels of Fig.~\ref{fig:modeoverlapsoftime} show diagonal terms $(s_{\ell m} | s_{\ell m})$, cross-terms involving the dominant quadrupolar mode $(s_{22} | s_{\ell m})$, and finally other cross-terms between subdominant modes. Results are displayed as a function of the time to merger, and we show separately the result obtained for the full post-merger signal.

We can draw several conclusions from Fig.~\ref{fig:modeoverlapsoftime}. The signal reaches a roughly detectable level of $\SNR=10$ about two days before merger. We see that the SNR accumulates rapidly in the last instants before merger, as shown in particular by reaching $\SNR/4$ only 7 minutes before coalescence. For a period following first detection most or all of the higher modes are not significant. Stopping the signal at a given time in the inspiral somewhat alters the hierarchy between subdominant modes that was seen in their power spectra: contrasting Fig.~\ref{fig:modeoverlapsoftime} with Fig.~\ref{fig:hctdiahm}, we see e.g that $h_{21}$ is now subdominant before merger (and increases significantly in strength when including the post-merger signal). This is because modes $h_{\ell m}$ with a higher $m$ will reach a higher frequency at any given time, while the weighting by the noise favors higher frequencies. This is also visible in Fig.~\ref{fig:hctdiahm} where ticks translate a cut in time into a different cut in frequency for different modes.

We also see that, since even the most subdominant modes contribute significantly to $\SNR^{2}$, our set $(22, 21, 33, 44, 55)$ seems to be incomplete; we will need waveforms with a richer set of higher harmonics to analyze real LISA data. Diagonal terms $(s_{\ell m} | s_{\ell m})$ and cross terms with $(\ell m) \neq (\ell' m')$ are qualitatively different. Diagonal terms accumulate coherently, while cross terms are oscillatory, as they feature two modes with different phasings. This oscillatory character tends to suppress the contribution of the cross terms; however, we see they are not negligible, in particular the ones involving the dominant harmonic and a subdominant harmonic.


\subsection{Decomposing the instrument response}
\label{subsec:MBHBresponseapprox}

\begin{figure}
  \centering
  \includegraphics[width=.98\linewidth]{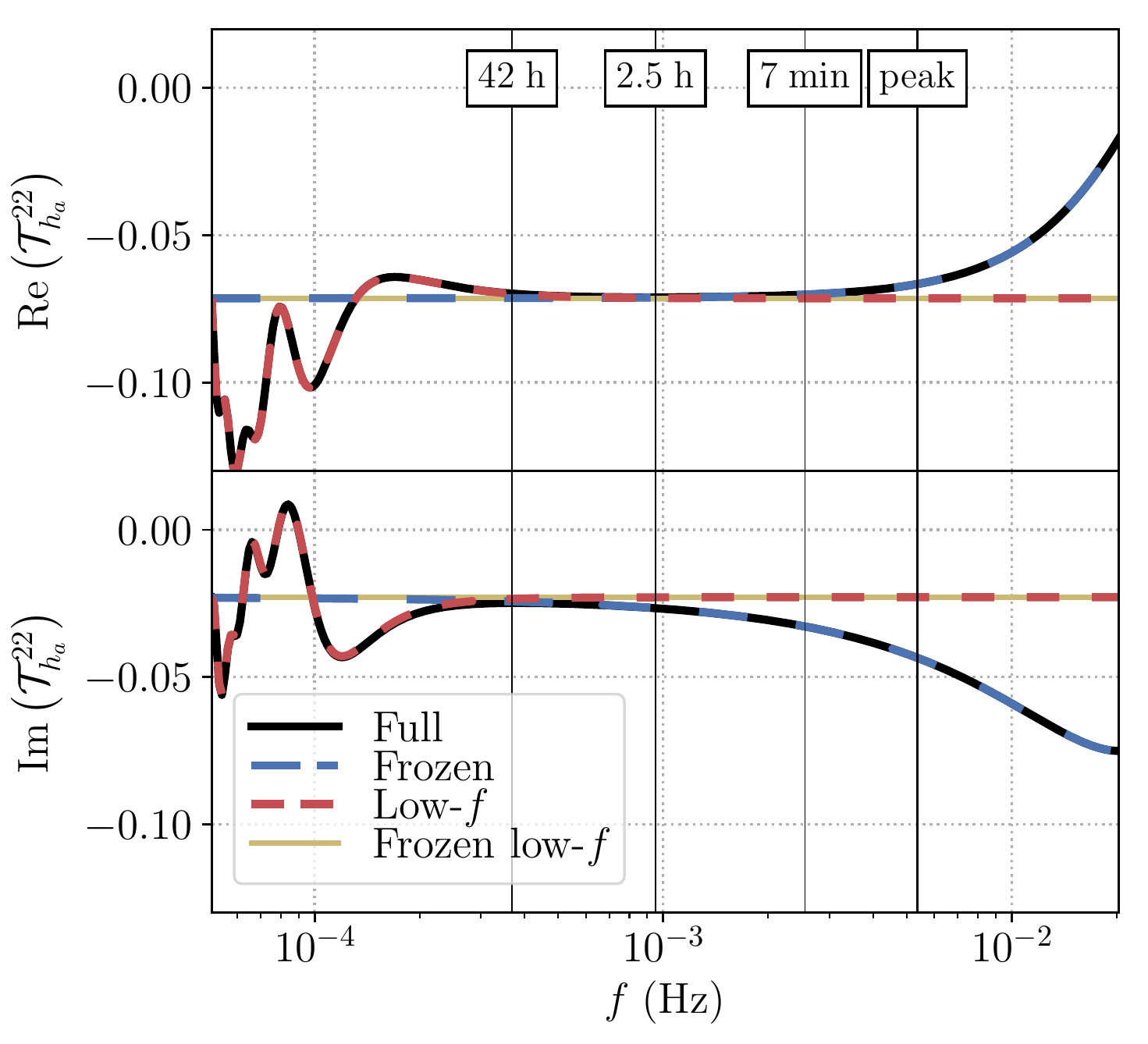}
  \caption{Real and imaginary part of the transfer function $\calT^{22}_{h_{a}}$~\eqref{eq:defhaet} in different response approximations. The four lines stand for the four approximation levels (i)-(iv) defined in~\ref{subsec:MBHBresponseapprox}. Vertical lines indicate the frequencies of the time cuts of~\ref{subsec:MBHBacctime} and the peak frequency.}
  \label{fig:responseapproxtdia22}
\end{figure}

The LISA instrument response, as recalled in Section~\ref{sec:response}, is both time- and frequency-dependent.
As we will see, these features play an important role in breaking degeneracies in the parameter estimation of the source, notably allowing us to localize the source in the sky. It is therefore important to understand these features, also in the light of the pre-merger accumulation of signal with time.

The time-dependency in the response follows the motion of the LISA constellation. In the low-frequency picture of~Sec.~\ref{subsec:lowfresponse}, the time dependency enters both in the Doppler phase term~\eqref{eq:defPhiR} (LISA moves in the wavefront) and in the time-dependent LISA-frame angles~\eqref{eq:Lframeangles} (the LISA arms change orientation). However, we have seen in the previous section that the MBHB signals we consider here are quite short (less than two days). This limits the effect of the LISA motion: between the time where the signal becomes detectable and the merger, LISA has barely changed in orientation and position. Moreover, the period when the signal accumulates most of its SNR is even shorter, as shown in Fig.~\ref{fig:modeoverlapsoftime}.

To disentangle these different physical effects, we distinguish four different approximations of the response:
\begin{enumerate}[(i)]
	\item \textit{Full}: in this case, we keep the full response of Secs.~\ref{subsec:FDresponse}-\ref{subsec:TDIresponse}, with its complete time and frequency dependency;
	\item \textit{Frozen}: we neglect the LISA motion by evaluating all time-dependent vectors at $\tf = t_{f}^{\rm peak}$, effectively freezing LISA in its orbit, while keeping the frequency dependency in the response;
	\item \textit{Low-$f$}: we implement the low-frequency (long-wavelength) approximation as in Sec.~\ref{subsec:lowfresponse}, while still keeping the time-dependency due to the motion;
	\item \textit{Frozen low-$f$}: we neglect both the time and frequency dependency in the response.
\end{enumerate}

Case (iii), \textit{Low-$f$}, has been extensively used in the past for the analysis of MBHB signals.
Since it is based on the $f \ll f_{L}$ approximation, it is more appropriate for inspiral-only signals (to which most previous studies were limited) than for a full IMR signal.

In the case (iv), \textit{Frozen low-$f$}, the response is equivalent to two LIGO-type detectors, lying motionless in the same location and rotated from each other by $\pi/4$, and there is no other information on the source's sky position other than the frequency-independent pattern functions of the two effective detectors. Contrarily to networks of ground-based observatories, we have no triangulation information from times of arrival at different detectors. This case is the most degenerate, and will be useful to get an analytical understanding of approximate degeneracies occuring when using the more complicated full response (i). This limit can also be a representative approximation for some short-duration premerger LISA MBHB observations.

The effect on the response of adopting these approximations is shown in Fig.~\ref{fig:responseapproxtdia22}. We display the strain transfer function $\calT_{h_{a}}^{22}(f)$ as defined in~\eqref{eq:defhaet} for the mode $h_{22}$, with the four lines showing the four approximations (i)-(iv). Vertical lines also indicate the times to merger highlighted in Sec.~\ref{subsec:MBHBacctime}, as well as the peak frequency. Relating this figure to Figs.~\ref{fig:hctdiahm} and~\ref{fig:modeoverlapsoftime}, we stress that most of the SNR is accumulated in the very last instants before merger and at the merger itself, due to the noise normalization not visible at the level of the transfer function.

The \textit{Frozen low-$f$} transfer function is just a constant factor, similarly to the LIGO response where $h_{+,\times}$ are simply multiplied by pattern functions. The \textit{Low-$f$} transfer function goes to the same constant at high frequencies, where the signal chirps so rapidly that the LISA motion is negligible. At low frequencies, however, modulations due to the LISA motion appear. The \textit{Frozen} transfer function asymptotes to the constant of the \textit{Frozen low-$f$} case at low frequencies, for $f \ll f_{L}$. At higher frequencies, we see a growing departure from this approximation. As noted in Sec.~\ref{subsec:lowfresponse}, when reaching the armlength transfer frequency $f_{L} = 0.12 \Hz$ the long-wavelength approximation has completely broken down; departures start to be significant at much lower frequencies. Finally, the \textit{Full} transfer function displays all the features we discussed.

We note finally a coincidence in Fig.~\ref{fig:responseapproxtdia22}: the frequency below which we see the imprint of the LISA motion and the frequency above which we see the breakdown of the long-wavelength approximation appear to be the same. This will not be true in general: the former is essentially a measure of the time-to-frequency correspondence $\tf$, with lower-mass signals having support and showing these features at higher frequencies, while the latter only marks the magnitude of $2\pi f L$ factors and is source-independent\footnote{The effect in the transfer functions is source-independent; however, the total SNR will determine the impact of these features on the parameter estimation.}.


\section{Massive black holes parameter estimation}
\label{sec:SMBHPE}

\subsection{Analysis of IMR signals}
\label{subsec:SMBHPEfull}

\begin{figure*}
  \centering
  \includegraphics[width=.98\linewidth]{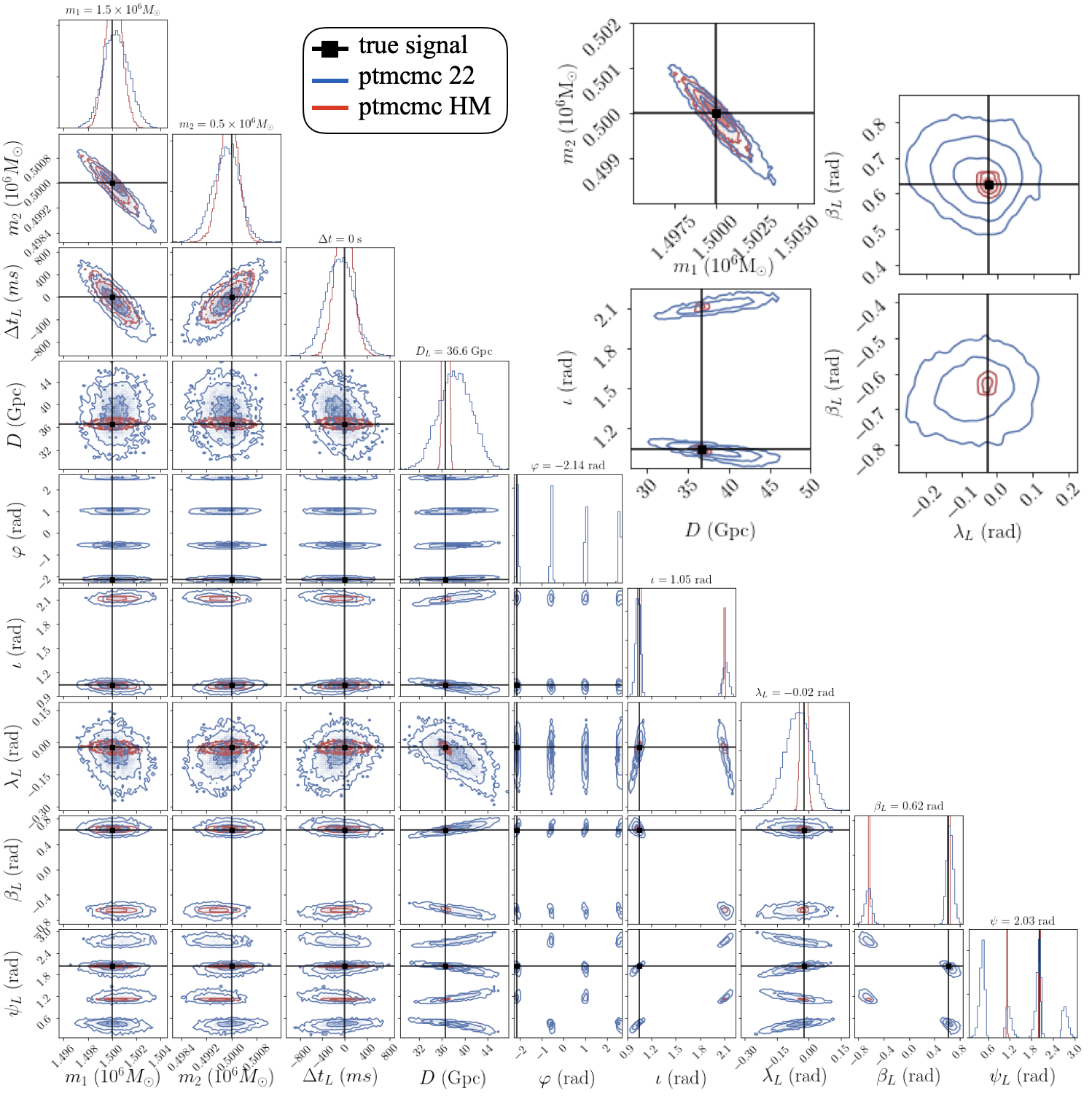}
  \caption{Inferred parameter posterior distribution for the MBHB system I of Table~\ref{tab:MBHBparams}, with $1$-$2$-$3$-$\sigma$ contours and the black cross indicating the true parameters. The result injecting and recovering with only the 22 mode is shown in blue, and the result injecting and recovering with higher harmonics is in red. The top right panels zoom on the masses, distance-inclination, and sky position marginalized posteriors. The time is centered so that $\Delta t=0$ corresponds to the injection. All extrinsic parameters are given in the LISA-frame as defined in App.~\ref{app:LISAframe}.}
  \label{fig:PEsmbh22hmCase0}
\end{figure*}

\begin{figure*}
  \centering
  \includegraphics[width=.98\linewidth]{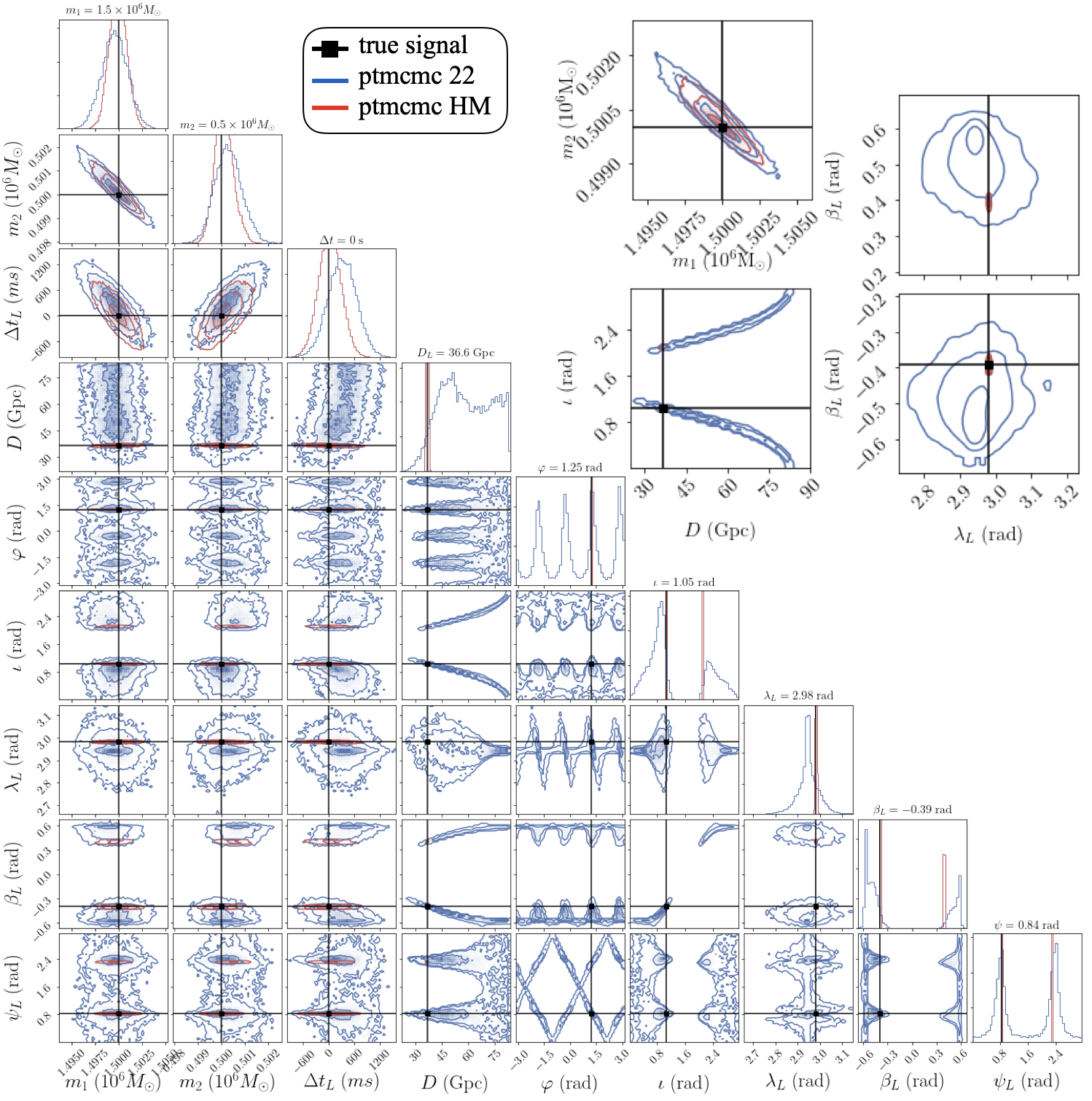}
  \caption{Same as Fig.~\ref{fig:PEsmbh22hmCase0}, for the MBHB system II of Table~\ref{tab:MBHBparams}.
           In the extrinsic parameters, the red 2D contours are barely visible on this scale.}
  \label{fig:PEsmbh22hmCase9}
\end{figure*}

\begin{figure*}
  \centering
  \includegraphics[width=.98\linewidth]{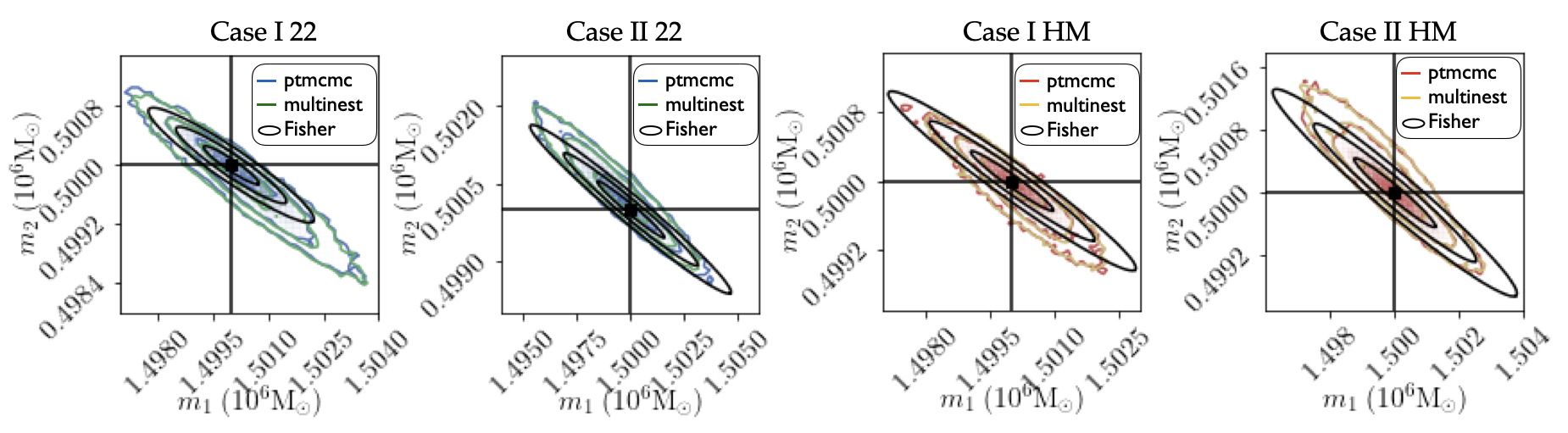}
  \caption{Comparisons between inferred posteriors obtained with our two different Bayesian samplers described in~\ref{subsec:samplers} and with the Fisher matrix approximation described in~\ref{subsec:Fisher}. Results are shown for the systems I and II of Table~\ref{tab:MBHBparams}, with the 22-mode only for the two left panels and with higher harmonics in the two right panels. Blue (22) and red (HM) show the $\texttt{ptmcmc}$ result, and green (22) and yellow (HM) the $\texttt{multinest}$ result. In all panels, the black ellipses show the Fisher matrix estimate for the posterior.}
  \label{fig:PEsmbhm1m2ptmcmcbambi}
\end{figure*}

\begin{figure*}
  \centering
  \includegraphics[width=.98\linewidth]{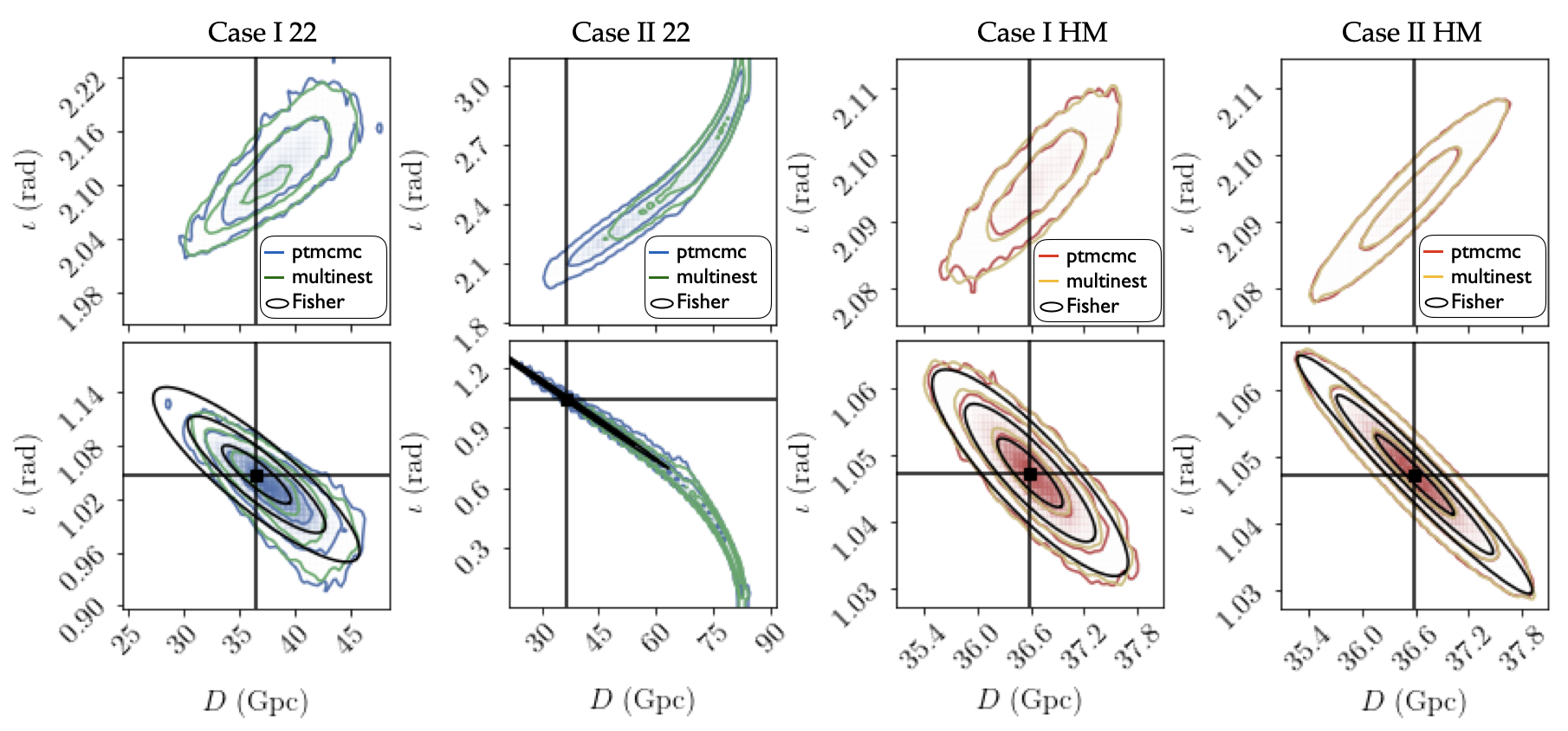}
  \caption{Same comparisons as in Fig.~\ref{fig:PEsmbhm1m2ptmcmcbambi}, for the distance and inclination parameters, zooming on the relevant regions.}
  \label{fig:PEsmbhDincptmcmcbambi}
\end{figure*}

\begin{figure*}
  \centering
  \includegraphics[width=.98\linewidth]{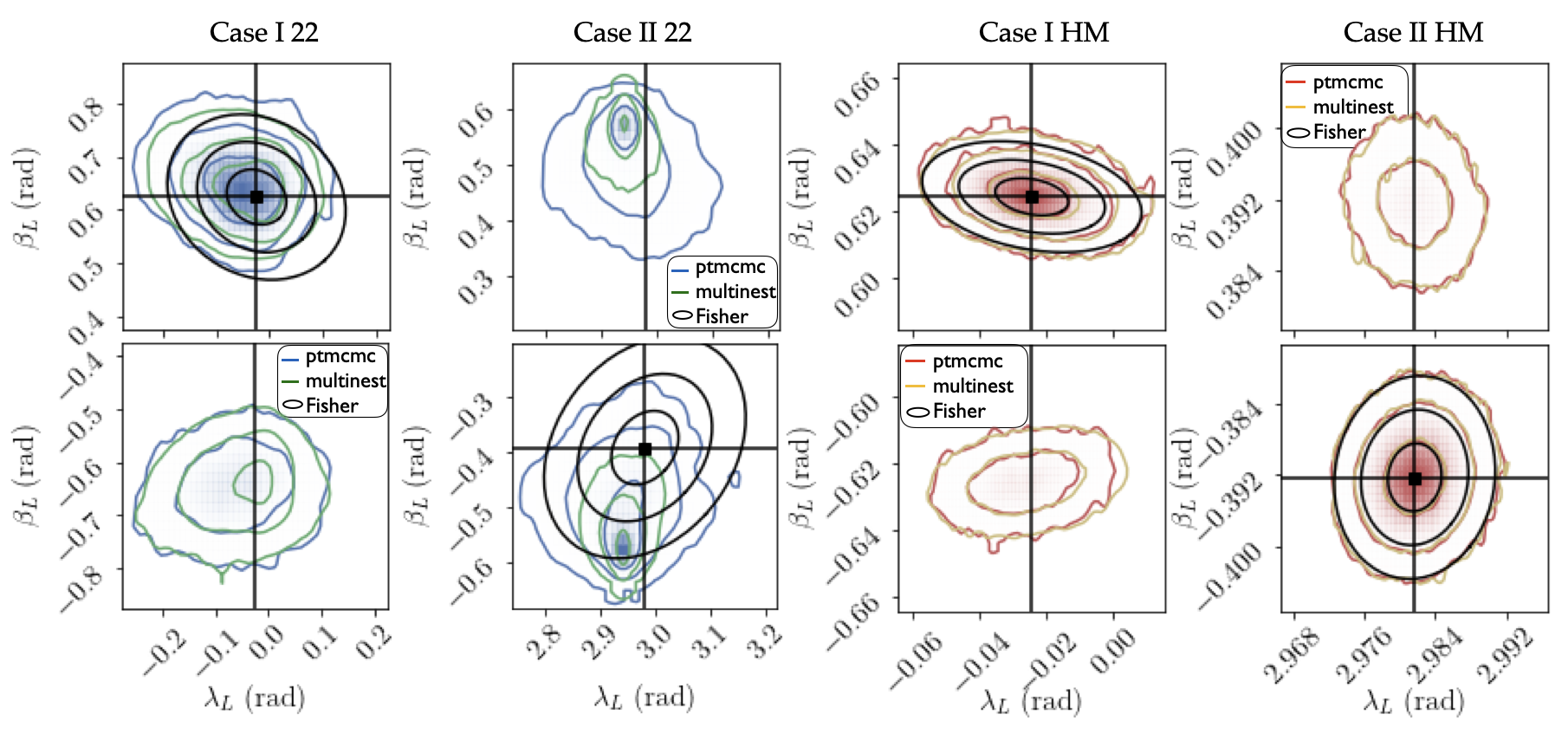}
  \caption{Same comparisons as in Fig.~\ref{fig:PEsmbhm1m2ptmcmcbambi}, for the sky position parameters, zooming on the relevant regions for inclination. The parameters shown are in the LISA-frame as defined in~\ref{app:LISAframe}.}
  \label{fig:PEsmbhskyptmcmcbambi}
\end{figure*}

Using the methodology summarized in Sec.~\ref{sec:method}, here we present the results of Bayesian parameter estimation for the two MBHB sources listed in Table~\ref{tab:MBHBparams}.
The priors used are logarithmic in mass, flat in luminosity distance, and uniform for the angles: flat on the sphere for the pairs of angles $(\iota, \varphi)$ and $(\lambda, \beta)$, and flat for the polarization $\psi$. We expect priors to be unimportant for the masses, since they are well determined in this very high SNR limit. The luminosity distance, as we will see, can be less well determined in the absence of higher harmonics, and one should keep in mind that the prior choice does affect the posterior there.

We find that both our samplers, \texttt{ptmcmc} and \texttt{multinest}, require between $\sim 10^{7}$ and $\sim 10^{8}$ likelihood evaluations to produce a final set of posterior samples.
Thanks to our fast likelihood implementation, this already represents a manageable computing cost.
We stress again that the settings of both samplers were not optimized by taking into account the characteristics of the specific problem or the expected correlations between parameters.
Therefore, we expect that the cost can be reduced with future optimizations.

To illustrate the role of higher harmonics in the analysis, we will present two classes of results: ``22'', where we inject and recover signals including only the dominant harmonic $h_{22}$, and ``HM'', where we both inject and recover with the higher harmonics $(h_{22}, h_{21}, h_{33}, h_{44}, h_{55})$. We do not attempt to recover an injection with higher modes using 22-mode waveforms; one would expect parameter biases in this case, induced by the inadequacy of the simplified waveforms. A related question would be to assess the mode content necessary for waveform models to mitigate systematic biases when analyzing full GR signals. We leave such waveform systematics studies for future work; we only note that, as shown in Fig.~\ref{fig:modeoverlapsoftime}, from SNR only the set of five modes we include in the present study is quite obviously incomplete.

While the 22-mode-only signals here are non-physical, studying them is instructive not only because  22-only waveforms have been a common approximation in previous work but also because (as seen in Fig.~\ref{fig:modeoverlapsoftime}) upon first detection LISA MBHB signals will often be fairly approximated by 22-mode-only signals.

Figs.~\ref{fig:PEsmbh22hmCase0} and~\ref{fig:PEsmbh22hmCase9} show posterior distributions for all parameters for Cases I and II respectively.
We overlay the ``22'' and ``HM'' posteriors.
All results for MBHB systems are presented using LISA-frame parameters, as defined in App.~\ref{app:conventions}. We should keep in mind the difference in SNR of Cases I and II ($\SNR \simeq 945$ versus $\SNR \simeq 666$ with higher harmonics, see Table~\ref{tab:MBHBparams}), resulting from their different orientation, however their posteriors have qualitative differences that go beyond a simple scaling of the errors as $\sim 1/\SNR$.

In Case I, the ``22'' posterior appears to be well represented by a multimodal Gaussian, with multimodality for the angular parameters. The sky position, most notably, admits a degenerate mode at $\betaL \rightarrow -\betaL$, although it has less weight than the main mode around the injection. By contrast, in Case II, the ``22'' posterior is much more degenerate. The distance and inclination are very degenerate with each other, with a support extending all the way to $\iota=0$ (or $\iota=\pi$). The phase $\varphi$ and polarization $\psiL$ have a distinct extended degeneracy along lines of constant $\varphi + \psiL$ and $\varphi - \psiL$. Most notably, the sky position, if retaining the same overall bimodality $\betaL \rightarrow -\betaL$ as in Case I, shows here a curious feature: the marginalized sky posterior peaks away from the injected value. Looking at correlations, we see that the shifted peak corresponds to the region of high distance (and extremal inclination). This is a genuine feature of the multidimensional posterior, even without noise,  and will be explained in Sec.~\ref{subsec:MBHBPEdegen22}.

In both Cases I and II, including the higher harmonics has a major effect on the parameter recovery (as was already stressed in~\cite{Arun+07a, TriasSintes07, PorterCornish08, McWilliams+09}), much beyond what we would expect solely from the modest gain in total SNR shown in Table.~\ref{tab:MBHBparams} (and a scaling of statistical errors $\sim 1/\SNR$). The marginalized posterior of the masses is narrower, although not qualitatively different; the major change is in the extrinsic parameters.

We can understand the dramatic effect of the higher harmonics on distance and inclination by noting that, in a signal of the form $\sum_{\ell m} \sYlm(\iota, \varphi)h_{\ell m}$ where the $\sYlm$ all have a different dependency with the inclination $\iota$, measuring independently two separate contributions $(\ell m) \neq (\ell' m')$ gives us an independent measurement of $\iota$ by the relative amplitude of the two mode contributions.
When only the $h_{22}$ mode is included, both the luminosity distance and inclination determine the overall signal amplitude and they are therefore degenerate.

Higher harmonics also lead to an independent determination of the phase $\varphi$, which in turn breaks degeneracies in the other angular parameters. In both cases I and II, the sky localization in vastly improved, with a remaining multimodality that we will discuss in the next section.

We compare the Bayesian inference results obtained with our two samplers, \texttt{ptmcmc} and \texttt{multinest}, with the error estimates given by the Fisher matrix approximation~\ref{subsec:Fisher} in Fig.~\ref{fig:PEsmbhm1m2ptmcmcbambi} for the masses, Fig.~\ref{fig:PEsmbhDincptmcmcbambi} for distance/inclination, and Fig.~\ref{fig:PEsmbhskyptmcmcbambi} for the sky position. We find a good agreement between the two Bayesian samplers, except for Case II when using only 22-mode signals: there, \texttt{multinest} seems to fail to explore the full posterior, remaining stuck in the large-distance region of the very degenerate parameter space. We will futher explore the nature of this degeneracy in~Sec.\ref{subsec:MBHBPEdegen22} and in App.~\ref{app:samplingdegen}. The Fisher matrix approach focuses on the vicinity of the true signal's parameters and is by construction unable to handle multimodality. It is also insufficient in capturing the degeneracy features of Case II with $h_{22}$ signals. It gives good results, however, for the mass parameters and for the main mode of the posterior in non-degenerate cases, in particular when including higher harmonics.


\subsection{Degeneracies in the sky}
\label{subsec:MBHBPEdegen}

A remarkable feature in Figs.~\ref{fig:PEsmbh22hmCase0} and~\ref{fig:PEsmbh22hmCase9} is the existence of a degenerate mode (or secondary maximum) in the posterior distribution for the sky position $(\lambda_{L}, \beta_{L})$, located at $\beta_{L}^{*} = -\beta_{L}^{\rm inj}$, i.e. by reflecting the wave vector $k$ across the plane of the LISA constellation.
The secondary mode contains less probability than the ``correct'' mode close to the true parameters, but it is present in both simulations and it survives the inclusion of higher harmonics.

We can gain insight about degeneracies in the parameter space by considering the simple expressions representing the instrument response in the \textit{Frozen low-$f$} approximation described in Sec.~\ref{subsec:MBHBresponseapprox}, appropriate for the low-frequency limit and signals short enough that the LISA motion can be neglected; we will explore the respective influence of these sometimes neglected effects in Sec.~\ref{sec:MBHBPEacctime}.

The \textit{Frozen low-$f$} response is given by~\eqref{eq:FapcFepc}-\eqref{eq:transferlowfae}, ignoring the $k\cdot p_{0}$ delay as a mere constant delay, and neglecting time-dependency in the LISA frame angles $(\lambda_{L}, \beta_{L}, \psi_{L})$. The response is then summarized by the pattern functions for harmonics $F^{\ell m}_{a,e} (\iota, \varphi, \lambda_{L}, \beta_{L}, \psi_{L})$ given in~\eqref{eq:Flmae}, which we reproduce here:
\begin{align}\label{eq:Flmaerepeat}
	F_{a,e}^{\ell m} &= \frac{1}{2} \sYlm (\iota, \varphi) e^{-2 i \psi_{L}} \left( F_{a,e}^{+} + i F_{a,e}^{\times}\right) (\lambda_{L}, \beta_{L}) \nn\\
	& + \frac{1}{2} (-1)^{\ell} \sYlminusmstar (\iota, \varphi) e^{+2 i \psi_{L}} \left( F_{a,e}^{+} - i F_{a,e}^{\times} \right) (\lambda_{L}, \beta_{L}) \,.
\end{align}

Changing the sign of $\beta_{L}$ in the pattern functions $F_{a,e}^{+,\times}$ given in~\eqref{eq:FapcFepc} has no effect on $F_{a,e}^{+}$ and it changes the sign of $F_{a,e}^{\times}$ for both channels.
Thus, the $\left( F_{a,e}^{+} \pm i F_{a,e}^{\times}\right)$ factors in the two terms of Eq.~\eqref{eq:Flmaerepeat} are exchanged.
Moreover, since spin-weighted spherical harmonics~\cite{Goldberg+67} obey the relation
\be
	\sYlm (\pi - \iota, \varphi) = (-1)^{\ell} \sYlminusmstar (\iota, \varphi) \,,
\ee
we see that simultaneously changing $\beta_{L} \rightarrow -\beta_{L}$, $\iota \rightarrow \pi-\iota$ and $\psi_{L} \rightarrow \pi - \psi_{L}$ leaves $F_{a}^{\ell m}$, $F_{e}^{\ell m}$ unchanged.

This defines a transformation of extrinsic parameters yielding an exact degeneracy in the \textit{Frozen low-$f$} approximation, which we call the \textit{reflected} sky position (for a reflection with respect to the LISA plane):
\begin{align} \label{eq:degenreflection}
	\lambdaL^{*} &= \lambdaL \,, \nn \\
	\betaL^{*} &= -\betaL \,, \nn \\
	\psiL^{*} &= \pi - \psiL \,, \nn \\
	\iota^{*} &= \pi - \iota \,, \nn \\
	\varphi^{*} &= \varphi \,,
\end{align}
where we chose the transformation for $\psiL$ to keep this parameter in the range $[0, \pi]$.

From the structure of~\eqref{eq:Flmaerepeat}, other points in parameter space are degenerate in the low-frequency limit. First, $\lambdaL \rightarrow \lambdaL + \pi$ leaves the pattern functions $F_{a,e}^{+,\times}$ invariant. For $\lambdaL \rightarrow \lambdaL \pm \pi/2$, these pattern functions acquire an overall minus sign. Such an overall minus sign is readily compensated by a shift $\psiL \rightarrow \psiL \pm \pi/2$, so that we have the other transformation
\begin{align} \label{eq:degenlambdapi2}
	\lambdaL^{(k)} &= \lambdaL + \frac{k \pi}{2} \mod 2\pi \,, \, k = 0, \dots, 3 \,, \nn \\
	\psiL^{(k)} &= \psiL + \frac{k \pi}{2} \mod \pi \,, \, k = 0, \dots, 3 \,.
\end{align}

Combining~\eqref{eq:degenreflection} and~\eqref{eq:degenlambdapi2}, we arrive at eight different degenerate positions in the sky, equally spaced in $\lambdaL$ and symmetric above and below the LISA plane, with an inclination $\pi - \iota$ for the reflected positions and various values for the polarization $\psiL$. Among these eight secondary modes in the sky, two play special roles: first, the \textit{reflected} mode~\eqref{eq:degenreflection} already mentioned; and second, the \textit{antipodal} mode with
\begin{align} \label{eq:degenantipodal}
	\lambdaL^{(a)} &= \lambdaL + \pi \,, \nn \\
	\betaL^{(a)} &= -\betaL \,, \nn \\
	\psiL^{(a)} &= \pi - \psiL \,, \nn \\
	\iota^{(a)} &= \pi - \iota \,, \nn \\
	\varphi^{(a)} &= \varphi \,.
\end{align}
How does this situation change when considering a more complete instrument response, moving away from the \textit{Frozen low-$f$} approximation ? As in Sec IV.C
we separately consider relaxing each of the qualifiers \textit{Frozen} and \textit{Low-f}.

When adding back the frequency-dependence while keeping LISA motionless, in the \textit{Frozen} response, the \textit{reflected} mode is the only mode remaining degenerate. Indeed, as noted in Sec.~\ref{subsec:FDresponse}, in~\eqref{eq:Gslr} the frequency-dependent terms feature $k\cdot n_{l}$, $k \cdot p_{s,r}$ projections, that depend only on the projection of the wave vector $k$ in the plane of LISA, invariant for $\betaL \rightarrow -\betaL$. The LISA motion, however, will break this degeneracy in general.

When adding back the LISA motion while still ignoring the frequency-dependence in the response, in the \textit{Low-$f$} response, the \textit{antipodal} mode is the only one that keeps pattern functions that are exactly degenerate with the injection: all other modes will evolve with the time-dependent LISA frame. The antipodal mode is not moving, because the antipode of the true direction of the arriving signal  is defined as $k \rightarrow -k$ independently of the orientation of LISA; the only degeneracy-breaking term is then the orbital delay $k \cdot p_{0}$ in the Doppler phase~\eqref{eq:defPhiR}. When considering the \textit{Full} response, the frequency dependency terms in~\eqref{eq:Gslr} featuring $k\cdot n_{l}$, $k \cdot p_{s,r}$ will break this antipodal degeneracy.

\begin{table}
	\begin{tabularx}{.48\textwidth}{|X||c|c|c|c|}
		\hline
		Sky mode & \textit{Full} & \textit{Frozen} & \textit{Low-$f$} & \textit{Frozen low-$f$} \\
		\hline
		\textit{reflected}: \newline $-\betaL$, $\lambdaL$ & $t$-dep. & degen. & $t$-dep. & degen.  \\
		\hline
		\textit{antipodal}: \newline $-\betaL$, $\lambdaL + \pi$ & $f$-dep.$+\Delta \Phi_{R}$ & $f$-dep. & $\Delta \Phi_{R}$ & degen.  \\
		\hline
		$\betaL$, $\lambdaL + \pi/2$ & $t$-$f$-dep. & $f$-dep. & $t$-dep. & degen.  \\
		\hline
		$\betaL$, $\lambdaL + \pi$ & $t$-$f$-dep. & $f$-dep. & $t$-dep. & degen.  \\
		\hline
		$\betaL$, $\lambdaL - \pi/2$ & $t$-$f$-dep. & $f$-dep. & $t$-dep. & degen.  \\
		\hline
		$-\betaL$, $\lambdaL + \pi/2$ & $t$-$f$-dep. & $f$-dep. & $t$-dep. & degen.  \\
		\hline
		$-\betaL$, $\lambdaL - \pi/2$ & $t$-$f$-dep. & $f$-dep. & $t$-dep. & degen.  \\
		\hline
	\end{tabularx}
	\caption{Sketch of the degeneracy structure of the eight modes in the sky, with the cases where the degeneracy is exact (degen.), and with the qualitative effects that break the degeneracy with the injected signal: $t$-dependency, $f$-dependency, both, or only the Doppler phase $\Delta \Phi_{R}$~\eqref{eq:defPhiR}. This qualitative structure is the same with and without higher harmonics.}
	\label{tab:MBHBdegen}
\end{table}

Thus, we have found that in the \textit{Frozen low-$f$} approximation for the response we expect a pattern of eight degenerate positions in the sky (with certain rules for inclination and polarization). Table~\ref{tab:MBHBdegen} summarizes the qualitative picture of degeneracy breaking by the features of the response. We note that the recent work~\cite{Baibhav+20}, focusing on low-frequency ringdown-dominated signals for which the LISA motion can be neglected, remarked the same 8-modes sky degeneracy that we described in this section. We will see in Sec.~\ref{sec:MBHBPEacctime} that the eight-mode degeneracy pattern indeed appears when doing a pre-merger analysis; on the other hand, in results Figs.~\ref{fig:PEsmbh22hmCase0} and~\ref{fig:PEsmbh22hmCase9} for full IMR signals, out of the eight possible sky modes only the \textit{reflected} mode~\eqref{eq:degenreflection} survives. We will investigate in detail in Sec.~\ref{sec:MBHBPEacctime} how time-dependence and frequency-dependence in the response break part, but not all, of the degeneracies.

However, this analysis does not explain why one obtains, with a zero noise realization, marginalized posteriors for the sky positions that appear biased from the injected signal, when ignoring higher harmonics. This is the question that we will address in the next Section.


\subsection{Apparent sky position bias for 22-mode signals}
\label{subsec:MBHBPEdegen22}

\begin{figure}
  \centering
  \begin{minipage}{.49\linewidth}
  \includegraphics[width=.99\linewidth]{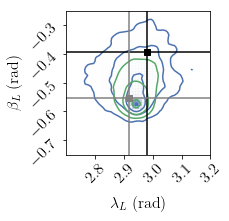}
  \end{minipage}
  \begin{minipage}{.49\linewidth}
  \includegraphics[width=.99\linewidth]{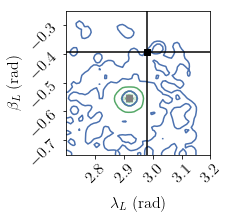}
  \end{minipage}
  \caption{Sky position posterior for System II in Table~\ref{tab:MBHBparams} in presence of strong degeneracies, zooming close to the injected value and discarding the secondary mode in the sky. The left panel shows the complete likelihood, while the right panel shows the simplified likelihood~\eqref{eq:simplelikelihoodhm} in the \textit{Frozen, low-$f$} approximation and with fixed masses and time. The results of \texttt{ptmcmc} (blue) and \texttt{multinest} (green) are superimposed. The black cross is the injected value, and the grey cross is the analytic prediction~\eqref{eq:soldegensky} for the offset peak created by degeneracies.}
  \label{fig:Skydownsmbh22hmSimpleLikeCase9}
\end{figure}

In this section, we investigate the cause of the apparent bias in sky position in the posterior distribution of System~II when injecting and recovering with 22-mode only waveforms, as shown in Figs.~\ref{fig:PEsmbh22hmCase9} (and in the middle left panel of~\ref{fig:PEsmbhskyptmcmcbambi}). The posterior for the sky forms a peak that appears shifted from the injection. This feature is surprising: since we set the noise realization to zero, the maximum likelihood is by construction reached at the injection. It occurs only when 22-mode only signals are used, and is present for both Bayesian samplers, although $\texttt{ptmcmc}$ and $\texttt{multinest}$ differ noticeably for this case with $\texttt{ptmcmc}$ recovering more parameter volume. We will show that we can understand this feature as a projection effect in the multidimensional degenerate posterior. Understanding the structure of these features could prove useful to inform Bayesian samplers, for instance by adapting jump proposals to speed up mixing of MCMC chains.

In the following we will aim at explaining these features analytically by building a simplified model for the response. In this \textit{simplified likelihood} approximation, we:
\begin{itemize}
	\item pin the masses and the LISA-frame coalescence time to their injected values;
	\item use the \textit{Frozen low-$f$} response, ignoring the LISA motion and the frequency-dependence in the response.
\end{itemize}
The first point allows us to focus only on the extrinsic parameters, as we find weak correlations between intrinsic and extrinsic parameters. Such a decoupling is also at play in LIGO/Virgo, and makes low-latency sky localization possible~\cite{SingerPrice15}. The second point is partly justified by the fact that our signal is short, as shown in Fig.~\ref{fig:modeoverlapsoftime}: for the MBHB system that we picked as an example the SNR accumulates in a matter of days, with $\mathrm{SNR}=10$ reached $\sim 40 \,\mathrm{h}$ before merger. Neglecting the high-frequency features is in fact a stronger approximation, as will be explored in Sec.~\ref{subsec:MBHBacctimebreakdegen}. In this limit we can apply the simple analytic expressions for the response given in~\ref{subsec:lowfresponse}.

Under all these simplifying assumptions, the likelihood becomes pure function of the extrinsic parameters $(D, \iota, \varphi, \lambdaL, \betaL, \psiL)$, and takes the form of trivial geometric factors multiplying constant mode overlaps that can be precomputed.  Eq.~\ref{eq:aetlowf} shows that the $T$-channel is negligible in this approximation, so that the likelihood~\eqref{eq:deflnLaet} is
\be
	\ln \calL = -\frac{1}{2} \left( \tilde{a} - \tilde{a}_{\rm inj} | \tilde{a} - \tilde{a}_{\rm inj} \right) -\frac{1}{2} \left( \tilde{e} - \tilde{e}_{\rm inj} | \tilde{e} - \tilde{e}_{\rm inj} \right) \,.
\ee
Having fixed the intrinsic parameters and the time, in~\eqref{eq:transferlowfae} the modes $h_{\ell m}$ are fixed as well, and it is convenient to introduce the notation
\be\label{eq:innerproductlmlpmp}
	\left\langle \ell m | \ell' m' \right\rangle = 4\mathrm{Re} \int \frac{df}{S_{n}^{a,e}} \left( 6 \pi f L\right)^{2} \tilde{h}_{\ell m} \tilde{h}_{\ell' m'}^{*} \,.
\ee
for mode overlaps that are constant factors and can be precomputed for a given system. Here, the noise PSD $S_{n}^{a,e}$ is given by~\eqref{eq:Snaet} and is identical between the channels $a$ and $e$.

Using the results of Sec.~\ref{subsec:lowfresponse}, ignoring in~\eqref{eq:transferlowfae} the factor $\exp [2 i \pi f k\cdot p_{0}]$ as a pure constant corresponding to a redefinition of time, we obtain
\begin{align}\label{eq:simplelikelihoodhm}
	\ln \calL &= -\frac{1}{2} \sum_{\ell m} \sum_{\ell' m'} \left[ \left( \vphantom{s_{a}^{\ell' m'} } s_{a}^{\ell m} - s_{a, \mathrm{inj}}^{\ell m}\right) \left( s_{a}^{\ell' m'} - s_{a, \mathrm{inj}}^{\ell' m'}\right)^{*} \right. \nn\\
	& \qquad \left. + \left( \vphantom{s_{a}^{\ell' m'} } s_{e}^{\ell m} - s_{e, \mathrm{inj}}^{\ell m}\right) \left( s_{e}^{\ell' m'} - s_{e, \mathrm{inj}}^{\ell' m'}\right)^{*} \right] \left\langle \ell m | \ell' m' \right\rangle \,,
\end{align}
where we introduced
\be\label{eq:defsaelm}
	s_{a,e}^{\ell m} = \frac{1}{d} F_{a,e}^{\ell m} \,,
\ee
with $d = D / D_{\rm inj}$ the dimensionless ratio of luminosity distances, and with mode transfer functions $F_{a,e}^{\ell m}$ given in~\eqref{eq:Flmae}. In each term the intrinsic/extrinsic parameter dependence is thus separated with the intrinsic parameters (that we keep fixed) intervening only in $\left\langle \ell m | \ell' m' \right\rangle$.

In the case where we only include the dominant harmonic $h_{22}$, the likelihood~\eqref{eq:simplelikelihoodhm} simplifies to
\be\label{eq:simplelikelihood22}
	\ln \calL = -\frac{1}{2} \left\langle 22 | 22 \right\rangle  \left[ \left| s_{a}^{22} - s_{a, \mathrm{inj}}^{22} \right|^{2} + \left| s_{e}^{22} - s_{e, \mathrm{inj}}^{22}\right|^{2} \right] \,.
\ee
For $(\ell m)=(22)$ the functions~\eqref{eq:defsaelm} take the explicit form
\begin{align}
	s_{a,e}^{22} &= \frac{1}{4d} \sqrt{\frac{5}{\pi}} \cos^{4}\frac{\iota}{2} e^{2i(\varphi-\psiL)} \left( F_{a,e}^{+} + i F_{a,e}^{\times} \right) \\
	&+ \frac{1}{4d} \sqrt{\frac{5}{\pi}} \sin^{4}\frac{\iota}{2} e^{2i(\varphi+\psiL)} \left( F_{a,e}^{+} - i F_{a,e}^{\times} \right) \,,
\end{align}
with the pattern functions for the channels $a$ and $e$ defined in \eqref{eq:defFapcFepc}-\eqref{eq:FapcFepc}.

In order to elucidate the degeneracies in the problem, it will be useful to introduce the following notation. First, since $\iota \in [0, \pi]$ while $\beta_{L} \in [-\pi/2, \pi/2]$, it will be more convenient to work with the colatitude $\theta_{L} = \pi/2 - \beta_{L}$. We will further abbreviate notation by using the variables
\be
	t_{\iota} \equiv \tan \frac{\iota}{2} \,, \quad t_{\theta} \equiv \tan \frac{\theta_{L}}{2} \,.
\ee
We also introduce the azimuthal angles $\lambda_{a} = \lambda_{L} - \pi/6$, $\lambda_{e} = \lambda_{L} + \pi/12 = \lambda_{a} + \pi/4$, and form the combinations
\be
	\sigma_{\pm} \equiv \frac{1}{2} \left[ s_{a}^{22} \pm i s_{e}^{22} \right] \,.
\ee
In this notation, we obtain
\bsub\label{eq:sigmapm}
\begin{align}
	\sigma_{+} &= \rho e^{2i\varphi} \left[ t_{\theta}^{4} e^{-2 i \psiL} + t_{\iota}^{4} e^{2 i \psiL} \right] e^{-2i \lambda_{a}} \,, \\
	\sigma_{-} &= \rho e^{2i\varphi} \left[ e^{-2 i \psiL} + t_{\theta}^{4} t_{\iota}^{4} e^{2 i \psiL} \right] e^{2i \lambda_{a}} \,,
\end{align}
\esub
with a common prefactor
\be\label{eq:sigmafactorrho}
	\rho(d, \iota, \theta_{L}) = \frac{1}{4d} \sqrt{\frac{5}{\pi}} \frac{1}{\left( 1 + t_{\iota}^{2} \right)^{2} \left(1 + t_{\theta}^{2} \right)^{2}} \,.
\ee
Finding points in the parameter space that are degenerate with the injection, {i.e.} with $\ln \calL \simeq 0$, amounts to finding choices for the parameters which obtain the same values as the injection for both quantities $\sigma_{+}$ and $\sigma_{-}$.

We can now use the simplified response written in the form~\eqref{eq:sigmapm}-\eqref{eq:sigmafactorrho} to look for symmetries and degeneracies. First, it is easy to check in this new notation that the transformations~\eqref{eq:degenreflection} and~\eqref{eq:degenlambdapi2} indeed leave the likelihood exactly invariant. There is also a symmetry based on exchanging $\iota \leftrightarrow \theta_{L}$. While leaving $\sigma_-$ unchanged, this conjugates the factor inside brackets for $\sigma_{+}$, which can be compensated using the phase term $\varphi - \lambda_{a}$. If $\Phi = \mathrm{Arg} \left[ t_{\theta}^{4} e^{-2 i \psiL} + t_{\iota}^{4} e^{2 i \psiL} \right] $, we obtain the symmetry
\begin{align}
	\iota' = \theta_{L} \,, \qquad \theta_{L}' = \iota \,, \nn\\
	\varphi' = \varphi + \frac{1}{2} \Phi \quad \mathrm{mod} \; \pi\,, \nn\\
	\lambda_{a}' = \lambda_{a} - \frac{1}{2} \Phi \quad \mathrm{mod} \; \pi\,.
\end{align}

Beyond these discrete symetries, since likelihood function dependence on the six extrinsic parameters is funneled through just two complex functions, we should expect a two-dimensional degenerate subspace. Indeed we can explicitly find a general solution for parameter values that solve $\sigma_{\pm}(D, t_\iota, t_\theta, \varphi, \lambda_a, \psiL)=\sigma_{\pm,{\rm inj}}$, which we just sketch here, as needed to explain features of the degeneracies.

Defining the ratio
\be
	r(\lambda_a, t_\iota, t_\theta, \psiL) = \frac{\sigma_{+}}{\sigma_{-}} = \frac{t_{\theta}^{4}  + t_{\iota}^{4} e^{4 i \psiL}}{1+ t_{\theta}^{4} t_{\iota}^{4} e^{4 i \psiL}} e^{-4i \lambda_{a}}\,,
\ee
makes clear that $r=r_{\rm inj}$ provides one complex condition on 4 real unknown parameters, eliminating parameters $D$ and $\varphi$ while retaining the features of the full degenerate subspace.  Going further, $|r|^2=|r_{\rm inj}|^2$ yields one condition on 3 parameters, eliminating $\lambda_a$. In fact, with a little rearrangement, this can be written as a quadratic expression for either $t_\iota^4$ or $t_\theta^4$ given the other variable ($t_\theta$ or $t_\iota$) and $\psiL$. Given a solution for $(t_\iota,t_\theta,\psi_L)$, the rest of the solution then proceeds backwards. An explicit expression for commensurate $\lambda$ comes from solving $r=r_{\rm inj}$, and then $D$ and $\varphi$ are obtained from solving, e.g., $\sigma_-=\sigma_{-}^{\rm inj}$.

We can now understand the degeneracies we saw by considering limits of the ratio $r$. For instance, when $t_{\iota}^{4} \ll 1$, we have simply $|r| \simeq t_{\theta}^{4}$. This means that in this limit all values of $\psiL$ are allowed, with $\theta_{L}$ fixed to a specific value. Similarly, for $t_{\iota}^{4} \gg 1$, we have $|r| \simeq t_{\theta}^{-4}$. A large part of the degenerate subspace volume then tends to be found with a parameter $\beta_L$ near these special values fixed by the modulus constraint, with a special value of $\lambda_{L}$ coming from the complex argument constraint. By contrast, intermediate values of $t_{\iota}^{4}$ will not give as much parameter space allowed for the degeneracy, as illustrated by the case $t_{\iota} = 1$ (i.e. $\iota = \pi/2$). In that case $|r|=1$ irrespective of $\theta_{L}$ and $\psiL$, and no solution exists if $|r_{\rm inj}| \neq 1$.

In terms of the original parameters, these special sky positions are
\bsub\label{eq:soldegensky}
\begin{align}
  \betaL &\approx \beta_{L}^{\dagger} = \pm\left(\frac{\pi}{2} - 2 \mathrm{Arctan} \; \left| r_{\rm inj}\right|^{1/4} \right)\,, \\
  \lambdaL &\approx \lambda_{L}^{\dagger} =\frac{\pi}{6} -\frac{1}{4} \mathrm{Arg} \; r_{\rm inj} \;\mathrm{mod}\; \frac{\pi}{2} \,,\\
  \rho e^{2i(\varphi-\psiL)} &\approx e^{-2i(\lambda_{L}^{\dagger} - \pi/6)} \sigma_{-}^{\rm inj} \,.
\end{align}
\esub
With the $\lambdaL \rightarrow \lambdaL + k \pi/2$ and $\betaL \rightarrow - \betaL$ symmetries corresponding to the eight-mode sky symmetry discussed in Sec.~\ref{subsec:MBHBPEdegen}. The degeneracy for the pair of parameters $(\varphi, \psiL)$ is exact, and the constraint $\rho = |\sigma_{-}^{\rm inj}|$ gives an approximate degeneracy for the pair $(d, \iota)$ as long as we remain in the regime $t_{\iota}^{4} \ll 1$ or $t_{\iota}^{4} \gg 1$ (which is quite extended, thanks to the quartic power). Thus, for each of these sky positions built from~\eqref{eq:soldegensky}, many different values of $(\varphi, \psiL)$ and $(d, \iota)$ produce a waveform very close to the injection, resulting in apparent peaks in the marginalized posterior distribution for the sky position, located at the special sky positions $(\lambda_{L}^{\dagger}, \theta_{L}^{\dagger})$, which are offset from the injected value.

Similarly, considering the limits $t_{\theta}^{4} \ll 1$, $t_{\theta}^{4} \gg 1$, we can expect a significant part of the degenerate parameter space near special values for inclination and correspondingly for $\lambda_L+\varphi$
\bsub\label{eq:soldegeniotaphi}
\begin{align}
	\iota&\approx\iota^{\dagger} = \frac{\pi}{2} \mp \left(\frac{\pi}{2} - 2 \mathrm{Arctan} \; \left| r_{\rm inj}\right|^{1/4} \right) \,, \\
	\lambda_L-\psiL&\approx (\lambda_{L} - \psiL)^{\dagger} = \frac{\pi}{6} -\frac{1}{4} \mathrm{Arg} \; r_{\rm inj} \;\mathrm{mod}\; \frac{\pi}{2} \,\\
        \rho e^{2i\varphi} &\approx e^{-2i ((\lambda_{L} - \psi_{L})^{\dagger} - \pi/6)} \sigma_{-}^{\rm inj} \,.
\end{align}
\esub

When considering more harmonics beyond the dominant mode $h_{22}$, such degeneracies will be broken easily. As in the case of the full likelihood, the simplified likelihood~\eqref{eq:simplelikelihoodhm} will have several terms with different inclination and phase dependencies. If the signal is loud enough for at least two modes to be detected, then the inclination $\iota$ and the phase $\varphi$ are fixed by the relative amplitude and phase of these modes. This will break the degeneracies $(d, \iota)$ and the degeneracy $(\varphi, \psiL)$.

We illustrate our findings in  by running a Bayesian parameter estimation for the simplified likelihood~\eqref{eq:simplelikelihood22} with our two samplers, \texttt{multinest} and \texttt{ptmcmc}. The results display the degeneracy structure that we discussed in this Section. Focusing on the sky position in the vicinity of the true source parameters, Fig.~\ref{fig:Skydownsmbh22hmSimpleLikeCase9} contrasts the posterior distribution obtained with the \textit{full} and the \textit{simplified} likelihoods. Although differences are visible, we see that both show a similar peak shifted from the true signal's sky position, that agrees well with our prediction~\eqref{eq:soldegensky}. More details on the full posterior distribution with the simplified likelihood are given in App.~\ref{app:samplingdegen}.

\section{Massive black holes: accumulation of information with time}
\label{sec:MBHBPEacctime}

\subsection{Pre-merger analysis}
\label{subsec:MBHBacctimePE}

The rate at which parameter information accumulates during the observation of an inspiral is crucial for establishing the LISA downlink and data processing requirements, as well as for planning multimessenger observations~\cite{ArmitageNatarajan02, DalCanton+19}.
In this Section, we explore how parameter information accumulates on approach to merger by performing parameter estimation studies with temporally truncated signals, as explained in~\ref{subsec:MBHBacctime}: the signals are cut at points where the accumulated SNR is about $\{10,42,166\}$ corresponding to $\{1/64,1/16,1/4\}$ of the total SNR=$666.0$, and times $\{41 \mathrm{hr}, 2.5 \mathrm{hr}, 7 \mathrm{min}\}$ before merger.
As shown in Fig.~\ref{fig:modeoverlapsoftime}, most of the SNR accumulates over the last few minutes of this signal.


We implement these temporal cuts in the Fourier domain, using the time-frequency correspondence~\eqref{eq:deftflm}, adapted for higher harmonics following~\eqref{eq:fscalinglm}. Our cuts are sufficiently early before merger for this relation to be a good approximation for temporal cuts, and consistent among the various harmonics, as shown in Fig.~\ref{fig:t-f-relation-lm}.

\begin{figure*}
  \centering
  \begin{minipage}{.32\linewidth}
      \includegraphics[width=.99\linewidth]{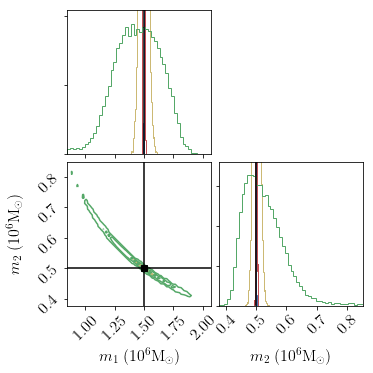}
   \end{minipage}
   \begin{minipage}{.32\linewidth}
      \includegraphics[width=.99\linewidth]{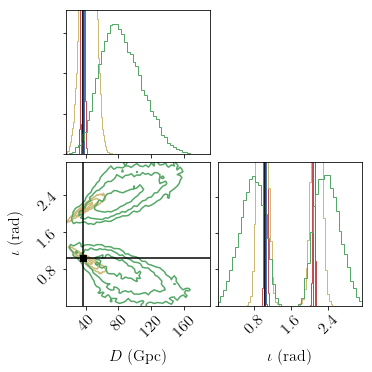}
   \end{minipage}
   \begin{minipage}{.32\linewidth}
      \includegraphics[width=.99\linewidth]{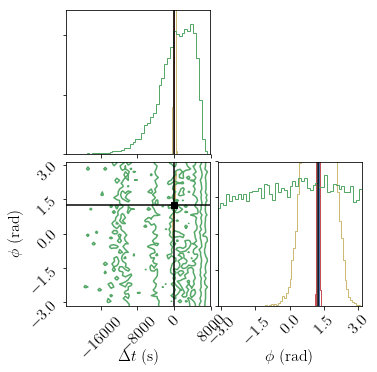}
   \end{minipage}
   \begin{minipage}{.32\linewidth}
      \includegraphics[width=.99\linewidth]{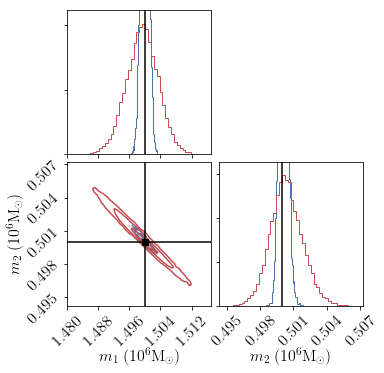}
   \end{minipage}
   \begin{minipage}{.32\linewidth}
      \includegraphics[width=.99\linewidth]{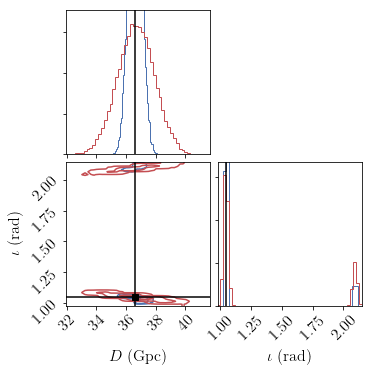}
   \end{minipage}
   \begin{minipage}{.32\linewidth}
      \includegraphics[width=.99\linewidth]{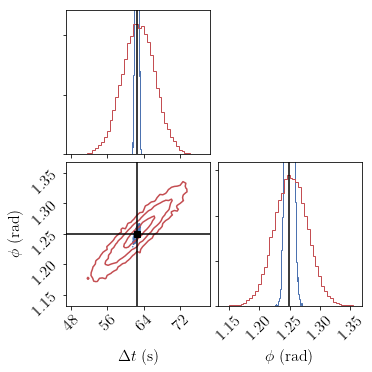}
   \end{minipage}
  \caption{Refinement of parameter inference over time for system II. The posteriors shown correspond to the time cuts defined in~\ref{subsec:MBHBacctime} at $\mathrm{SNR}/64$ (green), $\mathrm{SNR}/16$ (yellow), $\mathrm{SNR}/4$ (red) and post-merger (blue) for the masses (left), distance/inclination (center), and time/phase (right). On the first row, all four posteriors are shown, while in the second row we display only $\mathrm{SNR}/4$ and post-merger, zooming closer to the true parameters. We inject and recover with higher harmonics. The black cross shows the true parameters.}
  \label{fig:smbhCornerZoomHMCase9}
\end{figure*}

In Fig.~\ref{fig:smbhCornerZoomHMCase9} we show how parameter information accumulates at these time points. The results are shown for Case II, including the higher harmonics in the signal. Individual masses are loosely constrained, within a factor of 2, at first detection; the chirp mass combination is better determined. They start to be individually constrained to better than $10\%$ by 2.5 hours before merger. The luminosity distance is poorly determined, within a factor of 4, when reaching the detection threshold 41 hours before merger; the constraint improves to roughly $10\%$ at 2.5 hours prior to merger. Merger time is estimated with an uncertainty of roughly 2hrs at first detection, which improves to a few minutes by 2.5 hours before merger.

\begin{figure*}
  \centering
  \includegraphics[width=.8\linewidth]{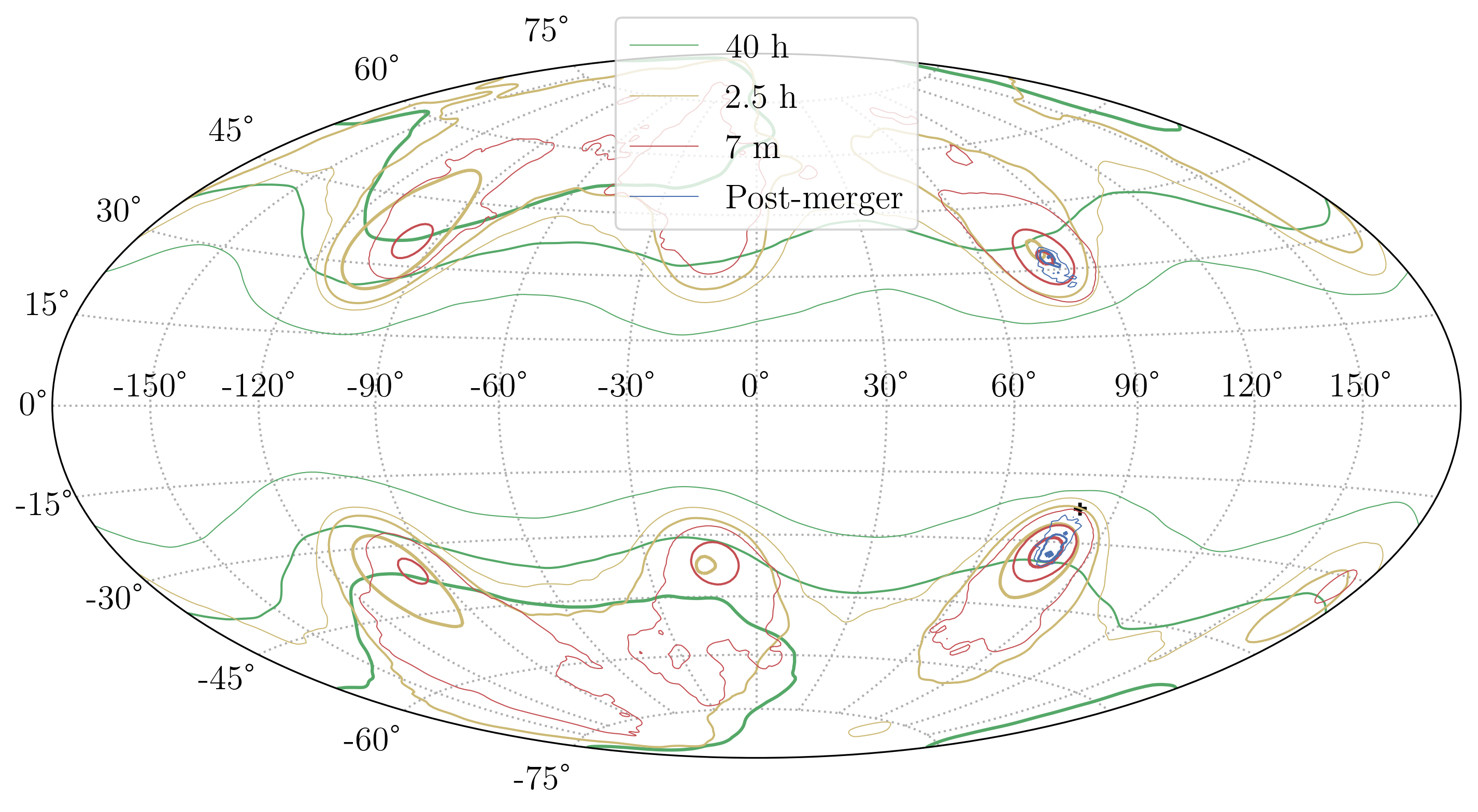}
  \includegraphics[width=.8\linewidth]{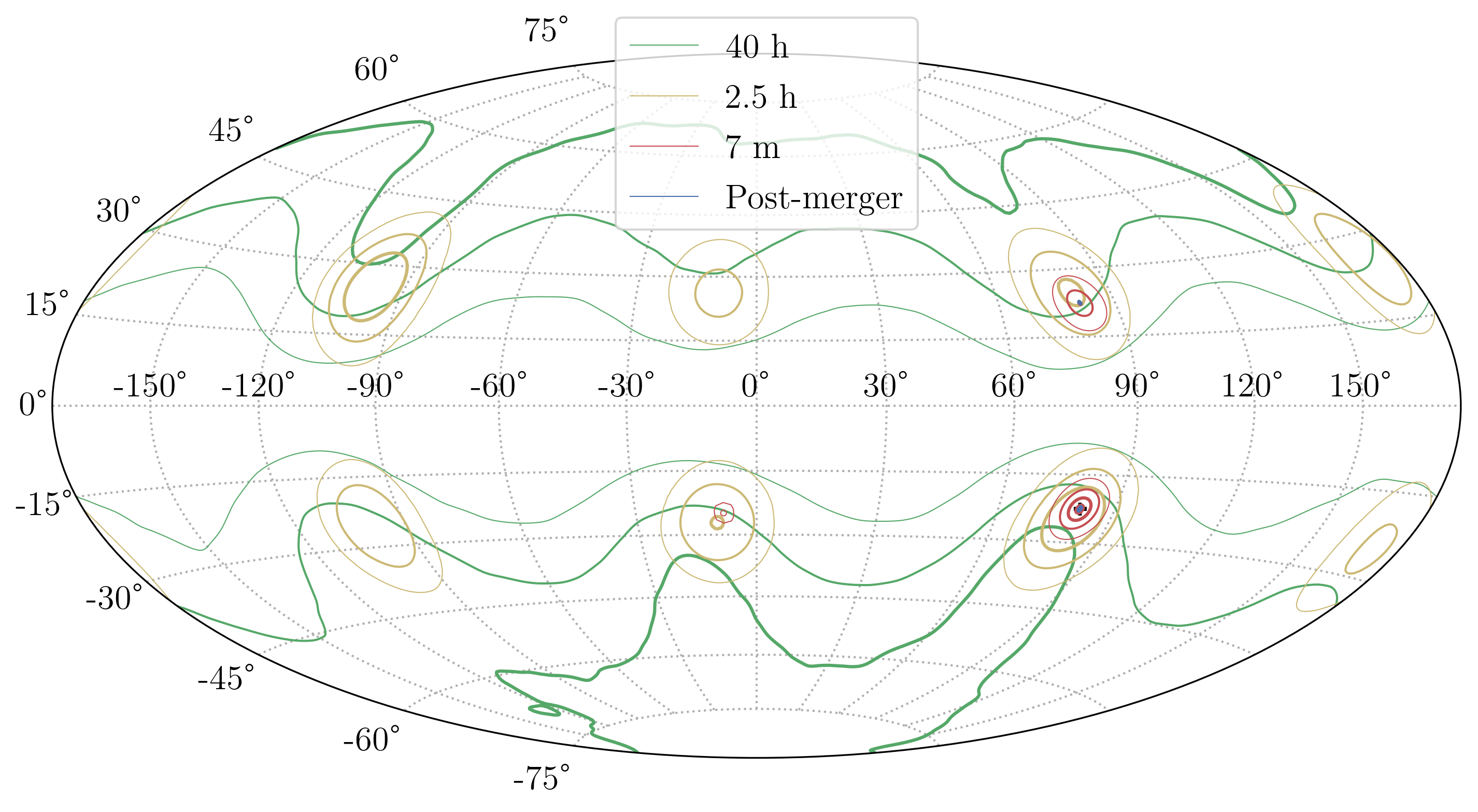}
  \caption{Refinement of the LISA-frame sky position inference over time for system II. The posteriors shown correspond to the time cuts defined in~\ref{subsec:MBHBacctime} at $\mathrm{SNR}/64$ (green), $\mathrm{SNR}/16$, $\mathrm{SNR}/4$ and post-merger (blue). For the top figure, we inject and recover with a 22-mode only waveform, while for the bottom figure we inject and recover with higher harmonics. The black cross is the injected sky position; the apparent offset in the 22-mode analysis corresponds to the phenomenon highlighted in~\ref{subsec:MBHBPEdegen22}. The blue contour for the analysis with the full signal corresponds to the posterior of Fig.~\ref{fig:PEsmbh22hmCase9} (where blue and red were used for 22 and HM); in the bottom figure with higher harmonics, is almost reduced to two dots on this scale.}
  \label{fig:MollweidesmbhTseriesCase922hm}
\end{figure*}

The corresponding sky position posteriors are shown in Fig.~\ref{fig:MollweidesmbhTseriesCase922hm}, using LISA frame angles~\ref{eq:Lframeangles}. The lines show the 1-2-3-$\sigma$ uncertainty contours on the sky, for each of our time cuts as well as for the full, post-merger signal, with the two panels differing by the inclusion of higher harmonics. We see clearly, in the pre-merger analysis, the 8-modes sky degeneracy introduced in~\eqref{eq:degenreflection}-\eqref{eq:degenlambdapi2}. The different cuts give us an idea of the continuous evolution from a badly determined sky position at first, when reaching detection, to an 8-modes degeneracy structure, to finally only two modes surviving when reaching merger, the true injection and the \textit{reflected} sky position.

\subsection{Degeneracy breaking by the time and frequency dependence in the response}
\label{subsec:MBHBacctimebreakdegen}

\begin{figure*}
  \centering
  \includegraphics[width=.99\linewidth]{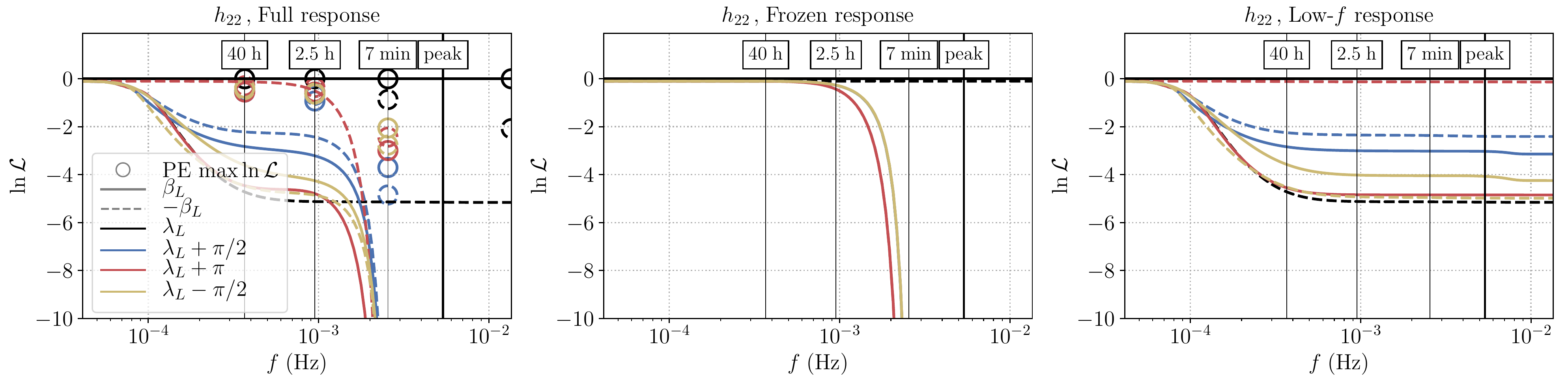}
  \includegraphics[width=.99\linewidth]{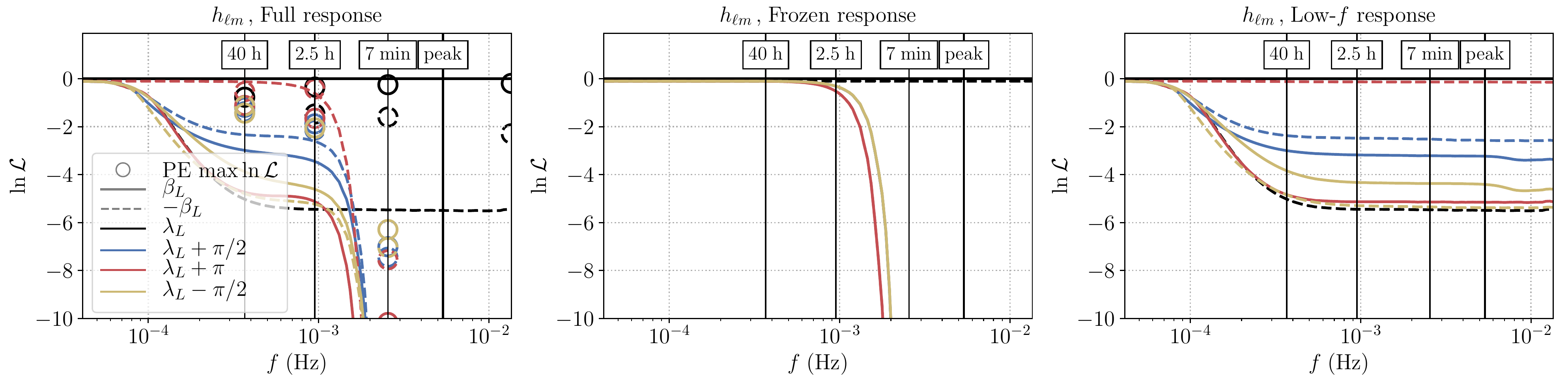}
  \caption{Log-likelihood~\eqref{eq:deflnLaet} at degenerate modes in the sky summarized in Table~\ref{tab:MBHBdegen}, for different response approximations (from left to right: \textit{Full}, \textit{Frozen}, \textit{Low-$f$}), with 22-mode only (top) and with higher harmonics (bottom). The log-likelihood is shown as a function of the maximal frequency where we cut the 22-mode signal (with the rescaling~\eqref{eq:fscalinglm} for the other modes), which can be thought of as accumulating signal with time (the vertical lines represent the time cuts of~\ref{subsec:MBHBacctime} and the merger). The lines represent the sky positions listed in Table~\ref{tab:MBHBdegen}: in terms of LISA-frame parameters full or dashed stands for the reflection in latitude, and the color stands for the $\pi/2$-shifted longitudes. The dashed black curve corresponds to the \textit{reflected} position, and the dashed red to the \textit{antipodal} position. At the true position (solid black), $\ln\calL$ is zero by construction while it is decreasing as a function of $f$ for other sky positions. Reaching a very negative $\ln\calL$ means that this sky position is excluded by the data accumulated thus far. In the left panel, the circles indicate the maximum log-likelihood found among the posterior samples, when the parameters are allowed to deviate slightly from the theoretical degenerate point.}
  \label{fig:lnLskymodesapprox}
\end{figure*}

The pre-merger analysis of the previous section made apparent the 8-modes degeneracy pattern that can be predicted analytically from the structure of the \textit{Frozen, low-$f$} response approximation (see~\eqref{eq:degenreflection}, \eqref{eq:degenlambdapi2}).
However, the analysis of IMR signals in~\ref{subsec:SMBHPEfull} shows that only the \textit{reflected} mode~\eqref{eq:degenreflection} survives post-merger. In this Section we explain how and why this transition occurs.

We have already presented a discussion of different qualitative effects in the instrument response in Sec.~\ref{subsec:MBHBresponseapprox} and Fig.~\ref{fig:responseapproxtdia22}: on one hand, the motion of LISA leaves an imprint (the time-dependency) on the low frequencies, where the time-frequency map is steep enough that a short interval in frequency maps to a large interval of time; on the other hand, the breakdown of the long-wavelength approximation leaves an imprint (the frequency-dependency) at high frequencies. Here we look at the quantitative importance of these features on the inference as a function of frequency (or as a function of time). We can readily select one or the other feature by using the \textit{Frozen} response, ignoring time dependency, or the \textit{Low-$f$} response, ignoring frequency dependency.

In Fig.~\ref{fig:lnLskymodesapprox} we show log-likelihood values obtained with either the \textit{Full}, \textit{Frozen} or \textit{Low-$f$} response, with and without higher harmonics, for each of the eight modes in the sky described by~\eqref{eq:degenreflection}-\eqref{eq:degenlambdapi2}. The likelihood~\eqref{eq:deflnLaet} is computed with the same response approximation for the injection and for the template. The results are shown as a function of frequency in a cumulative sense: we compute the likelihood by accumulating signal up to the frequency shown in the $x$-axis. When including higher harmonics, this cut in frequency is interpreted as a cut in time and propagated to other harmonics according to~\eqref{eq:fscalinglm}.

A value of $\ln \calL = 0$ means that the template signal with shifted parameters is identical to the injection, while a very negative $\ln \calL$ means that these parameter values are ruled out. The injection (full black line) has always $\ln \calL = 0$ by construction. The \textit{reflected} sky mode~\eqref{eq:degenreflection} is the dashed black line, and the \textit{antipodal} sky mode~\eqref{eq:degenantipodal} is the dashed red line.

With the \textit{Low-$f$} response, the \textit{antipodal} mode remains almost exactly degenerate with the injection, differing only by the Doppler phase~\eqref{eq:defPhiR}, which has a small effect on our short signals as shown in Fig.~\ref{fig:tranferSMBHCase9}. As explained below~\eqref{eq:degenantipodal}, other modes fail to reproduce the injected signal because of the time-dependence of the pattern functions. They acquire a moderate penalty at low frequency, but $\ln \calL$ then goes to a constant since the motion becomes negligible at high frequencies, as shown by the transfer functions in Fig.~\ref{fig:responseapproxtdia22}.

With the \textit{Frozen} response, the \textit{reflected} mode remains exactly degenerate as anticipated in Sec.~\ref{subsec:MBHBPEdegen}. All other modes see their $\ln \calL$ fall rapidly at around $2\mHz$ due to the onset of frequency-dependency in the response. Although the frequency dependency vanishes in the limit $f\ll f_L\approx 0.12$ Hz, note that the effect is crucial even for $f/f_L\approx 0.02$. Comparing to Fig.~\ref{fig:responseapproxtdia22}, we see that around $2\mHz$ the breaking of the long-wavelength approximation still appears mild; however, the SNR starts to accumulate a lot from $\SNR/16 \sim 40$ at $2.5\, \mathrm{h}$ to $\SNR/4 \sim 160$ at $7\min$, magnifying the effect of these features. Comparing the two rows of the figure, we see that the higher harmonics have an effect here, in making the elimination of secondary modes happen slightly earlier in frequency (or time). This is expected as they start to contribute significantly to the SNR (see Fig.\ref{fig:modeoverlapsoftime}) and reach higher in frequency than the $h_{22}$ harmonic according to~\eqref{eq:fscalinglm}.

The \textit{Full} response case is essentially a superposition of the previous two response approximations, with a successive onset of the time-dependency and frequency-dependency to break degeneracies. We note that overall, including the higher harmonics does not change this qualitative picture, despite an earlier onset of frequency-dependent degeneracy-breaking features.

Limitations of such explorations should be made clear: by using a pointwise estimate, we cannot make statistical statements and are missing volume effects discussed in Sec.~\ref{subsec:MBHBPEdegen22}; by transforming the angular parameters with an analytical prescription while keeping the other fixed, we are in a sense overconstraining the degeneracy. In particular, it is conceivable that one could find a better match to the injected signal in the vicinity of a degenerate sky mode, by adjusting slightly all parameters.

This is indeed what happens in multidimensional Bayesian parameter estimation, and is illustrated in Fig.~\ref{fig:lnLskymodesapprox}: in the first panel, for the three pre-merger analyses as well as for the post-merger analysis, we overlay circles indicating the best $\ln \calL$ found among the posterior samples, in each eighth of the sky corresponding to the eight sky modes (when no samples are present in this region of the sky, nothing is displayed). A sampler is not an optimizer, so the precise value achieved is not optimal and would vary when repeating runs; nevertheless, this measure gives a good proxy for how much closer to the injected signal we could get by slightly biasing the parameters.

We see that the best $\ln \calL$ among samples can be higher than the point estimate would tell us, especially for the analysis at $\sim 7 \min$ before merger, and higher harmonics make a visible difference. This has consequences for approximations like the Fisher analysis, if secondary peaks are shifted from the analytical predictions. We leave for future work the investigation of approximate representations of sky degeneracies.


\section{Stellar mass black holes}
\label{sec:SBHB}


\subsection{Signals and transfer functions}
\label{subsec:signaltransferSBHB}

\begin{figure*}
  \centering
  \includegraphics[width=.98\linewidth]{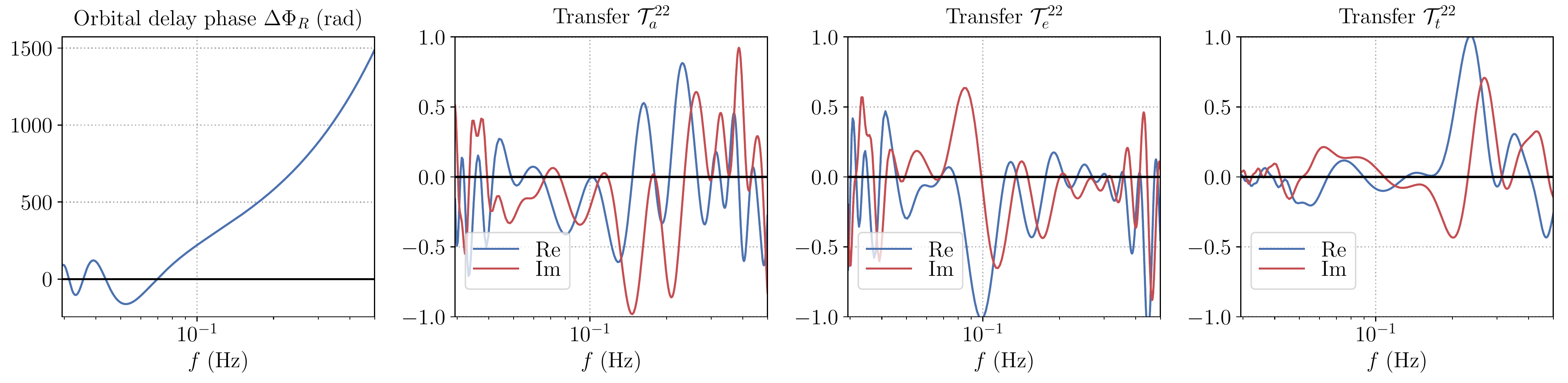}
  \caption{Example of instrument response for the SBHB system I in Table~\ref{tab:SBHBparams}. The first panel isolates the Doppler phase $\Delta\Phi_{R}$ as defined in~\eqref{eq:defPhiR}, while the three other panels show the TDI transfer functions for $A,E,T$. We show here the transfer functions for the 22-mode contribution to $a$, $e$, $t$ as defined in~\eqref{eq:defTlmaet}.}
  \label{fig:transferSBHBcaseI}
\end{figure*}

Stellar-mass black-hole binary (SBHB) inspirals have been recently recognized as a potentially
important source of LISA detections \cite{Sesana16, Vitale16}. Full Bayesian parameter
estimation studies for these LISA signals have not yet been developed, but the Fisher approach has been used in~\cite{Sesana16, Vitale16, Nishizawa+16a, Nishizawa+16b}. In this section we present two case
studies of parameter inference for such sources using the formalism and methods
described earlier, while limiting the analysis to the extrinsic parameters (excluding the masses, and fixing the time to coalescence). The parameters of the simulated mergers were drawn from the Radler LISA Data Challenge (LDC), which itself followed the population of~\cite{Sesana16}, and are listed in Table~\ref{tab:SBHBparams}.
System I is moderately massive, placed at a favorable close distance; with an SNR of $26.6$ it would be at the tail of the most favorable detections that we can hope for~\cite{Sesana16, Gerosa+19, Moore+19}.
System II is a higher-mass, moderate-SNR system.
Both systems happen to be close to the ecliptic plane.

Fig.~\ref{fig:transferSBHBcaseI} illustrates, for the system I in this table, the instrumental transfer functions~\eqref{eq:defTlmaet} as well as the Doppler phase~\eqref{eq:defPhiR}. Contrasting with the MBHB case, we see that the transfer functions have much more structure in this case. Both the time-dependency and the frequency-dependency in the response lead to oscillatory features in the transfer functions. The Doppler phase~\eqref{eq:defPhiR}, that was negligible for our short MBHB systems, now varies by tens of radians already at the lowest frequency.


\begin{table}
	\begin{tabular}{|c||c|c|}
		\hline
		Identifier 		& I 	& II   \\
		\hline
		\hline
		Mass 1 ($\Msol$) 	& $21.44$ & $63.51$ \\
		\hline
		Mass 2 ($\Msol$) 	& $20.09$ & $45.15$ \\
		\hline
		Redshift 		& $0.01$ & $0.07$ \\
		\hline
		Lum. Distance (Mpc) 		&  $49.1$ & $318.2$\\
		\hline
		Inclination (rad) 		& $0.65$ & $1.92$ \\
		\hline
		Phase (rad) 			& $0$	& $0$ \\
		\hline
		Ecliptic longitude (rad) 			& $3.44$	& $5.66$ \\
		\hline
		Ecliptic latitude (rad) 			& $-0.074$	& $-0.055$ \\
		\hline
		Polarization (rad) 			& $1.74$	& $2.20$ \\
		\hline
		Time to merger (yr) 			& $2$	& $2$ \\
		\hline
		\hline
		LISA SNR $h_{22}$			& 26.6 & 11.9 \\
		\hline
	\end{tabular}
	\caption{Parameters of the simulated SBHB inspiral signals.}
	\label{tab:SBHBparams}
\end{table}


\subsection{Bayesian inference of extrinsic parameters}
\label{subsec:PESBHB}

We caution the reader that our purpose here is to show the feasibility of
Bayesian inference for SBHB sources in LISA using the methodology presented in
this paper, and to illustrate the qualitative features of the posterior distribution for extrinsic parameters, in relation to our previous investigations for MBHB systems. To this end, the problem we pose is simplified
in several ways: we assume that we know the signal to be present, bypassing the challenges of detection itself~\cite{Moore+19}, we fix the masses and the time to merger to the injected value, and we neglect spins. The parameter recovery of the intrinsic parameters, masses, spins, is important to address as one can expect strong degeneracies in mass ratio and spin for these inspiral signals; it will be tackled in a forthcoming publication~\cite{Toubiana+20}.

The waveform model used to work with SBHB systems is a standard post-Newtonian
frequency-domain waveform (we use a version of the \texttt{TaylorF2} approximant as implemented in \texttt{LAL}~\cite{lal}). In practice, because LISA only observes the early
inspiral of SBHB systems, the signal is well within the domain of applicability of the post-Newtonian approximation~\cite{Mangiagli+18}.
In the early inspiral, harmonics beyond the dominant quadrupole give a negligible contribution, so we only include the $(2,2)$ mode.

As done in the previous section, we apply our parameter inference procedure to
these systems using both the \texttt{multinest} and \texttt{ptmcmc} samplers, and we
numerically evaluate the Fisher matrix at the parameters of the injection. In order to generate these results, our \texttt{ptmcmc} sampler requires between
$5\times 10^5$ and $3\times 10^6$ steps with 100 temperatures, taking on the
order of days on a single CPU core without careful optimization. The \texttt{multinest}
sampler, on the other hand, completes in a few hours only.

\begin{figure*}
	\includegraphics[width=0.9\textwidth]{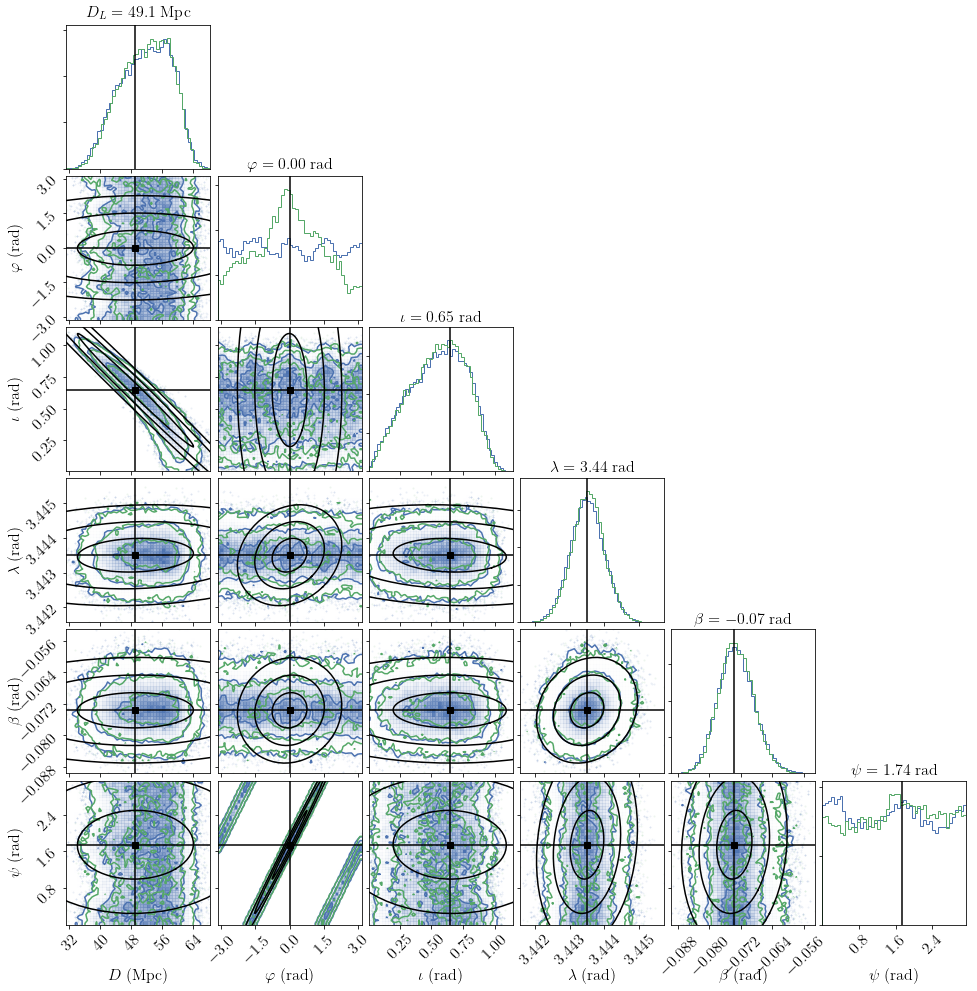}
	\caption{Inferred parameter posterior distribution for the SBHB system I of Table~\ref{tab:SBHBparams}, with $1$-$2$-$3$-$\sigma$ contours and the black cross indicating the true parameters. The black ellipses show the same contours for the Fisher matrix approximation. Waveforms include only the 22 mode. The results obtained with outr two different samples are shown, $\texttt{ptmcmc}$ in blue and $\texttt{multinest}$ in green. The initial frequency is fixed to obtain a time-to-merger of 2 years. All extrinsic parameters are given in the SSB-frame.}
	\label{fig:sbhb_corner_1}
\end{figure*}

\begin{figure*}
	\includegraphics[width=0.9\textwidth]{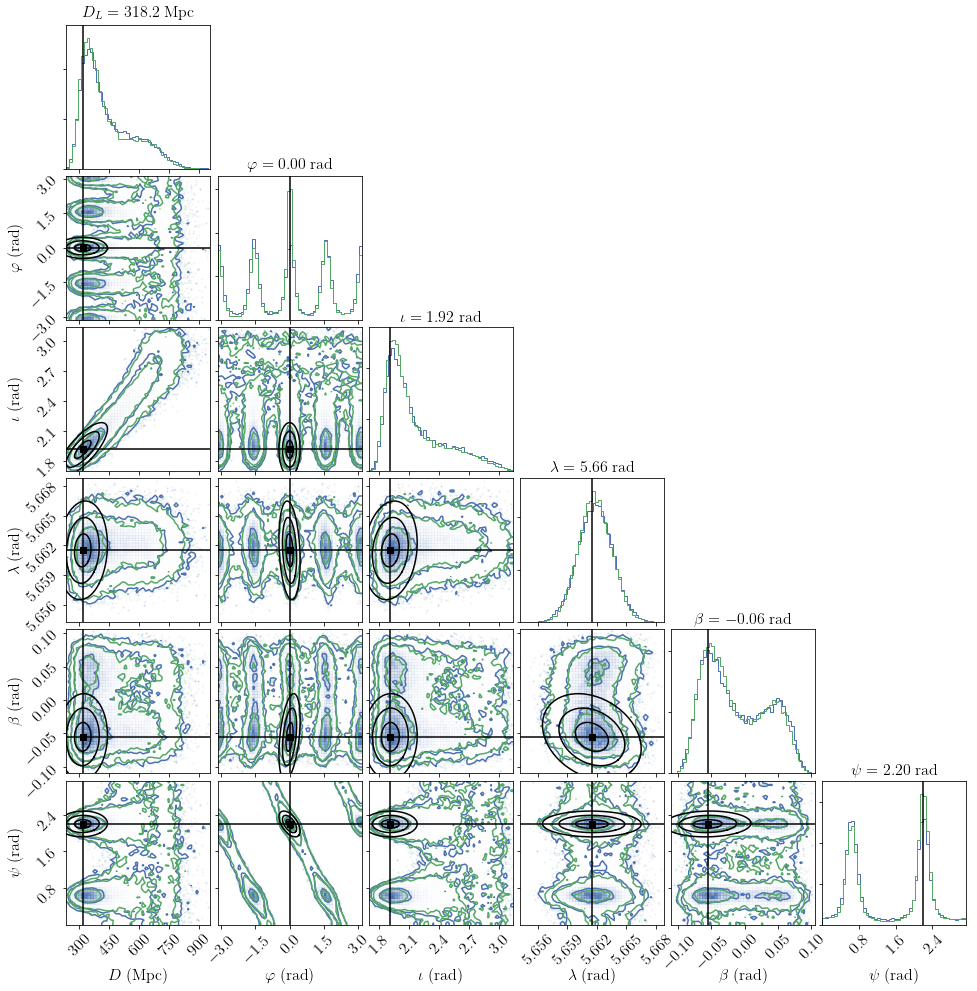}
	\caption{Same as Fig.~\ref{fig:sbhb_corner_1}, for the SBHB system II of Table~\ref{tab:SBHBparams}.}
	\label{fig:sbhb_corner_12}
\end{figure*}

The resulting posterior distributions are shown in Fig.~\ref{fig:sbhb_corner_1}
and \ref{fig:sbhb_corner_12}. We find close agreement between the \texttt{multinest} and
\texttt{ptmcmc} samplers. The joint distributions for extrinsic parameters are fairly simple, with extended degeneracies occuring only in the angular parameters $(\varphi, \psi)$. Thanks to the complexity of the instrument transfer functions shown in Fig.~\ref{fig:transferSBHBcaseI} carrying a lot of information, the sky position determination is very good. The typical degeneracy between inclination and distance
is evident, as there is no degeneracy breaking from higher harmonics. Even though the SNR values are much lower than for our MBHB systems, the Fisher matrix estimates represent quite well the marginal
distributions of most parameters. We note, however, some notable deviation from Gaussianity for the system II: the sky position is almost bimodal, with an approximate symmetry across the ecliptic plane (we recall here that the ecliptic plane is not the plane of the LISA instrument that was used to report MBHB results), owing probably to the localization being close to this plane; and the distance and inclination posteriors show a long tail extending way beyond the Fisher estimate.

\section{Summary and discussion}
\label{sec:summarydiscussion}

In this work, we have explored the Bayesian parameter estimation for full, inspiral-merger-ringdown MBHB signals including higher harmonics, albeit neglecting the spin degrees of freedom, and we have also produced posterior distributions for the extrinsic parameters of SBHB signals.

We improved the speed of likelihood computations by combining several ingredients: fast IMR Fourier-domain waveforms, a fast Fourier-domain treatment of the instrument response and accelerated overlap computations between amplitude/phase signals in the absence of noise. The resulting likelihood costs, of about 15ms for waveforms with higher harmonics and 1ms for waveforms including only the dominant harmonic, allowed us to perform easily Bayesian parameter estimation on high-SNR signals requiring up to $10^{8}$ likelihood evaluations with standard Bayesian samplers.

Using these tools to simulate the parameter recovery of two examples of MBHB signals, we produced multidimensional posterior distributions.
Because our study did not consider spins and their well-known degeneracy with masses, our posterior distributions for masses cannot yet be taken as representative of the actual LISA capabilities.
Nevertheless, we encountered and investigated two different kinds of degeneracies in the extrinsic parameter space.

Considering first IMR signals, we found a \textit{two-fold} degeneracy for the sky position corresponding to the \textit{reflected} sky localization with respect to the plane of the LISA constellation (opposite sign of the LISA-frame latitude). We also found that, when limiting the signals to their dominant harmonic $h_{22}$, strong degeneracies in distance-inclination can occur, causing the posterior distribution in the sky to appear shifted from the injected value. We gave an analytic explanation of this degeneracy by eliminating the time and frequency dependencies in the response, reducing it to an explicitly degenerate form. However, including higher harmonics in the analysis strongly breaks those degeneracies, and leads to a much better determination of both distance-inclination and sky position. Although the reflected sky localization survives with higher harmonics, we also note that the posteriors appear as much more Gaussian in all extrinsic parameters; using a simplified signal with only the dominant harmonic significantly complicates the posterior sampling.

Secondly, considering pre-merger analyses to investigate how information accumulates with time, we found an \textit{eight-fold} degeneracy for the sky position, obtained by rotating the original and reflected sky positions by integer multiples of $\pi/2$ around the axis of the LISA constellation. For our example systems, the posterior only collapses on the two-fold degeneracy, noticed previously for IMR signals, minutes before merger. By decomposing the LISA instrument response into different effects, separating its time-dependency from its frequency-dependency, we could pinpoint the role of both effects. The time-dependency of the response, consequence of the motion of LISA, leaves a moderate imprint at low frequencies, penalizing moderately other modes in the sky but leaving the \textit{antipodal} sky position degenerate. By contrast, the frequency-dependency of the response, not included in previous parameter studies, is responsible for the degeneracy breaking and the collapse from eight to two modes in the sky. Note that the relative importance of the time and frequency dependency of the response will change for heavier or lighter systems.

We also explored the recovery of extrinsic parameters of SBHB signals (fixing the masses and spin parameters as well as the merger time), using the same methodology as for MBHB signals to arrive at a likelihood cost of $\sim 2 \mathrm{ms}$. Investigating two example systems, we found that the distribution of extrinsic parameters is mostly Gaussian, with a bimodality in the sky position found for one source lying close to the ecliptic plane, and an extended tail in the distance posterior.

Comparing the parameter estimation results obtained with our two samplers, $\texttt{ptmcmc}$ and $\texttt{multinest}$, we found an overall good agreement. We noted however that \texttt{multinest} can fall short of exploring the full degeneracies in parameter space, in the degenerate case of a signal without higher harmonics; this limitation is further illustrated by the example of a fully degenerate response. We also found that using Fisher matrix estimates does not reproduce Bayesian results in general: the complicated degeneracies in the sky cannot be covered by a Fisher computation that is intrinsically unimodal, and neither can the specific features of the $22$-mode degenerate posteriors. That degeneracies occur even with the high SNRs of our MBHB signals is remarkable, and reminds us that the usual statement of the validity of the Fisher matrix approximation in the high-SNR limit is very dependent on the morphology of the signal. In many respects, the moderate-SNR posteriors for our SBHB systems are more Gaussian than our high-SNR MBHB $22$-mode posteriors.

Our exploratory work calls for a number of natural extensions that will help us to better understand the future capabilities of the LISA instrument.

First, we only investigated a handful of sources. In particular, we chose our MBHB systems to be representative of the bulk of the expected population; they are not ``golden sources'', i.e. they are not part of the tail of lower-redshift systems that will deliver the most interesting science outputs with LISA, notably in terms of electromagnetic counterparts. It is likely that the parameter recovery here will be different, with a longer detectable signal and the motion of LISA playing a role in the premerger localization. We also limited ourselves to a single redshifted mass, while the morphology of the signals will change strongly between the two ends of the LISA spectrum, from intermediate mass black holes ($M\sim 10^{3} M_{\odot}$) to massive systems ($M>10^{7} M_{\odot}$).

Second, it will be necessary to incorporate more physics in our waveform models.
In this study our signals were limited to non-spinning, quasicircular systems.
The spin components along the orbital angular momentum will be degenerate with the masses, but we do not expect this to qualitatively change the posterior distributions for the extrinsic parameters.
This will be investigated in a forthcoming publication~\cite{Katz+20}.
Orbital precession and eccentricity, however, will change the harmonic structure of the signals and we expect they will play a much more important role.

Third, we used zero-noise realizations in order to accelerate the likelihood computation.
Although we do not expect that the noise realization will cause drastic changes, with in particular the structure of degeneracies of signals in parameter space remaining relevant, fast likelihoods for pre-sampled noisy data will eventually be necessary.
Analyzing signals with a noise injection will also become more important when considering more realistic instrumental noise models and in prototyping algorithms for analysis of actual LISA data.

We also note that our sampling tools themselves (\texttt{multinest} and \texttt{ptmcmc}) were not tailored for the problem at hand; it is very possible that better sampling algorithm would require much less evaluations of the likelihood. The understanding we gained of the parameter space degeneracies will allow us to inform the sampler with e.g. tailored jump proposals, effectively telling MCMC chains where to look for degenerate signals in parameter space. For instance, parameter estimation tools used in ground-based gravitational-wave astronomy make use of an analytical knowledge of degeneracies~\cite{Veitch+14}.

Finally, we recall that many scientific questions about LISA rely on parameter estimation tools, and would benefit from proper Bayesian analyses with realistic signals: pre-merger localization of sources and advance warnings for electromagnetic instruments, cosmography using LISA sources as standard sirens, astrophysical inference from the source population, accuracy requirements on waveform models, trade-offs between instrumental design and scientific outputs. We will keep extending our tools to address all these applications.


\appendix


\section{Conventions for orbits and frames}
\label{app:conventions}

For completeness, we give in this Appendix the definitions and convention choices that were sketched in Section~\ref{sec:response} but were not essential for the discussion.


\subsection{Source frame, wave frame and SSB frame}
\label{app:wavessbframe}

We first introduce a source frame $(\hat{x}_{S}, \hat{y}_{S}, \hat{z}_{S})$ attached to the binary system emitting gravitational waves. Relating this frame to the physical configuration of the binary is understood as being part of the waveform model. For comparable-mass systems without spin or with aligned spins\footnote{For precessing systems, different choices are possible, like $\hat{z}_{S} = \hat{J}$, the direction of the total angular momentum.}, the natural choice is to take the normal to the orbital plane as $\hat{z}_{S}$, and we will assume that the remaining rotation around $\hat{z}_{S}$ is fixed by the phase convention of the waveform model.

Introducing $k$ the wave propagation unit vector going from the source towards the observer, we define the inclination $\iota$ and the observer phase $\varphi$ as its spherical angular coordinates in the source frame, so that in that frame
\be
	k_{S} = (\sin\iota \cos\varphi, \sin\iota \sin\varphi, \cos\iota) \,.
\ee

By convention, for our polarization vectors $p$ and $q$ we will use the spherical coordinate vectors $p = e_{\theta}^{S}$, $q = e_{\phi}^{S}$. In terms of only $\hat{z}_{S}$,
\bsub
\begin{align}
	p &= q \times k \,,\\
	q &= \hat{z}_{S} \times k / |\hat{z}_{S} \times k|\,.
\end{align}
\esub

The vectors $(p,q,k)$ form the wave frame, such that the gravitational wave takes in this frame the familiar form
\be
	H_{W} = \begin{pmatrix}
		h_{+} & h_{\times} & 0 \\
		h_{\times} & -h_{+} & 0 \\
		0 & 0 & 0
		\end{pmatrix} \,.
\ee

We can now relate the wave frame to a detector frame $(\hat{x}, \hat{y}, \hat{z})$, which will fix our convention for the sky position and polarization angle. We choose here this detector frame to be based on the plane of ecliptic, and centered on the Solar System Barycenter; we will call this frame the SSB frame. Below, we will introduce another detector frame more suitable for short-lived signals, the LISA frame.

The source position in the sky is given by $(\lambda, \beta)$, the ecliptic longitude and latitude in this SSB-frame. Like with the source frame, we can introduce spherical vectors $(e_{r}^{SSB}, e_{\theta}^{SSB}, e_{\phi}^{SSB})$. Since the propagation vector is $k = - e_{r}^{SSB}$,
\be
	k = (- \cos\beta\cos\lambda, - \cos\beta\sin\lambda, -\sin\beta) \,.
\ee

The last degree of freedom between the frames represents a rotation along the line-of-sight, parametrized by the polarization angle $\psi$. We introduce reference polarization vectors in the SSB-frame as
\bsub
\begin{align}
	u &= \hat{z} \times k / |\hat{z} \times k| \,,\\
	v &= k \times u \,.
\end{align}
\esub
In terms of spherical vectors, $(k, u, v) = (-e_{r}^{SSB}, -e_{\phi}^{SSB}, -e_{\theta}^{SSB})$.

In the following, we will use the notation $R(v,\varpi)$ to denote the matrix of an active rotation around the vector $v$ by an angle $\varpi$. The polarization angle $\psi$ is then defined such that $(p,q)$ are obtained by rotating $(u,v)$ by the angle $\psi$ around $k$, i.e.
\be
	(p,q) = R(k, \psi) \cdot (u,v) \,.
\ee

These relations can be summarized by the active rotation matrix $R_{W}$ from the SSB-frame to the wave frame:
\be\label{eq:RW}
	R_{W} = R(z, \lambda - \pi/2) \cdot R(x, \beta + \pi/2 ) \cdot R(z, \psi)\,.
\ee


\subsection{LISA trajectories}
\label{app:lisatraj}

The orbits of the three spacecrafts around the Sun can be chosen so that, at leading order in the eccentricity of the orbits, the constellation retains the shape of an equilateral triangle in a cartwheeling motion following the Earth orbit. We denote by $a$ the semi-major axis, $e$ the eccentricity and pose $\alpha = \Omega_{0} (t-t_{0}))$, with $t_{0}$ a reference time for the initial position ($t_{0} = 0$ in our case). The reference SSB frame is $(\hat{x}, \hat{y}, \hat{z})$.

Using the notation $c,s = \cos, \sin \alpha$, the trajectory of the center of the constellation is simply
\be
	p_{0} = ac \hat{x} + as \hat{y} \,,
\ee
and we take $a\equiv R = 1 \, \mathrm{au}$. We denote the positions of the spacecrafts as $p_{A}$ for $A=1,2,3$, and the positions relative to the constellation center as $p_{A}^{L} = p_{A} - p_{0}$. Setting $\beta_{A} = 2(A-1)\pi/3 + \beta_{0}$ (with $\beta_{0}$ and initial condition set to 0 in our case), the Cartesian coordinates in the SSB-frame of the position of the spacecrafts read
\bsub
\begin{align}
	p_{A}^{L} &= a e \left[ \sin \beta_{A} c s - \cos\beta_{A} \left( 1 + s^{2} \right) \right] \hat{x} \nn\\
	& + a e \left[ \cos \beta_{A} c s - \sin\beta_{A} \left( 1 + c^{2} \right) \right] \hat{y} \nn\\
	& - a e \sqrt{3} \cos(\alpha - \beta_{A}) \hat{z} \,,
\end{align}
\esub
The armlength, constant in this approximation, is related to the eccentricity and semi-major axis by
\be
	L = 2\sqrt{3} a e \,.
\ee
The rigid approximation for the constellation, at first order in $e$, can be seen as a first-order approximation in the small parameter $L/R \simeq 0.017$ for a $2.5 \, \mathrm{Gm}$ armlength.


\subsection{The LISA frame}
\label{app:LISAframe}

It will be very useful to introduce a time-dependent frame $(\hat{x}_{L}, \hat{y}_{L}, \hat{z}_{L})(t)$ following the detector in its motion. Specifically, we choose this LISA frame (L-frame for short) such that at any time,
\bsub
\begin{align}
	p_{1}^{L} &= - \frac{L}{\sqrt{3}} \hat{x}_{L} \,,\\
	p_{2}^{L} &= \frac{L}{2\sqrt{3}} \hat{x}_{L} - \frac{L}{2} \hat{y}_{L} \,,\\
	p_{3}^{L} &= \frac{L}{2\sqrt{3}} \hat{x}_{L} + \frac{L}{2} \hat{y}_{L} \,.\\
\end{align}
\esub

This frame also provides us with an equivalent representation of the trajectories making use of rotation matrices. Since in our rigid approximation the constellation remains an equilateral triangle in its cartwheeling motion around the Sun, the configuration of the constellation at a later time is given by a rotation around the Sun composed with a rotation of the constellation around its symmetry axis.

If we denote by $R_{L}$ the active rotation from the SSB-frame to the L-frame such that for each of the three basis vectors $(\hat{x}_{L}, \hat{y}_{L}, \hat{z}_{L}) = R_{L} (\hat{x}, \hat{y}, \hat{z})$, we have
\be\label{eq:RL}
	R_{L} = R(z, \alpha) \cdot R(y, -\pi/3) \cdot R(z, -\alpha) \,,
\ee
where we recall that $\alpha = \Omega_{0} (t-t_{0}) $. For all vectors $X$ among $p_{A}^{L}$, $n_{A}$ we have $X(t) = R_{L} \cdot X(t=t_{0})$. If a vector $X$ is given by its components in the SSB-frame, in the L-frame the components are $X_{L} = R_{L}^{-1} \cdot X$.

While long-lasting signals like SBHBs will see a strong imprint of the LISA orbital motion over the course of observation, MBHB signals are dominated in SNR by a short-lived burst of emission at merger. For such signals, it will be useful to use a parametrization based on the LISA frame at the time of merger instead of the SSB frame.

The new parameters are defined as playing the same role as the SSB parameters, but relative to the L-frame. Namely, we define $R_{LW}$ the active rotation matrix from the L-frame to the wave frame, expressed in the L-frame basis, so that for each basis vector expressed in the L-frame basis $(\hat{x}_{W}, \hat{y}_{W}, \hat{z}_{W})_{L} = R_{LW} \cdot (\hat{x}_{L}, \hat{y}_{L}, \hat{z}_{L})_{L}$. The defining condition on $(\lambda_{L}, \beta_{L}, \psi_{L})$ is
\be\label{eq:defRLW}
	R_{LW} = R(z, \lambda_{L} - \pi/2) \cdot R(x, \beta_{L} + \pi/2 ) \cdot R(z, \psi_{L}) \,.
\ee

Coming back to vectors expressed in the SSB-basis, $R_{L}^{-1} \cdot (\hat{x}_{W}, \hat{y}_{W}, \hat{z}_{W}) = R_{LW} \cdot R_{L}^{-1} \cdot (\hat{x}_{L}, \hat{y}_{L}, \hat{z}_{L})$. Since $ (\hat{x}_{W}, \hat{y}_{W}, \hat{z}_{W}) = R_{W} \cdot (\hat{x}, \hat{y}, \hat{z})$ and $ (\hat{x}_{L}, \hat{y}_{L}, \hat{z}_{L}) = R_{L} \cdot (\hat{x}, \hat{y}, \hat{z})$, we arrive at
\be\label{eq:RLW}
	R_{LW} = R_{L}^{-1} \cdot R_{W} \,.
\ee

Combining~\eqref{eq:RLW}, \eqref{eq:RL}, \eqref{eq:RW} and~\eqref{eq:defRLW} yields the following expressions for the L-frame angular parameters\footnote{The convention here is that the point of coordinates $(x,y)$ in the plane has for argument $\arctan\left[ x, y \right]$.}:
\bsub\label{eq:Lframeangles}
\begin{align}
	\beta_{L} &= \arcsin \left[ \cos \frac{\pi}{3} \sin \beta - \sin \frac{\pi}{3} \cos \beta \cos \left( \lambda - \alpha \right) \right] \,,\\
	\lambda_{L} &= \arctan \left[ \cos\beta \cos\lambda \left( \cos \frac{\pi}{3} \cos^{2}\alpha + \sin^{2}\alpha \right) \right. \nn\\
	& \qquad\quad\;\; \left. + \cos\beta \sin\lambda \cos\alpha \sin\alpha \left( \cos\frac{\pi}{3} - 1 \right) \right. \nn\\
	& \qquad\quad\;\; \left. + \sin\frac{\pi}{3} \sin\beta \cos\alpha , \right. \nn\\
	& \qquad\qquad\; \left. \cos\beta \sin\lambda \left( \cos \frac{\pi}{3} \sin^{2}\alpha + \cos^{2}\alpha \right) \right. \nn\\
	& \qquad\quad\;\; \left. + \cos\beta \cos\lambda \cos\alpha \sin\alpha \left( \cos\frac{\pi}{3} - 1 \right) \right. \nn\\
	& \qquad\quad\;\; \left. + \sin\frac{\pi}{3} \sin\beta \sin\alpha \right] \,,\\
	\psi_{L} &= \psi + \arctan \left[\cos\frac{\pi}{3}\cos\beta + \sin\frac{\pi}{3} \sin\beta \cos(\lambda - \alpha), \right. \nn\\
	&\qquad\qquad\quad\;\; \left.-\sin \frac{\pi}{3} \sin(\lambda - \alpha) \right] \,.
\end{align}
\esub
Through $\alpha = \Omega_{0} (t-t_{0}) $, these expressions are time-dependent, reflecting the rotation of this frame to follow LISA on its orbit.

Finally, it is useful to introduce a ``time-at-LISA'' parameter $t_{L}$, taken into account the propagation delay from the SSB to the center of the LISA constellation, as
\be\label{eq:deftL}
	t_{L} = t + k\cdot p_{0} \,.
\ee
Labeling short-lived MBHB signals by their time of arrival at the SSB is a bad parametrization. For a given measured arrival time at the LISA instrument $t_{L}$, different sky positions will give separate peaks in the ``time-at-SSB'' variable $t$. The use of $t_{L}$ eliminates this issue.


\subsection{LISA instrumental nosie}
\label{app:LISAnoise}

The instrumental noise we consider in this study is constructed from~\cite{LISANoiseBudget16}, which specifies the LISA noise budget for different subsystems. The numerical expressions we use are (with all quantities in SI units):
\begin{widetext}
\bsub
\begin{align}
	S^{\rm pm} (f) &= \left(\frac{1}{2\pi f c}\right)^{2} (3\cdot 10^{-15})^{2} \left[1 + 36 \left( \left(\frac{10^{-4}}{f}\right)^{2} + \left(\frac{3\cdot10^{-5}}{f}\right)^{10} \right)\right] + \frac{1}{4}\left(1.7\cdot 10^{-12}\right)^{2} \left(\frac{2\pi f}{c}\right)^{2} + S^{\rm WD}(f) \,,\\
	S^{\rm op} (f) &= \left(\frac{2\pi f}{c}\right)^{2} \left[\left(8.9\cdot 10^{-12}\right)^{2} + \left(1.7\cdot 10^{-12}\right)^{2} + \left(2\cdot 10^{-12}\right)^{2}\right] \,.
\end{align}
\esub
\end{widetext}


%


%
%


\section{Sampling the fully degenerate extrinsic posterior}
\label{app:samplingdegen}

\begin{figure*}
  \centering
  \includegraphics[width=.9\linewidth]{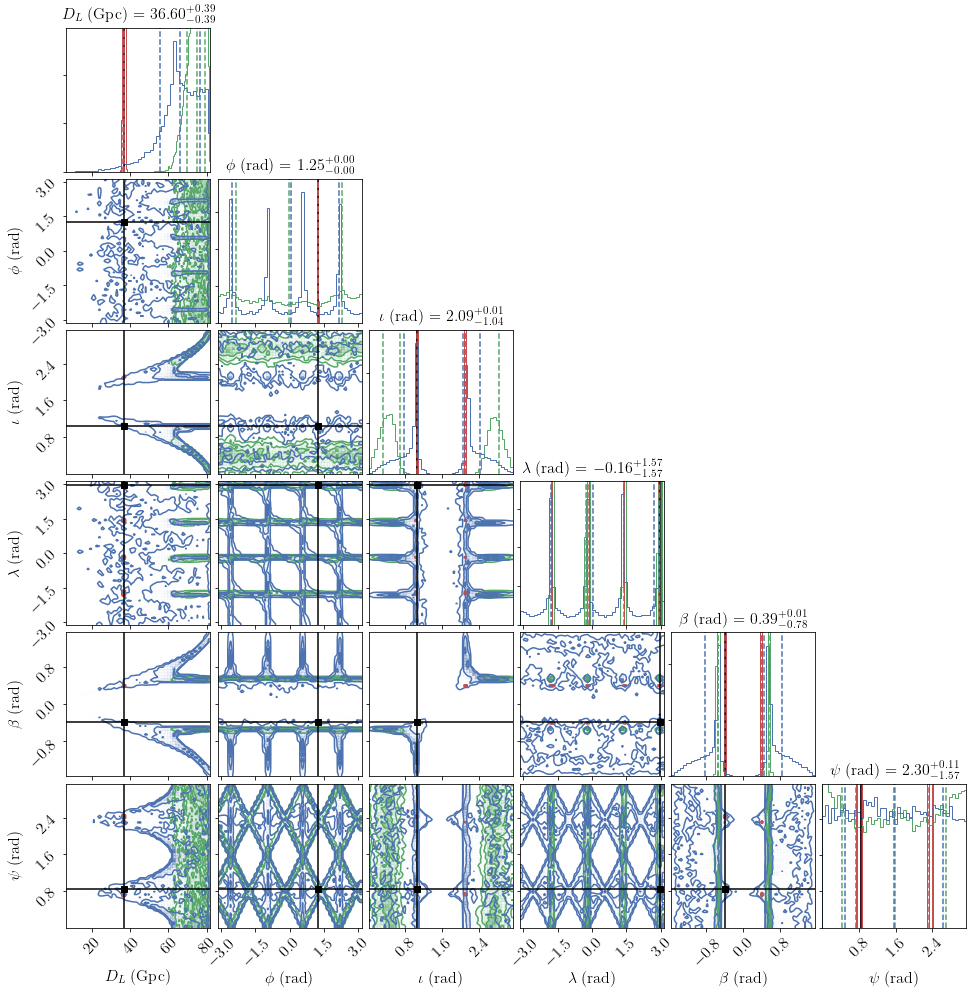}
  \caption{Posterior distributions obtained with the two samplers \texttt{ptmcmc} (blue) and $\texttt{multinest}$ (green) for the simplified likelihood~\eqref{eq:simplelikelihoodhm}, using only the dominant mode $(\ell, m) = (2,2)$. Masses and time are pinned to the injected value, and the motion of the detector is neglected. We also superimpose (red) the contours obtained with \texttt{multinest} by including all harmonics $(2,2)$, $(2,1)$, $(3,3)$, $(4,4)$, $(5,5)$.}
  \label{fig:PEsmbh22hmSimpleLikeCase9}
\end{figure*}

In this Appendix, we further illustrate our findings of Sec.~\ref{subsec:MBHBPEdegen22} by running our Bayesian parameter estimation codes with the \textit{simplified likelihood}~\eqref{eq:simplelikelihood22} consisting in pinning intrinsic parameters (masses and time) and using the \textit{Frozen, low-$f$} response approximation.

Since the inner products~\eqref{eq:innerproductlmlpmp} are constants computed only once, the external factors become trivially inexpensive, and the likelihood is very fast to compute. We take advantage of this to run \texttt{ptmcmc} for a very large number of steps, ensuring that we do not miss degenerate regions of the parameter space. The resulting posterior distribution is extremely degenerate, as shown in Fig.~\ref{fig:PEsmbh22hmSimpleLikeCase9}. The sampler \texttt{multinest} fails to resolve this structure, stopping after having only explored the degenerate region $\iota \rightarrow 0$ and $\iota \rightarrow \pi$ and the eight sky positions~\eqref{eq:soldegensky}. In accordance with~\eqref{eq:soldegensky}, this corresponds to lines in $\varphi \pm \psiL = \mathrm{const}$. The sampler \texttt{ptmcmc} finds more degenerate regions, in particular it finds similar structures with the roles of $\iota$ and $\betaL$ exchanged following~\eqref{eq:soldegeniotaphi} (as shown notably by the $(d,\iota)$ and $(d,\betaL)$ panels), as well as intermediate regions connecting all these features.

Fig.~\ref{fig:PEsmbh22hmSimpleLikeCase9} also shows the result obtained when including all available harmonics~\eqref{eq:listmodes} in~\eqref{eq:simplelikelihoodhm}. The degenerate structures collapse to leave the eight modes corresponding to $\betaL \rightarrow - \betaL$ and $\lambdaL \rightarrow \lambdaL + k\pi/2$, this time centered around the injected value and not the shifted value $(\lambda_{L}^{\dagger}, \beta_{L}^{\dagger})$. We found that this degeneracy-breaking occurs already when including a single subdominant harmonic (for instance $h_{21}$).

These results also give us a hint to explain the results of Sec.~\ref{subsec:SMBHPEfull}, and in particular the difference between Bayesian samplers shown in the center left panels of Figs.~\ref{fig:PEsmbhDincptmcmcbambi} and~\ref{fig:PEsmbhskyptmcmcbambi}. Since making the likelihood more degenerate in the simplified likelihood limit~\eqref{eq:simplelikelihood22} magnifies the difference between \texttt{ptmcmc} and \texttt{multinest}, with \texttt{ptmcmc} exploring a much larger volume in parameter space, their disagreement with the full likelihood can likely be attributed to an apparent shortcoming of \texttt{multinest}\footnote{We only used standard values for the metaparameters for this sampler, and we do not exclude that it would be possible to improve results with minor changes.} for resolving extended degenerate regions.


%
%
%


\begin{acknowledgments}

We would like to thank Stanislav Babak, Michael Katz, Antoine Klein, Vivien Raymond and Alexandre Toubiana for useful discussions.
We are indebted to Philip Graff for his help with the \texttt{bambi} sampler.
TDC was supported by an appointment to the NASA Postdoctoral Program at the Goddard Space Flight Center, administered by Universities Space Research Association under contract with NASA, during part of this work.

\end{acknowledgments}

\bibliography{references.bib}

\begin{thebibliography}{101}%
\makeatletter
\providecommand \@ifxundefined [1]{%
 \@ifx{#1\undefined}
}%
\providecommand \@ifnum [1]{%
 \ifnum #1\expandafter \@firstoftwo
 \else \expandafter \@secondoftwo
 \fi
}%
\providecommand \@ifx [1]{%
 \ifx #1\expandafter \@firstoftwo
 \else \expandafter \@secondoftwo
 \fi
}%
\providecommand \natexlab [1]{#1}%
\providecommand \enquote  [1]{``#1''}%
\providecommand \bibnamefont  [1]{#1}%
\providecommand \bibfnamefont [1]{#1}%
\providecommand \citenamefont [1]{#1}%
\providecommand \href@noop [0]{\@secondoftwo}%
\providecommand \href [0]{\begingroup \@sanitize@url \@href}%
\providecommand \@href[1]{\@@startlink{#1}\@@href}%
\providecommand \@@href[1]{\endgroup#1\@@endlink}%
\providecommand \@sanitize@url [0]{\catcode `\\12\catcode `\$12\catcode
  `\&12\catcode `\#12\catcode `\^12\catcode `\_12\catcode `\%12\relax}%
\providecommand \@@startlink[1]{}%
\providecommand \@@endlink[0]{}%
\providecommand \url  [0]{\begingroup\@sanitize@url \@url }%
\providecommand \@url [1]{\endgroup\@href {#1}{\urlprefix }}%
\providecommand \urlprefix  [0]{URL }%
\providecommand \Eprint [0]{\href }%
\providecommand \doibase [0]{http://dx.doi.org/}%
\providecommand \selectlanguage [0]{\@gobble}%
\providecommand \bibinfo  [0]{\@secondoftwo}%
\providecommand \bibfield  [0]{\@secondoftwo}%
\providecommand \translation [1]{[#1]}%
\providecommand \BibitemOpen [0]{}%
\providecommand \bibitemStop [0]{}%
\providecommand \bibitemNoStop [0]{.\EOS\space}%
\providecommand \EOS [0]{\spacefactor3000\relax}%
\providecommand \BibitemShut  [1]{\csname bibitem#1\endcsname}%
\let\auto@bib@innerbib\@empty
\bibitem [{\citenamefont {Abbott}\ \emph {et~al.}(2019)\citenamefont {Abbott}
  \emph {et~al.}}]{GWTC-1}%
  \BibitemOpen
  \bibfield  {author} {\bibinfo {author} {\bibfnamefont {B.~P.}\ \bibnamefont
  {Abbott}} \emph {et~al.} (\bibinfo {collaboration} {LIGO Scientific,
  Virgo}),\ }\href {\doibase 10.1103/PhysRevX.9.031040} {\bibfield  {journal}
  {\bibinfo  {journal} {Phys. Rev.}\ }\textbf {\bibinfo {volume} {X9}},\
  \bibinfo {pages} {031040} (\bibinfo {year} {2019})},\ \Eprint
  {http://arxiv.org/abs/1811.12907} {arXiv:1811.12907 [astro-ph.HE]}
  \BibitemShut {NoStop}%
\bibitem [{\citenamefont {Aasi}\ \emph {et~al.}(2015)\citenamefont {Aasi} \emph
  {et~al.}}]{LIGO}%
  \BibitemOpen
  \bibfield  {author} {\bibinfo {author} {\bibfnamefont {J.}~\bibnamefont
  {Aasi}} \emph {et~al.} (\bibinfo {collaboration} {LIGO Scientific}),\ }\href
  {\doibase 10.1088/0264-9381/32/7/074001} {\bibfield  {journal} {\bibinfo
  {journal} {Class. Quant. Grav.}\ }\textbf {\bibinfo {volume} {32}},\ \bibinfo
  {pages} {074001} (\bibinfo {year} {2015})},\ \Eprint
  {http://arxiv.org/abs/1411.4547} {arXiv:1411.4547 [gr-qc]} \BibitemShut
  {NoStop}%
\bibitem [{\citenamefont {Acernese}\ \emph {et~al.}(2015)\citenamefont
  {Acernese} \emph {et~al.}}]{Virgo}%
  \BibitemOpen
  \bibfield  {author} {\bibinfo {author} {\bibfnamefont {F.}~\bibnamefont
  {Acernese}} \emph {et~al.} (\bibinfo {collaboration} {VIRGO}),\ }\href
  {\doibase 10.1088/0264-9381/32/2/024001} {\bibfield  {journal} {\bibinfo
  {journal} {Class. Quant. Grav.}\ }\textbf {\bibinfo {volume} {32}},\ \bibinfo
  {pages} {024001} (\bibinfo {year} {2015})},\ \Eprint
  {http://arxiv.org/abs/1408.3978} {arXiv:1408.3978 [gr-qc]} \BibitemShut
  {NoStop}%
\bibitem [{\citenamefont {Aso}\ \emph {et~al.}(2013)\citenamefont {Aso},
  \citenamefont {Michimura}, \citenamefont {Somiya}, \citenamefont {Ando},
  \citenamefont {Miyakawa}, \citenamefont {Sekiguchi}, \citenamefont
  {Tatsumi},\ and\ \citenamefont {Yamamoto}}]{KAGRA}%
  \BibitemOpen
  \bibfield  {author} {\bibinfo {author} {\bibfnamefont {Y.}~\bibnamefont
  {Aso}}, \bibinfo {author} {\bibfnamefont {Y.}~\bibnamefont {Michimura}},
  \bibinfo {author} {\bibfnamefont {K.}~\bibnamefont {Somiya}}, \bibinfo
  {author} {\bibfnamefont {M.}~\bibnamefont {Ando}}, \bibinfo {author}
  {\bibfnamefont {O.}~\bibnamefont {Miyakawa}}, \bibinfo {author}
  {\bibfnamefont {T.}~\bibnamefont {Sekiguchi}}, \bibinfo {author}
  {\bibfnamefont {D.}~\bibnamefont {Tatsumi}}, \ and\ \bibinfo {author}
  {\bibfnamefont {H.}~\bibnamefont {Yamamoto}} (\bibinfo {collaboration} {The
  KAGRA Collaboration}),\ }\href {\doibase 10.1103/PhysRevD.88.043007}
  {\bibfield  {journal} {\bibinfo  {journal} {Phys. Rev. D}\ }\textbf {\bibinfo
  {volume} {88}},\ \bibinfo {pages} {043007} (\bibinfo {year}
  {2013})}\BibitemShut {NoStop}%
\bibitem [{\citenamefont {{Amaro-Seoane}}\ and\ \citenamefont
  {et~al.}(2017)}]{LISA2017}%
  \BibitemOpen
  \bibfield  {author} {\bibinfo {author} {\bibfnamefont {P.}~\bibnamefont
  {{Amaro-Seoane}}}\ and\ \bibinfo {author} {\bibnamefont {et~al.}},\
  }\href@noop {} {\bibfield  {journal} {\bibinfo  {journal} {ArXiv e-prints}\ }
  (\bibinfo {year} {2017})},\ \Eprint {http://arxiv.org/abs/1702.00786}
  {arXiv:1702.00786 [astro-ph.IM]} \BibitemShut {NoStop}%
\bibitem [{\citenamefont {{Klein}}\ \emph {et~al.}(2016)\citenamefont
  {{Klein}}, \citenamefont {{Barausse}}, \citenamefont {{Sesana}},
  \citenamefont {{Petiteau}}, \citenamefont {{Berti}}, \citenamefont {{Babak}},
  \citenamefont {{Gair}}, \citenamefont {{Aoudia}}, \citenamefont {{Hinder}},
  \citenamefont {{Ohme}},\ and\ \citenamefont {{Wardell}}}]{Klein+15}%
  \BibitemOpen
  \bibfield  {author} {\bibinfo {author} {\bibfnamefont {A.}~\bibnamefont
  {{Klein}}}, \bibinfo {author} {\bibfnamefont {E.}~\bibnamefont {{Barausse}}},
  \bibinfo {author} {\bibfnamefont {A.}~\bibnamefont {{Sesana}}}, \bibinfo
  {author} {\bibfnamefont {A.}~\bibnamefont {{Petiteau}}}, \bibinfo {author}
  {\bibfnamefont {E.}~\bibnamefont {{Berti}}}, \bibinfo {author} {\bibfnamefont
  {S.}~\bibnamefont {{Babak}}}, \bibinfo {author} {\bibfnamefont
  {J.}~\bibnamefont {{Gair}}}, \bibinfo {author} {\bibfnamefont
  {S.}~\bibnamefont {{Aoudia}}}, \bibinfo {author} {\bibfnamefont
  {I.}~\bibnamefont {{Hinder}}}, \bibinfo {author} {\bibfnamefont
  {F.}~\bibnamefont {{Ohme}}}, \ and\ \bibinfo {author} {\bibfnamefont
  {B.}~\bibnamefont {{Wardell}}},\ }\href {\doibase 10.1103/PhysRevD.93.024003}
  {\bibfield  {journal} {\bibinfo  {journal} {Phys. Rev. D}\ }\textbf {\bibinfo
  {volume} {93}},\ \bibinfo {eid} {024003} (\bibinfo {year} {2016})},\ \Eprint
  {http://arxiv.org/abs/1511.05581} {arXiv:1511.05581 [gr-qc]} \BibitemShut
  {NoStop}%
\bibitem [{\citenamefont {{Sesana}}(2016)}]{Sesana16}%
  \BibitemOpen
  \bibfield  {author} {\bibinfo {author} {\bibfnamefont {A.}~\bibnamefont
  {{Sesana}}},\ }\href {\doibase 10.1103/PhysRevLett.116.231102} {\bibfield
  {journal} {\bibinfo  {journal} {Physical Review Letters}\ }\textbf {\bibinfo
  {volume} {116}},\ \bibinfo {eid} {231102} (\bibinfo {year} {2016})},\ \Eprint
  {http://arxiv.org/abs/1602.06951} {arXiv:1602.06951 [gr-qc]} \BibitemShut
  {NoStop}%
\bibitem [{\citenamefont {{Nelemans}}\ \emph {et~al.}(2001)\citenamefont
  {{Nelemans}}, \citenamefont {{Yungelson}},\ and\ \citenamefont {{Portegies
  Zwart}}}]{Nelemans+01}%
  \BibitemOpen
  \bibfield  {author} {\bibinfo {author} {\bibfnamefont {G.}~\bibnamefont
  {{Nelemans}}}, \bibinfo {author} {\bibfnamefont {L.~R.}\ \bibnamefont
  {{Yungelson}}}, \ and\ \bibinfo {author} {\bibfnamefont {S.~F.}\ \bibnamefont
  {{Portegies Zwart}}},\ }\href {\doibase 10.1051/0004-6361:20010683}
  {\bibfield  {journal} {\bibinfo  {journal} {Astron. Astrophys.}\ }\textbf
  {\bibinfo {volume} {375}},\ \bibinfo {pages} {890} (\bibinfo {year}
  {2001})},\ \Eprint {http://arxiv.org/abs/astro-ph/0105221} {astro-ph/0105221}
  \BibitemShut {NoStop}%
\bibitem [{\citenamefont {{Armitage}}\ and\ \citenamefont
  {{Natarajan}}(2002)}]{ArmitageNatarajan02}%
  \BibitemOpen
  \bibfield  {author} {\bibinfo {author} {\bibfnamefont {P.~J.}\ \bibnamefont
  {{Armitage}}}\ and\ \bibinfo {author} {\bibfnamefont {P.}~\bibnamefont
  {{Natarajan}}},\ }\href {\doibase 10.1086/339770} {\bibfield  {journal}
  {\bibinfo  {journal} {Astrophys. J. Letters}\ }\textbf {\bibinfo {volume}
  {567}},\ \bibinfo {pages} {L9} (\bibinfo {year} {2002})},\ \Eprint
  {http://arxiv.org/abs/astro-ph/0201318} {arXiv:astro-ph/0201318 [astro-ph]}
  \BibitemShut {NoStop}%
\bibitem [{\citenamefont {{Dal Canton}}\ \emph {et~al.}(2019)\citenamefont
  {{Dal Canton}}, \citenamefont {{Mangiagli}}, \citenamefont {{Noble}},
  \citenamefont {{Schnittman}}, \citenamefont {{Ptak}}, \citenamefont
  {{Klein}}, \citenamefont {{Sesana}},\ and\ \citenamefont
  {{Camp}}}]{DalCanton+19}%
  \BibitemOpen
  \bibfield  {author} {\bibinfo {author} {\bibfnamefont {T.}~\bibnamefont {{Dal
  Canton}}}, \bibinfo {author} {\bibfnamefont {A.}~\bibnamefont {{Mangiagli}}},
  \bibinfo {author} {\bibfnamefont {S.~C.}\ \bibnamefont {{Noble}}}, \bibinfo
  {author} {\bibfnamefont {J.}~\bibnamefont {{Schnittman}}}, \bibinfo {author}
  {\bibfnamefont {A.}~\bibnamefont {{Ptak}}}, \bibinfo {author} {\bibfnamefont
  {A.}~\bibnamefont {{Klein}}}, \bibinfo {author} {\bibfnamefont
  {A.}~\bibnamefont {{Sesana}}}, \ and\ \bibinfo {author} {\bibfnamefont
  {J.}~\bibnamefont {{Camp}}},\ }\href {\doibase 10.3847/1538-4357/ab505a}
  {\bibfield  {journal} {\bibinfo  {journal} {Astrophys. J.}\ }\textbf
  {\bibinfo {volume} {886}},\ \bibinfo {eid} {146} (\bibinfo {year} {2019})},\
  \Eprint {http://arxiv.org/abs/1902.01538} {arXiv:1902.01538 [astro-ph.HE]}
  \BibitemShut {NoStop}%
\bibitem [{\citenamefont {{Schutz}}(1986)}]{Schutz86}%
  \BibitemOpen
  \bibfield  {author} {\bibinfo {author} {\bibfnamefont {B.~F.}\ \bibnamefont
  {{Schutz}}},\ }\href {\doibase 10.1038/323310a0} {\bibfield  {journal}
  {\bibinfo  {journal} {Nature}\ }\textbf {\bibinfo {volume} {323}},\ \bibinfo
  {pages} {310} (\bibinfo {year} {1986})}\BibitemShut {NoStop}%
\bibitem [{\citenamefont {{Tamanini}}\ \emph {et~al.}(2016)\citenamefont
  {{Tamanini}}, \citenamefont {{Caprini}}, \citenamefont {{Barausse}},
  \citenamefont {{Sesana}}, \citenamefont {{Klein}},\ and\ \citenamefont
  {{Petiteau}}}]{Tamanini+16}%
  \BibitemOpen
  \bibfield  {author} {\bibinfo {author} {\bibfnamefont {N.}~\bibnamefont
  {{Tamanini}}}, \bibinfo {author} {\bibfnamefont {C.}~\bibnamefont
  {{Caprini}}}, \bibinfo {author} {\bibfnamefont {E.}~\bibnamefont
  {{Barausse}}}, \bibinfo {author} {\bibfnamefont {A.}~\bibnamefont
  {{Sesana}}}, \bibinfo {author} {\bibfnamefont {A.}~\bibnamefont {{Klein}}}, \
  and\ \bibinfo {author} {\bibfnamefont {A.}~\bibnamefont {{Petiteau}}},\
  }\href {\doibase 10.1088/1475-7516/2016/04/002} {\bibfield  {journal}
  {\bibinfo  {journal} {JCAP}\ }\textbf {\bibinfo {volume} {2016}},\ \bibinfo
  {eid} {002} (\bibinfo {year} {2016})},\ \Eprint
  {http://arxiv.org/abs/1601.07112} {arXiv:1601.07112 [astro-ph.CO]}
  \BibitemShut {NoStop}%
\bibitem [{\citenamefont {{Cutler}}(1998)}]{Cutler97}%
  \BibitemOpen
  \bibfield  {author} {\bibinfo {author} {\bibfnamefont {C.}~\bibnamefont
  {{Cutler}}},\ }\href {\doibase 10.1103/PhysRevD.57.7089} {\bibfield
  {journal} {\bibinfo  {journal} {Phys. Rev. D}\ }\textbf {\bibinfo {volume}
  {57}},\ \bibinfo {pages} {7089} (\bibinfo {year} {1998})},\ \Eprint
  {http://arxiv.org/abs/gr-qc/9703068} {gr-qc/9703068} \BibitemShut {NoStop}%
\bibitem [{\citenamefont {{Larson}}\ \emph {et~al.}(2000)\citenamefont
  {{Larson}}, \citenamefont {{Hiscock}},\ and\ \citenamefont
  {{Hellings}}}]{Larson+99}%
  \BibitemOpen
  \bibfield  {author} {\bibinfo {author} {\bibfnamefont {S.~L.}\ \bibnamefont
  {{Larson}}}, \bibinfo {author} {\bibfnamefont {W.~A.}\ \bibnamefont
  {{Hiscock}}}, \ and\ \bibinfo {author} {\bibfnamefont {R.~W.}\ \bibnamefont
  {{Hellings}}},\ }\href {\doibase 10.1103/PhysRevD.62.062001} {\bibfield
  {journal} {\bibinfo  {journal} {Phys. Rev. D}\ }\textbf {\bibinfo {volume}
  {62}},\ \bibinfo {eid} {062001} (\bibinfo {year} {2000})},\ \Eprint
  {http://arxiv.org/abs/gr-qc/9909080} {gr-qc/9909080} \BibitemShut {NoStop}%
\bibitem [{\citenamefont {Cornish}\ and\ \citenamefont {Rubbo}(2003)}]{CR02}%
  \BibitemOpen
  \bibfield  {author} {\bibinfo {author} {\bibfnamefont {N.~J.}\ \bibnamefont
  {Cornish}}\ and\ \bibinfo {author} {\bibfnamefont {L.~J.}\ \bibnamefont
  {Rubbo}},\ }\href {\doibase 10.1103/PhysRevD.67.022001,
  10.1103/PhysRevD.67.029905} {\bibfield  {journal} {\bibinfo  {journal} {Phys.
  Rev. D}\ }\textbf {\bibinfo {volume} {67}},\ \bibinfo {pages} {022001}
  (\bibinfo {year} {2003})},\ \bibinfo {note} {[Erratum: Phys.
  Rev.D67,029905(2003)]},\ \Eprint {http://arxiv.org/abs/gr-qc/0209011}
  {arXiv:gr-qc/0209011 [gr-qc]} \BibitemShut {NoStop}%
\bibitem [{\citenamefont {{Marsat}}\ and\ \citenamefont
  {{Baker}}(2018)}]{MB18}%
  \BibitemOpen
  \bibfield  {author} {\bibinfo {author} {\bibfnamefont {S.}~\bibnamefont
  {{Marsat}}}\ and\ \bibinfo {author} {\bibfnamefont {J.~G.}\ \bibnamefont
  {{Baker}}},\ }\href@noop {} {\bibfield  {journal} {\bibinfo  {journal} {ArXiv
  e-prints}\ ,\ \bibinfo {eid} {arXiv:1806.10734}} (\bibinfo {year} {2018})},\
  \Eprint {http://arxiv.org/abs/1806.10734} {arXiv:1806.10734 [gr-qc]}
  \BibitemShut {NoStop}%
\bibitem [{\citenamefont {{Vecchio}}(2004)}]{Vecchio03}%
  \BibitemOpen
  \bibfield  {author} {\bibinfo {author} {\bibfnamefont {A.}~\bibnamefont
  {{Vecchio}}},\ }\href {\doibase 10.1103/PhysRevD.70.042001} {\bibfield
  {journal} {\bibinfo  {journal} {Phys. Rev. D}\ }\textbf {\bibinfo {volume}
  {70}},\ \bibinfo {eid} {042001} (\bibinfo {year} {2004})},\ \Eprint
  {http://arxiv.org/abs/astro-ph/0304051} {arXiv:astro-ph/0304051 [astro-ph]}
  \BibitemShut {NoStop}%
\bibitem [{\citenamefont {{Arun}}(2006)}]{Arun06}%
  \BibitemOpen
  \bibfield  {author} {\bibinfo {author} {\bibfnamefont {K.~G.}\ \bibnamefont
  {{Arun}}},\ }\href {\doibase 10.1103/PhysRevD.74.024025} {\bibfield
  {journal} {\bibinfo  {journal} {Phys. Rev. D}\ }\textbf {\bibinfo {volume}
  {74}},\ \bibinfo {eid} {024025} (\bibinfo {year} {2006})},\ \Eprint
  {http://arxiv.org/abs/gr-qc/0605021} {gr-qc/0605021} \BibitemShut {NoStop}%
\bibitem [{\citenamefont {{Berti}}\ \emph {et~al.}(2005)\citenamefont
  {{Berti}}, \citenamefont {{Buonanno}},\ and\ \citenamefont
  {{Will}}}]{Berti+04}%
  \BibitemOpen
  \bibfield  {author} {\bibinfo {author} {\bibfnamefont {E.}~\bibnamefont
  {{Berti}}}, \bibinfo {author} {\bibfnamefont {A.}~\bibnamefont {{Buonanno}}},
  \ and\ \bibinfo {author} {\bibfnamefont {C.~M.}\ \bibnamefont {{Will}}},\
  }\href {\doibase 10.1103/PhysRevD.71.084025} {\bibfield  {journal} {\bibinfo
  {journal} {Phys. Rev. D}\ }\textbf {\bibinfo {volume} {71}},\ \bibinfo {eid}
  {084025} (\bibinfo {year} {2005})},\ \Eprint
  {http://arxiv.org/abs/gr-qc/0411129} {gr-qc/0411129} \BibitemShut {NoStop}%
\bibitem [{\citenamefont {{Lang}}\ and\ \citenamefont
  {{Hughes}}(2006)}]{LangHughes06}%
  \BibitemOpen
  \bibfield  {author} {\bibinfo {author} {\bibfnamefont {R.~N.}\ \bibnamefont
  {{Lang}}}\ and\ \bibinfo {author} {\bibfnamefont {S.~A.}\ \bibnamefont
  {{Hughes}}},\ }\href {\doibase 10.1103/PhysRevD.74.122001} {\bibfield
  {journal} {\bibinfo  {journal} {Phys. Rev. D}\ }\textbf {\bibinfo {volume}
  {74}},\ \bibinfo {eid} {122001} (\bibinfo {year} {2006})},\ \Eprint
  {http://arxiv.org/abs/gr-qc/0608062} {arXiv:gr-qc/0608062 [gr-qc]}
  \BibitemShut {NoStop}%
\bibitem [{\citenamefont {{Babak}}\ \emph {et~al.}(2010)\citenamefont {{Babak}}
  \emph {et~al.}}]{MLDC09}%
  \BibitemOpen
  \bibfield  {author} {\bibinfo {author} {\bibfnamefont {S.}~\bibnamefont
  {{Babak}}} \emph {et~al.},\ }\href {\doibase 10.1088/0264-9381/27/8/084009}
  {\bibfield  {journal} {\bibinfo  {journal} {Classical and Quantum Gravity}\
  }\textbf {\bibinfo {volume} {27}},\ \bibinfo {eid} {084009} (\bibinfo {year}
  {2010})},\ \Eprint {http://arxiv.org/abs/0912.0548} {arXiv:0912.0548 [gr-qc]}
  \BibitemShut {NoStop}%
\bibitem [{\citenamefont {{Brown}}\ \emph {et~al.}(2007)\citenamefont
  {{Brown}}, \citenamefont {{Crowder}}, \citenamefont {{Cutler}}, \citenamefont
  {{Mand el}},\ and\ \citenamefont {{Vallisneri}}}]{Brown+07}%
  \BibitemOpen
  \bibfield  {author} {\bibinfo {author} {\bibfnamefont {D.~A.}\ \bibnamefont
  {{Brown}}}, \bibinfo {author} {\bibfnamefont {J.}~\bibnamefont {{Crowder}}},
  \bibinfo {author} {\bibfnamefont {C.}~\bibnamefont {{Cutler}}}, \bibinfo
  {author} {\bibfnamefont {I.}~\bibnamefont {{Mand el}}}, \ and\ \bibinfo
  {author} {\bibfnamefont {M.}~\bibnamefont {{Vallisneri}}},\ }\href {\doibase
  10.1088/0264-9381/24/19/S22} {\bibfield  {journal} {\bibinfo  {journal}
  {Classical and Quantum Gravity}\ }\textbf {\bibinfo {volume} {24}},\ \bibinfo
  {pages} {S595} (\bibinfo {year} {2007})},\ \Eprint
  {http://arxiv.org/abs/0704.2447} {arXiv:0704.2447 [gr-qc]} \BibitemShut
  {NoStop}%
\bibitem [{\citenamefont {{Cornish}}\ and\ \citenamefont
  {{Porter}}(2006)}]{CornishPorter06a}%
  \BibitemOpen
  \bibfield  {author} {\bibinfo {author} {\bibfnamefont {N.~J.}\ \bibnamefont
  {{Cornish}}}\ and\ \bibinfo {author} {\bibfnamefont {E.~K.}\ \bibnamefont
  {{Porter}}},\ }\href {\doibase 10.1088/0264-9381/23/19/S15} {\bibfield
  {journal} {\bibinfo  {journal} {Classical and Quantum Gravity}\ }\textbf
  {\bibinfo {volume} {23}},\ \bibinfo {pages} {S761} (\bibinfo {year}
  {2006})},\ \Eprint {http://arxiv.org/abs/gr-qc/0605085} {arXiv:gr-qc/0605085
  [gr-qc]} \BibitemShut {NoStop}%
\bibitem [{\citenamefont {{Crowder}}\ \emph {et~al.}(2006)\citenamefont
  {{Crowder}}, \citenamefont {{Cornish}},\ and\ \citenamefont
  {{Reddinger}}}]{Crowder+06}%
  \BibitemOpen
  \bibfield  {author} {\bibinfo {author} {\bibfnamefont {J.}~\bibnamefont
  {{Crowder}}}, \bibinfo {author} {\bibfnamefont {N.~J.}\ \bibnamefont
  {{Cornish}}}, \ and\ \bibinfo {author} {\bibfnamefont {J.~L.}\ \bibnamefont
  {{Reddinger}}},\ }\href {\doibase 10.1103/PhysRevD.73.063011} {\bibfield
  {journal} {\bibinfo  {journal} {Phys. Rev. D}\ }\textbf {\bibinfo {volume}
  {73}},\ \bibinfo {eid} {063011} (\bibinfo {year} {2006})},\ \Eprint
  {http://arxiv.org/abs/gr-qc/0601036} {arXiv:gr-qc/0601036 [gr-qc]}
  \BibitemShut {NoStop}%
\bibitem [{\citenamefont {{Wickham}}\ \emph {et~al.}(2006)\citenamefont
  {{Wickham}}, \citenamefont {{Stroeer}},\ and\ \citenamefont
  {{Vecchio}}}]{Wickham+06}%
  \BibitemOpen
  \bibfield  {author} {\bibinfo {author} {\bibfnamefont {E.~D.~L.}\
  \bibnamefont {{Wickham}}}, \bibinfo {author} {\bibfnamefont {A.}~\bibnamefont
  {{Stroeer}}}, \ and\ \bibinfo {author} {\bibfnamefont {A.}~\bibnamefont
  {{Vecchio}}},\ }\href {\doibase 10.1088/0264-9381/23/19/S20} {\bibfield
  {journal} {\bibinfo  {journal} {Classical and Quantum Gravity}\ }\textbf
  {\bibinfo {volume} {23}},\ \bibinfo {pages} {S819} (\bibinfo {year}
  {2006})},\ \Eprint {http://arxiv.org/abs/gr-qc/0605071} {gr-qc/0605071}
  \BibitemShut {NoStop}%
\bibitem [{\citenamefont {{R{\"o}ver}}\ \emph {et~al.}(2007)\citenamefont
  {{R{\"o}ver}}, \citenamefont {{Stroeer}}, \citenamefont {{Bloomer}},
  \citenamefont {{Christensen}}, \citenamefont {{Clark}}, \citenamefont
  {{Hendry}}, \citenamefont {{Messenger}}, \citenamefont {{Meyer}},
  \citenamefont {{Pitkin}}, \citenamefont {{Toher}}, \citenamefont
  {{Umst{\"a}tter}}, \citenamefont {{Vecchio}}, \citenamefont {{Veitch}},\ and\
  \citenamefont {{Woan}}}]{Roever+07}%
  \BibitemOpen
  \bibfield  {author} {\bibinfo {author} {\bibfnamefont {C.}~\bibnamefont
  {{R{\"o}ver}}}, \bibinfo {author} {\bibfnamefont {A.}~\bibnamefont
  {{Stroeer}}}, \bibinfo {author} {\bibfnamefont {E.}~\bibnamefont
  {{Bloomer}}}, \bibinfo {author} {\bibfnamefont {N.}~\bibnamefont
  {{Christensen}}}, \bibinfo {author} {\bibfnamefont {J.}~\bibnamefont
  {{Clark}}}, \bibinfo {author} {\bibfnamefont {M.}~\bibnamefont {{Hendry}}},
  \bibinfo {author} {\bibfnamefont {C.}~\bibnamefont {{Messenger}}}, \bibinfo
  {author} {\bibfnamefont {R.}~\bibnamefont {{Meyer}}}, \bibinfo {author}
  {\bibfnamefont {M.}~\bibnamefont {{Pitkin}}}, \bibinfo {author}
  {\bibfnamefont {J.}~\bibnamefont {{Toher}}}, \bibinfo {author} {\bibfnamefont
  {R.}~\bibnamefont {{Umst{\"a}tter}}}, \bibinfo {author} {\bibfnamefont
  {A.}~\bibnamefont {{Vecchio}}}, \bibinfo {author} {\bibfnamefont
  {J.}~\bibnamefont {{Veitch}}}, \ and\ \bibinfo {author} {\bibfnamefont
  {G.}~\bibnamefont {{Woan}}},\ }\href {\doibase 10.1088/0264-9381/24/19/S15}
  {\bibfield  {journal} {\bibinfo  {journal} {Classical and Quantum Gravity}\
  }\textbf {\bibinfo {volume} {24}},\ \bibinfo {pages} {S521} (\bibinfo {year}
  {2007})},\ \Eprint {http://arxiv.org/abs/0707.3969} {arXiv:0707.3969 [gr-qc]}
  \BibitemShut {NoStop}%
\bibitem [{\citenamefont {{Feroz}}\ \emph {et~al.}(2009)\citenamefont
  {{Feroz}}, \citenamefont {{Gair}}, \citenamefont {{Hobson}},\ and\
  \citenamefont {{Porter}}}]{Feroz+09}%
  \BibitemOpen
  \bibfield  {author} {\bibinfo {author} {\bibfnamefont {F.}~\bibnamefont
  {{Feroz}}}, \bibinfo {author} {\bibfnamefont {J.~R.}\ \bibnamefont {{Gair}}},
  \bibinfo {author} {\bibfnamefont {M.~P.}\ \bibnamefont {{Hobson}}}, \ and\
  \bibinfo {author} {\bibfnamefont {E.~K.}\ \bibnamefont {{Porter}}},\ }\href
  {\doibase 10.1088/0264-9381/26/21/215003} {\bibfield  {journal} {\bibinfo
  {journal} {Classical and Quantum Gravity}\ }\textbf {\bibinfo {volume}
  {26}},\ \bibinfo {eid} {215003} (\bibinfo {year} {2009})},\ \Eprint
  {http://arxiv.org/abs/0904.1544} {arXiv:0904.1544 [gr-qc]} \BibitemShut
  {NoStop}%
\bibitem [{\citenamefont {{Gair}}\ and\ \citenamefont
  {{Porter}}(2009)}]{GairPorter09}%
  \BibitemOpen
  \bibfield  {author} {\bibinfo {author} {\bibfnamefont {J.~R.}\ \bibnamefont
  {{Gair}}}\ and\ \bibinfo {author} {\bibfnamefont {E.~K.}\ \bibnamefont
  {{Porter}}},\ }\href {\doibase 10.1088/0264-9381/26/22/225004} {\bibfield
  {journal} {\bibinfo  {journal} {Classical and Quantum Gravity}\ }\textbf
  {\bibinfo {volume} {26}},\ \bibinfo {eid} {225004} (\bibinfo {year}
  {2009})},\ \Eprint {http://arxiv.org/abs/0903.3733} {arXiv:0903.3733 [gr-qc]}
  \BibitemShut {NoStop}%
\bibitem [{\citenamefont {{Petiteau}}\ \emph {et~al.}(2009)\citenamefont
  {{Petiteau}}, \citenamefont {{Shang}},\ and\ \citenamefont
  {{Babak}}}]{Petiteau+09}%
  \BibitemOpen
  \bibfield  {author} {\bibinfo {author} {\bibfnamefont {A.}~\bibnamefont
  {{Petiteau}}}, \bibinfo {author} {\bibfnamefont {Y.}~\bibnamefont {{Shang}}},
  \ and\ \bibinfo {author} {\bibfnamefont {S.}~\bibnamefont {{Babak}}},\ }\href
  {\doibase 10.1088/0264-9381/26/20/204011} {\bibfield  {journal} {\bibinfo
  {journal} {Classical and Quantum Gravity}\ }\textbf {\bibinfo {volume}
  {26}},\ \bibinfo {eid} {204011} (\bibinfo {year} {2009})},\ \Eprint
  {http://arxiv.org/abs/0905.1785} {arXiv:0905.1785 [gr-qc]} \BibitemShut
  {NoStop}%
\bibitem [{\citenamefont {{Porter}}\ and\ \citenamefont
  {{Carr{\'e}}}(2014)}]{PorterCarre13}%
  \BibitemOpen
  \bibfield  {author} {\bibinfo {author} {\bibfnamefont {E.~K.}\ \bibnamefont
  {{Porter}}}\ and\ \bibinfo {author} {\bibfnamefont {J.}~\bibnamefont
  {{Carr{\'e}}}},\ }\href {\doibase 10.1088/0264-9381/31/14/145004} {\bibfield
  {journal} {\bibinfo  {journal} {Classical and Quantum Gravity}\ }\textbf
  {\bibinfo {volume} {31}},\ \bibinfo {eid} {145004} (\bibinfo {year}
  {2014})},\ \Eprint {http://arxiv.org/abs/1311.7539} {arXiv:1311.7539 [gr-qc]}
  \BibitemShut {NoStop}%
\bibitem [{\citenamefont {{Porter}}\ and\ \citenamefont
  {{Cornish}}(2015)}]{PorterCornish15}%
  \BibitemOpen
  \bibfield  {author} {\bibinfo {author} {\bibfnamefont {E.~K.}\ \bibnamefont
  {{Porter}}}\ and\ \bibinfo {author} {\bibfnamefont {N.~J.}\ \bibnamefont
  {{Cornish}}},\ }\href {\doibase 10.1103/PhysRevD.91.104001} {\bibfield
  {journal} {\bibinfo  {journal} {Phys. Rev. D}\ }\textbf {\bibinfo {volume}
  {91}},\ \bibinfo {eid} {104001} (\bibinfo {year} {2015})},\ \Eprint
  {http://arxiv.org/abs/1502.05735} {arXiv:1502.05735 [gr-qc]} \BibitemShut
  {NoStop}%
\bibitem [{\citenamefont {{Arun}}\ \emph {et~al.}(2007)\citenamefont {{Arun}},
  \citenamefont {{Iyer}}, \citenamefont {{Sathyaprakash}}, \citenamefont
  {{Sinha}},\ and\ \citenamefont {{van den Broeck}}}]{Arun+07a}%
  \BibitemOpen
  \bibfield  {author} {\bibinfo {author} {\bibfnamefont {K.~G.}\ \bibnamefont
  {{Arun}}}, \bibinfo {author} {\bibfnamefont {B.~R.}\ \bibnamefont {{Iyer}}},
  \bibinfo {author} {\bibfnamefont {B.~S.}\ \bibnamefont {{Sathyaprakash}}},
  \bibinfo {author} {\bibfnamefont {S.}~\bibnamefont {{Sinha}}}, \ and\
  \bibinfo {author} {\bibfnamefont {C.}~\bibnamefont {{van den Broeck}}},\
  }\href {\doibase 10.1103/PhysRevD.76.104016} {\bibfield  {journal} {\bibinfo
  {journal} {Phys. Rev. D}\ }\textbf {\bibinfo {volume} {76}},\ \bibinfo {eid}
  {104016} (\bibinfo {year} {2007})},\ \Eprint {http://arxiv.org/abs/0707.3920}
  {arXiv:0707.3920 [astro-ph]} \BibitemShut {NoStop}%
\bibitem [{\citenamefont {{Trias}}\ and\ \citenamefont
  {{Sintes}}(2008)}]{TriasSintes07}%
  \BibitemOpen
  \bibfield  {author} {\bibinfo {author} {\bibfnamefont {M.}~\bibnamefont
  {{Trias}}}\ and\ \bibinfo {author} {\bibfnamefont {A.~M.}\ \bibnamefont
  {{Sintes}}},\ }\href {\doibase 10.1103/PhysRevD.77.024030} {\bibfield
  {journal} {\bibinfo  {journal} {Phys. Rev. D}\ }\textbf {\bibinfo {volume}
  {77}},\ \bibinfo {eid} {024030} (\bibinfo {year} {2008})},\ \Eprint
  {http://arxiv.org/abs/0707.4434} {arXiv:0707.4434 [gr-qc]} \BibitemShut
  {NoStop}%
\bibitem [{\citenamefont {{Porter}}\ and\ \citenamefont
  {{Cornish}}(2008)}]{PorterCornish08}%
  \BibitemOpen
  \bibfield  {author} {\bibinfo {author} {\bibfnamefont {E.~K.}\ \bibnamefont
  {{Porter}}}\ and\ \bibinfo {author} {\bibfnamefont {N.~J.}\ \bibnamefont
  {{Cornish}}},\ }\href {\doibase 10.1103/PhysRevD.78.064005} {\bibfield
  {journal} {\bibinfo  {journal} {Phys. Rev. D}\ }\textbf {\bibinfo {volume}
  {78}},\ \bibinfo {eid} {064005} (\bibinfo {year} {2008})},\ \Eprint
  {http://arxiv.org/abs/0804.0332} {arXiv:0804.0332 [gr-qc]} \BibitemShut
  {NoStop}%
\bibitem [{\citenamefont {{McWilliams}}\ \emph
  {et~al.}(2010{\natexlab{a}})\citenamefont {{McWilliams}}, \citenamefont
  {{Thorpe}}, \citenamefont {{Baker}},\ and\ \citenamefont
  {{Kelly}}}]{McWilliams+09}%
  \BibitemOpen
  \bibfield  {author} {\bibinfo {author} {\bibfnamefont {S.~T.}\ \bibnamefont
  {{McWilliams}}}, \bibinfo {author} {\bibfnamefont {J.~I.}\ \bibnamefont
  {{Thorpe}}}, \bibinfo {author} {\bibfnamefont {J.~G.}\ \bibnamefont
  {{Baker}}}, \ and\ \bibinfo {author} {\bibfnamefont {B.~J.}\ \bibnamefont
  {{Kelly}}},\ }\href {\doibase 10.1103/PhysRevD.81.064014} {\bibfield
  {journal} {\bibinfo  {journal} {Phys. Rev. D}\ }\textbf {\bibinfo {volume}
  {81}},\ \bibinfo {eid} {064014} (\bibinfo {year} {2010}{\natexlab{a}})},\
  \Eprint {http://arxiv.org/abs/0911.1078} {arXiv:0911.1078 [gr-qc]}
  \BibitemShut {NoStop}%
\bibitem [{\citenamefont {{Thorpe}}\ \emph {et~al.}(2009)\citenamefont
  {{Thorpe}}, \citenamefont {{McWilliams}}, \citenamefont {{Kelly}},
  \citenamefont {{Fahey}}, \citenamefont {{Arnaud}},\ and\ \citenamefont
  {{Baker}}}]{Thorpe+08}%
  \BibitemOpen
  \bibfield  {author} {\bibinfo {author} {\bibfnamefont {J.~I.}\ \bibnamefont
  {{Thorpe}}}, \bibinfo {author} {\bibfnamefont {S.~T.}\ \bibnamefont
  {{McWilliams}}}, \bibinfo {author} {\bibfnamefont {B.~J.}\ \bibnamefont
  {{Kelly}}}, \bibinfo {author} {\bibfnamefont {R.~P.}\ \bibnamefont
  {{Fahey}}}, \bibinfo {author} {\bibfnamefont {K.}~\bibnamefont {{Arnaud}}}, \
  and\ \bibinfo {author} {\bibfnamefont {J.~G.}\ \bibnamefont {{Baker}}},\
  }\href {\doibase 10.1088/0264-9381/26/9/094026} {\bibfield  {journal}
  {\bibinfo  {journal} {Classical and Quantum Gravity}\ }\textbf {\bibinfo
  {volume} {26}},\ \bibinfo {eid} {094026} (\bibinfo {year} {2009})},\ \Eprint
  {http://arxiv.org/abs/0811.0833} {arXiv:0811.0833 [astro-ph]} \BibitemShut
  {NoStop}%
\bibitem [{\citenamefont {{McWilliams}}\ \emph
  {et~al.}(2010{\natexlab{b}})\citenamefont {{McWilliams}}, \citenamefont
  {{Kelly}},\ and\ \citenamefont {{Baker}}}]{McWilliams+10}%
  \BibitemOpen
  \bibfield  {author} {\bibinfo {author} {\bibfnamefont {S.~T.}\ \bibnamefont
  {{McWilliams}}}, \bibinfo {author} {\bibfnamefont {B.~J.}\ \bibnamefont
  {{Kelly}}}, \ and\ \bibinfo {author} {\bibfnamefont {J.~G.}\ \bibnamefont
  {{Baker}}},\ }\href {\doibase 10.1103/PhysRevD.82.024014} {\bibfield
  {journal} {\bibinfo  {journal} {Phys. Rev. D}\ }\textbf {\bibinfo {volume}
  {82}},\ \bibinfo {eid} {024014} (\bibinfo {year} {2010}{\natexlab{b}})},\
  \Eprint {http://arxiv.org/abs/1004.0961} {arXiv:1004.0961 [gr-qc]}
  \BibitemShut {NoStop}%
\bibitem [{\citenamefont {{McWilliams}}\ \emph {et~al.}(2011)\citenamefont
  {{McWilliams}}, \citenamefont {{Lang}}, \citenamefont {{Baker}},\ and\
  \citenamefont {{Thorpe}}}]{McWilliams+11}%
  \BibitemOpen
  \bibfield  {author} {\bibinfo {author} {\bibfnamefont {S.~T.}\ \bibnamefont
  {{McWilliams}}}, \bibinfo {author} {\bibfnamefont {R.~N.}\ \bibnamefont
  {{Lang}}}, \bibinfo {author} {\bibfnamefont {J.~G.}\ \bibnamefont {{Baker}}},
  \ and\ \bibinfo {author} {\bibfnamefont {J.~I.}\ \bibnamefont {{Thorpe}}},\
  }\href {\doibase 10.1103/PhysRevD.84.064003} {\bibfield  {journal} {\bibinfo
  {journal} {Phys. Rev. D}\ }\textbf {\bibinfo {volume} {84}},\ \bibinfo {eid}
  {064003} (\bibinfo {year} {2011})},\ \Eprint {http://arxiv.org/abs/1104.5650}
  {arXiv:1104.5650 [gr-qc]} \BibitemShut {NoStop}%
\bibitem [{\citenamefont {{Babak}}\ \emph {et~al.}(2008)\citenamefont
  {{Babak}}, \citenamefont {{Hannam}}, \citenamefont {{Husa}},\ and\
  \citenamefont {{Schutz}}}]{Babak+08}%
  \BibitemOpen
  \bibfield  {author} {\bibinfo {author} {\bibfnamefont {S.}~\bibnamefont
  {{Babak}}}, \bibinfo {author} {\bibfnamefont {M.}~\bibnamefont {{Hannam}}},
  \bibinfo {author} {\bibfnamefont {S.}~\bibnamefont {{Husa}}}, \ and\ \bibinfo
  {author} {\bibfnamefont {B.}~\bibnamefont {{Schutz}}},\ }\href@noop {}
  {\bibfield  {journal} {\bibinfo  {journal} {arXiv e-prints}\ ,\ \bibinfo
  {eid} {arXiv:0806.1591}} (\bibinfo {year} {2008})},\ \Eprint
  {http://arxiv.org/abs/0806.1591} {arXiv:0806.1591 [gr-qc]} \BibitemShut
  {NoStop}%
\bibitem [{\citenamefont {{eLISA Consortium}}\ \emph
  {et~al.}(2013)\citenamefont {{eLISA Consortium}}, \citenamefont {{Amaro
  Seoane}}, \citenamefont {{Aoudia}}, \citenamefont {{Audley}}, \citenamefont
  {{Auger}}, \citenamefont {{Babak}}, \citenamefont {{Baker}}, \citenamefont
  {{Barausse}}, \citenamefont {{Barke}}, \citenamefont {{Bassan}},
  \citenamefont {{Beckmann}} \emph {et~al.}}]{elisa13}%
  \BibitemOpen
  \bibfield  {author} {\bibinfo {author} {\bibnamefont {{eLISA Consortium}}},
  \bibinfo {author} {\bibfnamefont {P.}~\bibnamefont {{Amaro Seoane}}},
  \bibinfo {author} {\bibfnamefont {S.}~\bibnamefont {{Aoudia}}}, \bibinfo
  {author} {\bibfnamefont {H.}~\bibnamefont {{Audley}}}, \bibinfo {author}
  {\bibfnamefont {G.}~\bibnamefont {{Auger}}}, \bibinfo {author} {\bibfnamefont
  {S.}~\bibnamefont {{Babak}}}, \bibinfo {author} {\bibfnamefont
  {J.}~\bibnamefont {{Baker}}}, \bibinfo {author} {\bibfnamefont
  {E.}~\bibnamefont {{Barausse}}}, \bibinfo {author} {\bibfnamefont
  {S.}~\bibnamefont {{Barke}}}, \bibinfo {author} {\bibfnamefont
  {M.}~\bibnamefont {{Bassan}}}, \bibinfo {author} {\bibfnamefont
  {V.}~\bibnamefont {{Beckmann}}},  \emph {et~al.},\ }\href@noop {} {\bibfield
  {journal} {\bibinfo  {journal} {ArXiv e-prints}\ } (\bibinfo {year}
  {2013})},\ \Eprint {http://arxiv.org/abs/1305.5720} {arXiv:1305.5720
  [astro-ph.CO]} \BibitemShut {NoStop}%
\bibitem [{\citenamefont {{Baibhav}}\ \emph {et~al.}(2020)\citenamefont
  {{Baibhav}}, \citenamefont {{Berti}},\ and\ \citenamefont
  {{Cardoso}}}]{Baibhav+20}%
  \BibitemOpen
  \bibfield  {author} {\bibinfo {author} {\bibfnamefont {V.}~\bibnamefont
  {{Baibhav}}}, \bibinfo {author} {\bibfnamefont {E.}~\bibnamefont {{Berti}}},
  \ and\ \bibinfo {author} {\bibfnamefont {V.}~\bibnamefont {{Cardoso}}},\
  }\href@noop {} {\bibfield  {journal} {\bibinfo  {journal} {arXiv e-prints}\
  ,\ \bibinfo {eid} {arXiv:2001.10011}} (\bibinfo {year} {2020})},\ \Eprint
  {http://arxiv.org/abs/2001.10011} {arXiv:2001.10011 [gr-qc]} \BibitemShut
  {NoStop}%
\bibitem [{\citenamefont {Khan}\ \emph {et~al.}(2016)\citenamefont {Khan},
  \citenamefont {Husa}, \citenamefont {Hannam}, \citenamefont {Ohme},
  \citenamefont {P\"urrer}, \citenamefont {Jim\'enez~Forteza},\ and\
  \citenamefont {Boh{\'e}}}]{Khan+15}%
  \BibitemOpen
  \bibfield  {author} {\bibinfo {author} {\bibfnamefont {S.}~\bibnamefont
  {Khan}}, \bibinfo {author} {\bibfnamefont {S.}~\bibnamefont {Husa}}, \bibinfo
  {author} {\bibfnamefont {M.}~\bibnamefont {Hannam}}, \bibinfo {author}
  {\bibfnamefont {F.}~\bibnamefont {Ohme}}, \bibinfo {author} {\bibfnamefont
  {M.}~\bibnamefont {P\"urrer}}, \bibinfo {author} {\bibfnamefont
  {X.}~\bibnamefont {Jim\'enez~Forteza}}, \ and\ \bibinfo {author}
  {\bibfnamefont {A.}~\bibnamefont {Boh{\'e}}},\ }\href {\doibase
  10.1103/PhysRevD.93.044007} {\bibfield  {journal} {\bibinfo  {journal} {Phys.
  Rev. D}\ }\textbf {\bibinfo {volume} {D93}},\ \bibinfo {pages} {044007}
  (\bibinfo {year} {2016})},\ \Eprint {http://arxiv.org/abs/1508.07253}
  {arXiv:1508.07253 [gr-qc]} \BibitemShut {NoStop}%
\bibitem [{\citenamefont {{London}}\ \emph {et~al.}(2018)\citenamefont
  {{London}}, \citenamefont {{Khan}}, \citenamefont {{Fauchon-Jones}},
  \citenamefont {{Garc{\'{\i}}a}}, \citenamefont {{Hannam}}, \citenamefont
  {{Husa}}, \citenamefont {{Jim{\'e}nez-Forteza}}, \citenamefont
  {{Kalaghatgi}}, \citenamefont {{Ohme}},\ and\ \citenamefont
  {{Pannarale}}}]{London+17}%
  \BibitemOpen
  \bibfield  {author} {\bibinfo {author} {\bibfnamefont {L.}~\bibnamefont
  {{London}}}, \bibinfo {author} {\bibfnamefont {S.}~\bibnamefont {{Khan}}},
  \bibinfo {author} {\bibfnamefont {E.}~\bibnamefont {{Fauchon-Jones}}},
  \bibinfo {author} {\bibfnamefont {C.}~\bibnamefont {{Garc{\'{\i}}a}}},
  \bibinfo {author} {\bibfnamefont {M.}~\bibnamefont {{Hannam}}}, \bibinfo
  {author} {\bibfnamefont {S.}~\bibnamefont {{Husa}}}, \bibinfo {author}
  {\bibfnamefont {X.}~\bibnamefont {{Jim{\'e}nez-Forteza}}}, \bibinfo {author}
  {\bibfnamefont {C.}~\bibnamefont {{Kalaghatgi}}}, \bibinfo {author}
  {\bibfnamefont {F.}~\bibnamefont {{Ohme}}}, \ and\ \bibinfo {author}
  {\bibfnamefont {F.}~\bibnamefont {{Pannarale}}},\ }\href {\doibase
  10.1103/PhysRevLett.120.161102} {\bibfield  {journal} {\bibinfo  {journal}
  {Physical Review Letters}\ }\textbf {\bibinfo {volume} {120}},\ \bibinfo
  {eid} {161102} (\bibinfo {year} {2018})},\ \Eprint
  {http://arxiv.org/abs/1708.00404} {arXiv:1708.00404 [gr-qc]} \BibitemShut
  {NoStop}%
\bibitem [{\citenamefont {{P{\"u}rrer}}(2014)}]{Puerrer14}%
  \BibitemOpen
  \bibfield  {author} {\bibinfo {author} {\bibfnamefont {M.}~\bibnamefont
  {{P{\"u}rrer}}},\ }\href {\doibase 10.1088/0264-9381/31/19/195010} {\bibfield
   {journal} {\bibinfo  {journal} {Classical and Quantum Gravity}\ }\textbf
  {\bibinfo {volume} {31}},\ \bibinfo {eid} {195010} (\bibinfo {year}
  {2014})},\ \Eprint {http://arxiv.org/abs/1402.4146} {arXiv:1402.4146 [gr-qc]}
  \BibitemShut {NoStop}%
\bibitem [{\citenamefont {{Boh{\'e}}}\ \emph {et~al.}(2016)\citenamefont
  {{Boh{\'e}}}, \citenamefont {{Shao}}, \citenamefont {{Taracchini}},
  \citenamefont {{Buonanno}}, \citenamefont {{Babak}}, \citenamefont {{Harry}},
  \citenamefont {{Hinder}}, \citenamefont {{Ossokine}}, \citenamefont
  {{P{\"u}rrer}}, \citenamefont {{Raymond}}, \citenamefont {{Chu}},
  \citenamefont {{Fong}}, \citenamefont {{Kumar}}, \citenamefont {{Pfeiffer}},
  \citenamefont {{Boyle}}, \citenamefont {{Hemberger}}, \citenamefont
  {{Kidder}}, \citenamefont {{Lovelace}}, \citenamefont {{Scheel}},\ and\
  \citenamefont {{Szil{\'a}gyi}}}]{Bohe+16}%
  \BibitemOpen
  \bibfield  {author} {\bibinfo {author} {\bibfnamefont {A.}~\bibnamefont
  {{Boh{\'e}}}}, \bibinfo {author} {\bibfnamefont {L.}~\bibnamefont {{Shao}}},
  \bibinfo {author} {\bibfnamefont {A.}~\bibnamefont {{Taracchini}}}, \bibinfo
  {author} {\bibfnamefont {A.}~\bibnamefont {{Buonanno}}}, \bibinfo {author}
  {\bibfnamefont {S.}~\bibnamefont {{Babak}}}, \bibinfo {author} {\bibfnamefont
  {I.~W.}\ \bibnamefont {{Harry}}}, \bibinfo {author} {\bibfnamefont
  {I.}~\bibnamefont {{Hinder}}}, \bibinfo {author} {\bibfnamefont
  {S.}~\bibnamefont {{Ossokine}}}, \bibinfo {author} {\bibfnamefont
  {M.}~\bibnamefont {{P{\"u}rrer}}}, \bibinfo {author} {\bibfnamefont
  {V.}~\bibnamefont {{Raymond}}}, \bibinfo {author} {\bibfnamefont
  {T.}~\bibnamefont {{Chu}}}, \bibinfo {author} {\bibfnamefont
  {H.}~\bibnamefont {{Fong}}}, \bibinfo {author} {\bibfnamefont
  {P.}~\bibnamefont {{Kumar}}}, \bibinfo {author} {\bibfnamefont {H.~P.}\
  \bibnamefont {{Pfeiffer}}}, \bibinfo {author} {\bibfnamefont
  {M.}~\bibnamefont {{Boyle}}}, \bibinfo {author} {\bibfnamefont {D.~A.}\
  \bibnamefont {{Hemberger}}}, \bibinfo {author} {\bibfnamefont {L.~E.}\
  \bibnamefont {{Kidder}}}, \bibinfo {author} {\bibfnamefont {G.}~\bibnamefont
  {{Lovelace}}}, \bibinfo {author} {\bibfnamefont {M.~A.}\ \bibnamefont
  {{Scheel}}}, \ and\ \bibinfo {author} {\bibfnamefont {B.}~\bibnamefont
  {{Szil{\'a}gyi}}},\ }\href@noop {} {\bibfield  {journal} {\bibinfo  {journal}
  {ArXiv e-prints}\ } (\bibinfo {year} {2016})},\ \Eprint
  {http://arxiv.org/abs/1611.03703} {arXiv:1611.03703 [gr-qc]} \BibitemShut
  {NoStop}%
\bibitem [{\citenamefont {{Blackman}}\ \emph {et~al.}(2017)\citenamefont
  {{Blackman}}, \citenamefont {{Field}}, \citenamefont {{Scheel}},
  \citenamefont {{Galley}}, \citenamefont {{Ott}}, \citenamefont {{Boyle}},
  \citenamefont {{Kidder}}, \citenamefont {{Pfeiffer}},\ and\ \citenamefont
  {{Szil{\'a}gyi}}}]{Blackman+17b}%
  \BibitemOpen
  \bibfield  {author} {\bibinfo {author} {\bibfnamefont {J.}~\bibnamefont
  {{Blackman}}}, \bibinfo {author} {\bibfnamefont {S.~E.}\ \bibnamefont
  {{Field}}}, \bibinfo {author} {\bibfnamefont {M.~A.}\ \bibnamefont
  {{Scheel}}}, \bibinfo {author} {\bibfnamefont {C.~R.}\ \bibnamefont
  {{Galley}}}, \bibinfo {author} {\bibfnamefont {C.~D.}\ \bibnamefont {{Ott}}},
  \bibinfo {author} {\bibfnamefont {M.}~\bibnamefont {{Boyle}}}, \bibinfo
  {author} {\bibfnamefont {L.~E.}\ \bibnamefont {{Kidder}}}, \bibinfo {author}
  {\bibfnamefont {H.~P.}\ \bibnamefont {{Pfeiffer}}}, \ and\ \bibinfo {author}
  {\bibfnamefont {B.}~\bibnamefont {{Szil{\'a}gyi}}},\ }\href {\doibase
  10.1103/PhysRevD.96.024058} {\bibfield  {journal} {\bibinfo  {journal} {Phys.
  Rev. D}\ }\textbf {\bibinfo {volume} {96}},\ \bibinfo {eid} {024058}
  (\bibinfo {year} {2017})},\ \Eprint {http://arxiv.org/abs/1705.07089}
  {arXiv:1705.07089 [gr-qc]} \BibitemShut {NoStop}%
\bibitem [{\citenamefont {Varma}\ \emph {et~al.}(2019)\citenamefont {Varma},
  \citenamefont {Field}, \citenamefont {Scheel}, \citenamefont {Blackman},
  \citenamefont {Kidder},\ and\ \citenamefont {Pfeiffer}}]{Varma+18}%
  \BibitemOpen
  \bibfield  {author} {\bibinfo {author} {\bibfnamefont {V.}~\bibnamefont
  {Varma}}, \bibinfo {author} {\bibfnamefont {S.~E.}\ \bibnamefont {Field}},
  \bibinfo {author} {\bibfnamefont {M.~A.}\ \bibnamefont {Scheel}}, \bibinfo
  {author} {\bibfnamefont {J.}~\bibnamefont {Blackman}}, \bibinfo {author}
  {\bibfnamefont {L.~E.}\ \bibnamefont {Kidder}}, \ and\ \bibinfo {author}
  {\bibfnamefont {H.~P.}\ \bibnamefont {Pfeiffer}},\ }\href {\doibase
  10.1103/PhysRevD.99.064045} {\bibfield  {journal} {\bibinfo  {journal} {Phys.
  Rev.}\ }\textbf {\bibinfo {volume} {D99}},\ \bibinfo {pages} {064045}
  (\bibinfo {year} {2019})},\ \Eprint {http://arxiv.org/abs/1812.07865}
  {arXiv:1812.07865 [gr-qc]} \BibitemShut {NoStop}%
\bibitem [{\citenamefont {{Vitale}}(2016)}]{Vitale16}%
  \BibitemOpen
  \bibfield  {author} {\bibinfo {author} {\bibfnamefont {S.}~\bibnamefont
  {{Vitale}}},\ }\href {\doibase 10.1103/PhysRevLett.117.051102} {\bibfield
  {journal} {\bibinfo  {journal} {Phys. Rev. Letters}\ }\textbf {\bibinfo
  {volume} {117}},\ \bibinfo {eid} {051102} (\bibinfo {year} {2016})},\ \Eprint
  {http://arxiv.org/abs/1605.01037} {arXiv:1605.01037 [gr-qc]} \BibitemShut
  {NoStop}%
\bibitem [{\citenamefont {{Nishizawa}}\ \emph {et~al.}(2016)\citenamefont
  {{Nishizawa}}, \citenamefont {{Berti}}, \citenamefont {{Klein}},\ and\
  \citenamefont {{Sesana}}}]{Nishizawa+16a}%
  \BibitemOpen
  \bibfield  {author} {\bibinfo {author} {\bibfnamefont {A.}~\bibnamefont
  {{Nishizawa}}}, \bibinfo {author} {\bibfnamefont {E.}~\bibnamefont
  {{Berti}}}, \bibinfo {author} {\bibfnamefont {A.}~\bibnamefont {{Klein}}}, \
  and\ \bibinfo {author} {\bibfnamefont {A.}~\bibnamefont {{Sesana}}},\ }\href
  {\doibase 10.1103/PhysRevD.94.064020} {\bibfield  {journal} {\bibinfo
  {journal} {Phys. Rev. D}\ }\textbf {\bibinfo {volume} {94}},\ \bibinfo {eid}
  {064020} (\bibinfo {year} {2016})},\ \Eprint
  {http://arxiv.org/abs/1605.01341} {arXiv:1605.01341 [gr-qc]} \BibitemShut
  {NoStop}%
\bibitem [{\citenamefont {{Nishizawa}}\ \emph {et~al.}(2017)\citenamefont
  {{Nishizawa}}, \citenamefont {{Sesana}}, \citenamefont {{Berti}},\ and\
  \citenamefont {{Klein}}}]{Nishizawa+16b}%
  \BibitemOpen
  \bibfield  {author} {\bibinfo {author} {\bibfnamefont {A.}~\bibnamefont
  {{Nishizawa}}}, \bibinfo {author} {\bibfnamefont {A.}~\bibnamefont
  {{Sesana}}}, \bibinfo {author} {\bibfnamefont {E.}~\bibnamefont {{Berti}}}, \
  and\ \bibinfo {author} {\bibfnamefont {A.}~\bibnamefont {{Klein}}},\ }\href
  {\doibase 10.1093/mnras/stw2993} {\bibfield  {journal} {\bibinfo  {journal}
  {Mon. Not. Roy. Astron. Soc.}\ }\textbf {\bibinfo {volume} {465}},\ \bibinfo
  {pages} {4375} (\bibinfo {year} {2017})},\ \Eprint
  {http://arxiv.org/abs/1606.09295} {arXiv:1606.09295 [astro-ph.HE]}
  \BibitemShut {NoStop}%
\bibitem [{\citenamefont {{Goldberg}}\ \emph {et~al.}(1967)\citenamefont
  {{Goldberg}}, \citenamefont {{Macfarlane}}, \citenamefont {{Newman}},
  \citenamefont {{Rohrlich}},\ and\ \citenamefont {{Sudarshan}}}]{Goldberg+67}%
  \BibitemOpen
  \bibfield  {author} {\bibinfo {author} {\bibfnamefont {J.~N.}\ \bibnamefont
  {{Goldberg}}}, \bibinfo {author} {\bibfnamefont {A.~J.}\ \bibnamefont
  {{Macfarlane}}}, \bibinfo {author} {\bibfnamefont {E.~T.}\ \bibnamefont
  {{Newman}}}, \bibinfo {author} {\bibfnamefont {F.}~\bibnamefont
  {{Rohrlich}}}, \ and\ \bibinfo {author} {\bibfnamefont {E.~C.~G.}\
  \bibnamefont {{Sudarshan}}},\ }\href {\doibase 10.1063/1.1705135} {\bibfield
  {journal} {\bibinfo  {journal} {Journal of Mathematical Physics}\ }\textbf
  {\bibinfo {volume} {8}},\ \bibinfo {pages} {2155} (\bibinfo {year}
  {1967})}\BibitemShut {NoStop}%
\bibitem [{\citenamefont {Blanchet}(2014)}]{BlanchetLiving}%
  \BibitemOpen
  \bibfield  {author} {\bibinfo {author} {\bibfnamefont {L.}~\bibnamefont
  {Blanchet}},\ }\href {\doibase 10.12942/lrr-2014-2} {\bibfield  {journal}
  {\bibinfo  {journal} {Living Rev. Rel.}\ }\textbf {\bibinfo {volume} {17}},\
  \bibinfo {pages} {2} (\bibinfo {year} {2014})},\ \Eprint
  {http://arxiv.org/abs/1310.1528} {arXiv:1310.1528 [gr-qc]} \BibitemShut
  {NoStop}%
\bibitem [{lal()}]{lal}%
  \BibitemOpen
  \href@noop {} {\enquote {\bibinfo {title} {{LSC} algorithm library},}\
  }\bibinfo {note} {{\tt
  \url{http://www.lsc-group.phys.uwm.edu/lal}}}\BibitemShut {NoStop}%
\bibitem [{\citenamefont {{Estabrook}}\ and\ \citenamefont
  {{Wahlquist}}(1975)}]{EW75}%
  \BibitemOpen
  \bibfield  {author} {\bibinfo {author} {\bibfnamefont {F.~B.}\ \bibnamefont
  {{Estabrook}}}\ and\ \bibinfo {author} {\bibfnamefont {H.~D.}\ \bibnamefont
  {{Wahlquist}}},\ }\href {\doibase 10.1007/BF00762449} {\bibfield  {journal}
  {\bibinfo  {journal} {General Relativity and Gravitation}\ }\textbf {\bibinfo
  {volume} {6}},\ \bibinfo {pages} {439} (\bibinfo {year} {1975})}\BibitemShut
  {NoStop}%
\bibitem [{\citenamefont {Rubbo}\ \emph {et~al.}(2004)\citenamefont {Rubbo},
  \citenamefont {Cornish},\ and\ \citenamefont {Poujade}}]{RCP04}%
  \BibitemOpen
  \bibfield  {author} {\bibinfo {author} {\bibfnamefont {L.~J.}\ \bibnamefont
  {Rubbo}}, \bibinfo {author} {\bibfnamefont {N.~J.}\ \bibnamefont {Cornish}},
  \ and\ \bibinfo {author} {\bibfnamefont {O.}~\bibnamefont {Poujade}},\ }\href
  {\doibase 10.1103/PhysRevD.69.082003} {\bibfield  {journal} {\bibinfo
  {journal} {Phys. Rev. D}\ }\textbf {\bibinfo {volume} {69}},\ \bibinfo
  {pages} {082003} (\bibinfo {year} {2004})},\ \Eprint
  {http://arxiv.org/abs/gr-qc/0311069} {arXiv:gr-qc/0311069 [gr-qc]}
  \BibitemShut {NoStop}%
\bibitem [{\citenamefont {Vallisneri}(2005)}]{Vallisneri04}%
  \BibitemOpen
  \bibfield  {author} {\bibinfo {author} {\bibfnamefont {M.}~\bibnamefont
  {Vallisneri}},\ }\href {\doibase 10.1103/PhysRevD.71.022001} {\bibfield
  {journal} {\bibinfo  {journal} {Phys. Rev. D}\ }\textbf {\bibinfo {volume}
  {71}},\ \bibinfo {pages} {022001} (\bibinfo {year} {2005})},\ \Eprint
  {http://arxiv.org/abs/gr-qc/0407102} {arXiv:gr-qc/0407102 [gr-qc]}
  \BibitemShut {NoStop}%
\bibitem [{\citenamefont {Krolak}\ \emph {et~al.}(2004)\citenamefont {Krolak},
  \citenamefont {Tinto},\ and\ \citenamefont {Vallisneri}}]{Krolak+04}%
  \BibitemOpen
  \bibfield  {author} {\bibinfo {author} {\bibfnamefont {A.}~\bibnamefont
  {Krolak}}, \bibinfo {author} {\bibfnamefont {M.}~\bibnamefont {Tinto}}, \
  and\ \bibinfo {author} {\bibfnamefont {M.}~\bibnamefont {Vallisneri}},\
  }\href {\doibase 10.1103/PhysRevD.70.022003, 10.1103/PhysRevD.76.069901}
  {\bibfield  {journal} {\bibinfo  {journal} {Phys. Rev. D}\ }\textbf {\bibinfo
  {volume} {70}},\ \bibinfo {pages} {022003} (\bibinfo {year} {2004})},\
  \bibinfo {note} {[Erratum: Phys. Rev.D76,069901(2007)]},\ \Eprint
  {http://arxiv.org/abs/gr-qc/0401108} {arXiv:gr-qc/0401108 [gr-qc]}
  \BibitemShut {NoStop}%
\bibitem [{\citenamefont {{Tinto}}\ and\ \citenamefont
  {{Armstrong}}(1999)}]{TintoArmstrong99}%
  \BibitemOpen
  \bibfield  {author} {\bibinfo {author} {\bibfnamefont {M.}~\bibnamefont
  {{Tinto}}}\ and\ \bibinfo {author} {\bibfnamefont {J.~W.}\ \bibnamefont
  {{Armstrong}}},\ }\href {\doibase 10.1103/PhysRevD.59.102003} {\bibfield
  {journal} {\bibinfo  {journal} {Phys. Rev. D}\ }\textbf {\bibinfo {volume}
  {59}},\ \bibinfo {eid} {102003} (\bibinfo {year} {1999})}\BibitemShut
  {NoStop}%
\bibitem [{\citenamefont {{Armstrong}}\ \emph {et~al.}(1999)\citenamefont
  {{Armstrong}}, \citenamefont {{Estabrook}},\ and\ \citenamefont
  {{Tinto}}}]{Armstrong+99}%
  \BibitemOpen
  \bibfield  {author} {\bibinfo {author} {\bibfnamefont {J.~W.}\ \bibnamefont
  {{Armstrong}}}, \bibinfo {author} {\bibfnamefont {F.~B.}\ \bibnamefont
  {{Estabrook}}}, \ and\ \bibinfo {author} {\bibfnamefont {M.}~\bibnamefont
  {{Tinto}}},\ }\href {\doibase 10.1086/308110} {\bibfield  {journal} {\bibinfo
   {journal} {Astrophys. J}\ }\textbf {\bibinfo {volume} {527}},\ \bibinfo
  {pages} {814} (\bibinfo {year} {1999})}\BibitemShut {NoStop}%
\bibitem [{\citenamefont {{Estabrook}}\ \emph {et~al.}(2000)\citenamefont
  {{Estabrook}}, \citenamefont {{Tinto}},\ and\ \citenamefont
  {{Armstrong}}}]{Estabrook+00}%
  \BibitemOpen
  \bibfield  {author} {\bibinfo {author} {\bibfnamefont {F.~B.}\ \bibnamefont
  {{Estabrook}}}, \bibinfo {author} {\bibfnamefont {M.}~\bibnamefont
  {{Tinto}}}, \ and\ \bibinfo {author} {\bibfnamefont {J.~W.}\ \bibnamefont
  {{Armstrong}}},\ }\href {\doibase 10.1103/PhysRevD.62.042002} {\bibfield
  {journal} {\bibinfo  {journal} {Phys. Rev. D}\ }\textbf {\bibinfo {volume}
  {62}},\ \bibinfo {eid} {042002} (\bibinfo {year} {2000})}\BibitemShut
  {NoStop}%
\bibitem [{\citenamefont {{Dhurandhar}}\ \emph {et~al.}(2002)\citenamefont
  {{Dhurandhar}}, \citenamefont {{Nayak}},\ and\ \citenamefont
  {{Vinet}}}]{Dhurandhar+02}%
  \BibitemOpen
  \bibfield  {author} {\bibinfo {author} {\bibfnamefont {S.~V.}\ \bibnamefont
  {{Dhurandhar}}}, \bibinfo {author} {\bibfnamefont {K.~R.}\ \bibnamefont
  {{Nayak}}}, \ and\ \bibinfo {author} {\bibfnamefont {J.~Y.}\ \bibnamefont
  {{Vinet}}},\ }\href {\doibase 10.1103/PhysRevD.65.102002} {\bibfield
  {journal} {\bibinfo  {journal} {Phys. Rev. D}\ }\textbf {\bibinfo {volume}
  {65}},\ \bibinfo {eid} {102002} (\bibinfo {year} {2002})}\BibitemShut
  {NoStop}%
\bibitem [{\citenamefont {{Tinto}}\ and\ \citenamefont
  {{Dhurandhar}}(2005)}]{Tintoliving}%
  \BibitemOpen
  \bibfield  {author} {\bibinfo {author} {\bibfnamefont {M.}~\bibnamefont
  {{Tinto}}}\ and\ \bibinfo {author} {\bibfnamefont {S.~V.}\ \bibnamefont
  {{Dhurandhar}}},\ }\href {\doibase 10.12942/lrr-2005-4} {\bibfield  {journal}
  {\bibinfo  {journal} {Living Reviews in Relativity}\ }\textbf {\bibinfo
  {volume} {8}},\ \bibinfo {eid} {4} (\bibinfo {year} {2005})},\ \Eprint
  {http://arxiv.org/abs/gr-qc/0409034} {gr-qc/0409034} \BibitemShut {NoStop}%
\bibitem [{\citenamefont {{Shaddock}}(2004)}]{Shaddock03}%
  \BibitemOpen
  \bibfield  {author} {\bibinfo {author} {\bibfnamefont {D.~A.}\ \bibnamefont
  {{Shaddock}}},\ }\href {\doibase 10.1103/PhysRevD.69.022001} {\bibfield
  {journal} {\bibinfo  {journal} {\prd}\ }\textbf {\bibinfo {volume} {69}},\
  \bibinfo {eid} {022001} (\bibinfo {year} {2004})},\ \Eprint
  {http://arxiv.org/abs/gr-qc/0306125} {arXiv:gr-qc/0306125 [gr-qc]}
  \BibitemShut {NoStop}%
\bibitem [{\citenamefont {{Cornish}}\ and\ \citenamefont
  {{Hellings}}(2003)}]{CornishHellings03}%
  \BibitemOpen
  \bibfield  {author} {\bibinfo {author} {\bibfnamefont {N.~J.}\ \bibnamefont
  {{Cornish}}}\ and\ \bibinfo {author} {\bibfnamefont {R.~W.}\ \bibnamefont
  {{Hellings}}},\ }\href {\doibase 10.1088/0264-9381/20/22/009} {\bibfield
  {journal} {\bibinfo  {journal} {Classical and Quantum Gravity}\ }\textbf
  {\bibinfo {volume} {20}},\ \bibinfo {pages} {4851} (\bibinfo {year}
  {2003})},\ \Eprint {http://arxiv.org/abs/gr-qc/0306096} {arXiv:gr-qc/0306096
  [gr-qc]} \BibitemShut {NoStop}%
\bibitem [{\citenamefont {{Shaddock}}\ \emph {et~al.}(2003)\citenamefont
  {{Shaddock}}, \citenamefont {{Tinto}}, \citenamefont {{Estabrook}},\ and\
  \citenamefont {{Armstrong}}}]{Shaddock+03}%
  \BibitemOpen
  \bibfield  {author} {\bibinfo {author} {\bibfnamefont {D.~A.}\ \bibnamefont
  {{Shaddock}}}, \bibinfo {author} {\bibfnamefont {M.}~\bibnamefont {{Tinto}}},
  \bibinfo {author} {\bibfnamefont {F.~B.}\ \bibnamefont {{Estabrook}}}, \ and\
  \bibinfo {author} {\bibfnamefont {J.~W.}\ \bibnamefont {{Armstrong}}},\
  }\href {\doibase 10.1103/PhysRevD.68.061303} {\bibfield  {journal} {\bibinfo
  {journal} {Phys. Rev. D}\ }\textbf {\bibinfo {volume} {68}},\ \bibinfo {eid}
  {061303} (\bibinfo {year} {2003})},\ \Eprint
  {http://arxiv.org/abs/gr-qc/0307080} {arXiv:gr-qc/0307080 [gr-qc]}
  \BibitemShut {NoStop}%
\bibitem [{\citenamefont {{Tinto}}\ \emph {et~al.}(2004)\citenamefont
  {{Tinto}}, \citenamefont {{Estabrook}},\ and\ \citenamefont
  {{Armstrong}}}]{Tinto+04}%
  \BibitemOpen
  \bibfield  {author} {\bibinfo {author} {\bibfnamefont {M.}~\bibnamefont
  {{Tinto}}}, \bibinfo {author} {\bibfnamefont {F.~B.}\ \bibnamefont
  {{Estabrook}}}, \ and\ \bibinfo {author} {\bibfnamefont {J.~W.}\ \bibnamefont
  {{Armstrong}}},\ }\href {\doibase 10.1103/PhysRevD.69.082001} {\bibfield
  {journal} {\bibinfo  {journal} {Phys. Rev. D}\ }\textbf {\bibinfo {volume}
  {69}},\ \bibinfo {eid} {082001} (\bibinfo {year} {2004})},\ \Eprint
  {http://arxiv.org/abs/gr-qc/0310017} {arXiv:gr-qc/0310017 [gr-qc]}
  \BibitemShut {NoStop}%
\bibitem [{\citenamefont {{Prince}}\ \emph {et~al.}(2002)\citenamefont
  {{Prince}}, \citenamefont {{Tinto}}, \citenamefont {{Larson}},\ and\
  \citenamefont {{Armstrong}}}]{Prince+02}%
  \BibitemOpen
  \bibfield  {author} {\bibinfo {author} {\bibfnamefont {T.~A.}\ \bibnamefont
  {{Prince}}}, \bibinfo {author} {\bibfnamefont {M.}~\bibnamefont {{Tinto}}},
  \bibinfo {author} {\bibfnamefont {S.~L.}\ \bibnamefont {{Larson}}}, \ and\
  \bibinfo {author} {\bibfnamefont {J.~W.}\ \bibnamefont {{Armstrong}}},\
  }\href {\doibase 10.1103/PhysRevD.66.122002} {\bibfield  {journal} {\bibinfo
  {journal} {Phys. Rev. D}\ }\textbf {\bibinfo {volume} {66}},\ \bibinfo {eid}
  {122002} (\bibinfo {year} {2002})},\ \Eprint
  {http://arxiv.org/abs/gr-qc/0209039} {arXiv:gr-qc/0209039 [gr-qc]}
  \BibitemShut {NoStop}%
\bibitem [{\citenamefont {{Petiteau A, Hewitson M, Heinzel G, Fitzsimons E and
  Halloin H}}(2016)}]{LISANoiseBudget16}%
  \BibitemOpen
  \bibfield  {author} {\bibinfo {author} {\bibnamefont {{Petiteau A, Hewitson
  M, Heinzel G, Fitzsimons E and Halloin H}}},\ }\href@noop {} {\emph {\bibinfo
  {title} {Lisa noise budget}}},\ \bibinfo {type} {LISA Tech. report}\ \bibinfo
  {number} {lISA-CST-TN-0001}\ (\bibinfo  {institution} {LISAConsortium},\
  \bibinfo {year} {2016})\BibitemShut {NoStop}%
\bibitem [{\citenamefont {{Flanagan}}\ and\ \citenamefont
  {{Hughes}}(2005)}]{FlanaganHughes05}%
  \BibitemOpen
  \bibfield  {author} {\bibinfo {author} {\bibfnamefont {{\'E}.~{\'E}.}\
  \bibnamefont {{Flanagan}}}\ and\ \bibinfo {author} {\bibfnamefont {S.~A.}\
  \bibnamefont {{Hughes}}},\ }\href {\doibase 10.1088/1367-2630/7/1/204}
  {\bibfield  {journal} {\bibinfo  {journal} {New Journal of Physics}\ }\textbf
  {\bibinfo {volume} {7}},\ \bibinfo {pages} {204} (\bibinfo {year} {2005})},\
  \Eprint {http://arxiv.org/abs/gr-qc/0501041} {arXiv:gr-qc/0501041 [astro-ph]}
  \BibitemShut {NoStop}%
\bibitem [{\citenamefont {Pan}\ \emph {et~al.}(2011)\citenamefont {Pan},
  \citenamefont {Buonanno}, \citenamefont {Boyle}, \citenamefont {Buchman},
  \citenamefont {Kidder}, \citenamefont {Pfeiffer},\ and\ \citenamefont
  {Scheel}}]{Pan+11}%
  \BibitemOpen
  \bibfield  {author} {\bibinfo {author} {\bibfnamefont {Y.}~\bibnamefont
  {Pan}}, \bibinfo {author} {\bibfnamefont {A.}~\bibnamefont {Buonanno}},
  \bibinfo {author} {\bibfnamefont {M.}~\bibnamefont {Boyle}}, \bibinfo
  {author} {\bibfnamefont {L.~T.}\ \bibnamefont {Buchman}}, \bibinfo {author}
  {\bibfnamefont {L.~E.}\ \bibnamefont {Kidder}}, \bibinfo {author}
  {\bibfnamefont {H.~P.}\ \bibnamefont {Pfeiffer}}, \ and\ \bibinfo {author}
  {\bibfnamefont {M.~A.}\ \bibnamefont {Scheel}},\ }\href {\doibase
  10.1103/PhysRevD.84.124052} {\bibfield  {journal} {\bibinfo  {journal} {Phys.
  Rev.}\ }\textbf {\bibinfo {volume} {D84}},\ \bibinfo {pages} {124052}
  (\bibinfo {year} {2011})},\ \Eprint {http://arxiv.org/abs/1106.1021}
  {arXiv:1106.1021 [gr-qc]} \BibitemShut {NoStop}%
\bibitem [{\citenamefont {Buonanno}\ and\ \citenamefont {Damour}(1999)}]{BD99}%
  \BibitemOpen
  \bibfield  {author} {\bibinfo {author} {\bibfnamefont {A.}~\bibnamefont
  {Buonanno}}\ and\ \bibinfo {author} {\bibfnamefont {T.}~\bibnamefont
  {Damour}},\ }\href@noop {} {\bibfield  {journal} {\bibinfo  {journal} {Phys.
  Rev. D}\ }\textbf {\bibinfo {volume} {59}},\ \bibinfo {pages} {084006}
  (\bibinfo {year} {1999})},\ \Eprint {http://arxiv.org/abs/gr-qc/9811091}
  {gr-qc/9811091} \BibitemShut {NoStop}%
\bibitem [{\citenamefont {Buonanno}\ and\ \citenamefont {Damour}(2000)}]{BD00}%
  \BibitemOpen
  \bibfield  {author} {\bibinfo {author} {\bibfnamefont {A.}~\bibnamefont
  {Buonanno}}\ and\ \bibinfo {author} {\bibfnamefont {T.}~\bibnamefont
  {Damour}},\ }\href@noop {} {\bibfield  {journal} {\bibinfo  {journal} {Phys.
  Rev. D}\ }\textbf {\bibinfo {volume} {62}},\ \bibinfo {pages} {064015}
  (\bibinfo {year} {2000})},\ \Eprint {http://arxiv.org/abs/gr-qc/0001013}
  {gr-qc/0001013} \BibitemShut {NoStop}%
\bibitem [{\citenamefont {{P{\"u}rrer}}(2016)}]{Puerrer15}%
  \BibitemOpen
  \bibfield  {author} {\bibinfo {author} {\bibfnamefont {M.}~\bibnamefont
  {{P{\"u}rrer}}},\ }\href {\doibase 10.1103/PhysRevD.93.064041} {\bibfield
  {journal} {\bibinfo  {journal} {Phys. Rev. D}\ }\textbf {\bibinfo {volume}
  {93}},\ \bibinfo {eid} {064041} (\bibinfo {year} {2016})},\ \Eprint
  {http://arxiv.org/abs/1512.02248} {arXiv:1512.02248 [gr-qc]} \BibitemShut
  {NoStop}%
\bibitem [{\citenamefont {Evans}\ \emph {et~al.}(2018)\citenamefont {Evans},
  \citenamefont {Sturani}, \citenamefont {Vitale},\ and\ \citenamefont
  {Hall}}]{UnofficialNoiseCurves18}%
  \BibitemOpen
  \bibfield  {author} {\bibinfo {author} {\bibfnamefont {M.}~\bibnamefont
  {Evans}}, \bibinfo {author} {\bibfnamefont {R.}~\bibnamefont {Sturani}},
  \bibinfo {author} {\bibfnamefont {S.}~\bibnamefont {Vitale}}, \ and\ \bibinfo
  {author} {\bibfnamefont {E.}~\bibnamefont {Hall}},\ }\href
  {https://dcc.ligo.org/LIGO-T1500293/public} {\emph {\bibinfo {title}
  {Unofficial sensitivity curves (ASD) for aLIGO, Kagra, Virgo, Voyager, Cosmic
  Explorer and ET}}},\ \bibinfo {type} {Tech. Rep.}\ (\bibinfo {year}
  {2018})\BibitemShut {NoStop}%
\bibitem [{\citenamefont {Cutler}\ and\ \citenamefont {Flanagan}(1994)}]{CF94}%
  \BibitemOpen
  \bibfield  {author} {\bibinfo {author} {\bibfnamefont {C.}~\bibnamefont
  {Cutler}}\ and\ \bibinfo {author} {\bibfnamefont {E.}~\bibnamefont
  {Flanagan}},\ }\href@noop {} {\bibfield  {journal} {\bibinfo  {journal}
  {Phys. Rev. D}\ }\textbf {\bibinfo {volume} {49}},\ \bibinfo {pages} {2658}
  (\bibinfo {year} {1994})}\BibitemShut {NoStop}%
\bibitem [{\citenamefont {{Poisson}}\ and\ \citenamefont
  {{Will}}(1995)}]{PW95}%
  \BibitemOpen
  \bibfield  {author} {\bibinfo {author} {\bibfnamefont {E.}~\bibnamefont
  {{Poisson}}}\ and\ \bibinfo {author} {\bibfnamefont {C.~M.}\ \bibnamefont
  {{Will}}},\ }\href {\doibase 10.1103/PhysRevD.52.848} {\bibfield  {journal}
  {\bibinfo  {journal} {Phys. Rev. D}\ }\textbf {\bibinfo {volume} {52}},\
  \bibinfo {pages} {848} (\bibinfo {year} {1995})},\ \Eprint
  {http://arxiv.org/abs/arXiv:gr-qc/9502040} {arXiv:gr-qc/9502040} \BibitemShut
  {NoStop}%
\bibitem [{\citenamefont {Baird}\ \emph {et~al.}(2013)\citenamefont {Baird},
  \citenamefont {Fairhurst}, \citenamefont {Hannam},\ and\ \citenamefont
  {Murphy}}]{Baird+2013}%
  \BibitemOpen
  \bibfield  {author} {\bibinfo {author} {\bibfnamefont {E.}~\bibnamefont
  {Baird}}, \bibinfo {author} {\bibfnamefont {S.}~\bibnamefont {Fairhurst}},
  \bibinfo {author} {\bibfnamefont {M.}~\bibnamefont {Hannam}}, \ and\ \bibinfo
  {author} {\bibfnamefont {P.}~\bibnamefont {Murphy}},\ }\href {\doibase
  10.1103/PhysRevD.87.024035} {\bibfield  {journal} {\bibinfo  {journal} {Phys.
  Rev. D}\ }\textbf {\bibinfo {volume} {87}},\ \bibinfo {pages} {024035}
  (\bibinfo {year} {2013})}\BibitemShut {NoStop}%
\bibitem [{\citenamefont {{Smith}}\ \emph {et~al.}(2014)\citenamefont
  {{Smith}}, \citenamefont {{Hanna}}, \citenamefont {{Mandel}},\ and\
  \citenamefont {{Vecchio}}}]{Smith+14}%
  \BibitemOpen
  \bibfield  {author} {\bibinfo {author} {\bibfnamefont {R.~J.~E.}\
  \bibnamefont {{Smith}}}, \bibinfo {author} {\bibfnamefont {C.}~\bibnamefont
  {{Hanna}}}, \bibinfo {author} {\bibfnamefont {I.}~\bibnamefont {{Mandel}}}, \
  and\ \bibinfo {author} {\bibfnamefont {A.}~\bibnamefont {{Vecchio}}},\ }\href
  {\doibase 10.1103/PhysRevD.90.044074} {\bibfield  {journal} {\bibinfo
  {journal} {Phys. Rev. D}\ }\textbf {\bibinfo {volume} {90}},\ \bibinfo {eid}
  {044074} (\bibinfo {year} {2014})},\ \Eprint {http://arxiv.org/abs/1305.3798}
  {arXiv:1305.3798 [astro-ph.HE]} \BibitemShut {NoStop}%
\bibitem [{\citenamefont {{Canizares}}\ \emph {et~al.}(2015)\citenamefont
  {{Canizares}}, \citenamefont {{Field}}, \citenamefont {{Gair}}, \citenamefont
  {{Raymond}}, \citenamefont {{Smith}},\ and\ \citenamefont
  {{Tiglio}}}]{Canizares+14}%
  \BibitemOpen
  \bibfield  {author} {\bibinfo {author} {\bibfnamefont {P.}~\bibnamefont
  {{Canizares}}}, \bibinfo {author} {\bibfnamefont {S.~E.}\ \bibnamefont
  {{Field}}}, \bibinfo {author} {\bibfnamefont {J.}~\bibnamefont {{Gair}}},
  \bibinfo {author} {\bibfnamefont {V.}~\bibnamefont {{Raymond}}}, \bibinfo
  {author} {\bibfnamefont {R.}~\bibnamefont {{Smith}}}, \ and\ \bibinfo
  {author} {\bibfnamefont {M.}~\bibnamefont {{Tiglio}}},\ }\href {\doibase
  10.1103/PhysRevLett.114.071104} {\bibfield  {journal} {\bibinfo  {journal}
  {Physical Review Letters}\ }\textbf {\bibinfo {volume} {114}},\ \bibinfo
  {eid} {071104} (\bibinfo {year} {2015})},\ \Eprint
  {http://arxiv.org/abs/1404.6284} {arXiv:1404.6284 [gr-qc]} \BibitemShut
  {NoStop}%
\bibitem [{\citenamefont {{Smith}}\ \emph {et~al.}(2016)\citenamefont
  {{Smith}}, \citenamefont {{Field}}, \citenamefont {{Blackburn}},
  \citenamefont {{Haster}}, \citenamefont {{P{\"u}rrer}}, \citenamefont
  {{Raymond}},\ and\ \citenamefont {{Schmidt}}}]{Smith+16}%
  \BibitemOpen
  \bibfield  {author} {\bibinfo {author} {\bibfnamefont {R.}~\bibnamefont
  {{Smith}}}, \bibinfo {author} {\bibfnamefont {S.~E.}\ \bibnamefont
  {{Field}}}, \bibinfo {author} {\bibfnamefont {K.}~\bibnamefont
  {{Blackburn}}}, \bibinfo {author} {\bibfnamefont {C.-J.}\ \bibnamefont
  {{Haster}}}, \bibinfo {author} {\bibfnamefont {M.}~\bibnamefont
  {{P{\"u}rrer}}}, \bibinfo {author} {\bibfnamefont {V.}~\bibnamefont
  {{Raymond}}}, \ and\ \bibinfo {author} {\bibfnamefont {P.}~\bibnamefont
  {{Schmidt}}},\ }\href {\doibase 10.1103/PhysRevD.94.044031} {\bibfield
  {journal} {\bibinfo  {journal} {Phys. Rev. D}\ }\textbf {\bibinfo {volume}
  {94}},\ \bibinfo {eid} {044031} (\bibinfo {year} {2016})},\ \Eprint
  {http://arxiv.org/abs/1604.08253} {arXiv:1604.08253 [gr-qc]} \BibitemShut
  {NoStop}%
\bibitem [{\citenamefont {{Cornish}}(2010)}]{Cornish10}%
  \BibitemOpen
  \bibfield  {author} {\bibinfo {author} {\bibfnamefont {N.~J.}\ \bibnamefont
  {{Cornish}}},\ }\href@noop {} {\bibfield  {journal} {\bibinfo  {journal}
  {arXiv e-prints}\ ,\ \bibinfo {eid} {arXiv:1007.4820}} (\bibinfo {year}
  {2010})},\ \Eprint {http://arxiv.org/abs/1007.4820} {arXiv:1007.4820 [gr-qc]}
  \BibitemShut {NoStop}%
\bibitem [{\citenamefont {{Porter}}(2014)}]{Porter14}%
  \BibitemOpen
  \bibfield  {author} {\bibinfo {author} {\bibfnamefont {E.~K.}\ \bibnamefont
  {{Porter}}},\ }\href@noop {} {\bibfield  {journal} {\bibinfo  {journal}
  {arXiv e-prints}\ ,\ \bibinfo {eid} {arXiv:1411.0598}} (\bibinfo {year}
  {2014})},\ \Eprint {http://arxiv.org/abs/1411.0598} {arXiv:1411.0598 [gr-qc]}
  \BibitemShut {NoStop}%
\bibitem [{\citenamefont {{Vinciguerra}}\ \emph {et~al.}(2017)\citenamefont
  {{Vinciguerra}}, \citenamefont {{Veitch}},\ and\ \citenamefont
  {{Mandel}}}]{Vinciguerra+17}%
  \BibitemOpen
  \bibfield  {author} {\bibinfo {author} {\bibfnamefont {S.}~\bibnamefont
  {{Vinciguerra}}}, \bibinfo {author} {\bibfnamefont {J.}~\bibnamefont
  {{Veitch}}}, \ and\ \bibinfo {author} {\bibfnamefont {I.}~\bibnamefont
  {{Mandel}}},\ }\href {\doibase 10.1088/1361-6382/aa6d44} {\bibfield
  {journal} {\bibinfo  {journal} {Classical and Quantum Gravity}\ }\textbf
  {\bibinfo {volume} {34}},\ \bibinfo {eid} {115006} (\bibinfo {year}
  {2017})},\ \Eprint {http://arxiv.org/abs/1703.02062} {arXiv:1703.02062
  [gr-qc]} \BibitemShut {NoStop}%
\bibitem [{\citenamefont {{Swendsen}}\ and\ \citenamefont
  {{Wang}}(1986)}]{SwendsenWang86}%
  \BibitemOpen
  \bibfield  {author} {\bibinfo {author} {\bibfnamefont {R.~H.}\ \bibnamefont
  {{Swendsen}}}\ and\ \bibinfo {author} {\bibfnamefont {J.-S.}\ \bibnamefont
  {{Wang}}},\ }\href {\doibase 10.1103/PhysRevLett.57.2607} {\bibfield
  {journal} {\bibinfo  {journal} {Phys. Rev. Letters}\ }\textbf {\bibinfo
  {volume} {57}},\ \bibinfo {pages} {2607} (\bibinfo {year}
  {1986})}\BibitemShut {NoStop}%
\bibitem [{\citenamefont {{Littenberg}}\ and\ \citenamefont
  {{Cornish}}(2009)}]{LittenbergCornish09}%
  \BibitemOpen
  \bibfield  {author} {\bibinfo {author} {\bibfnamefont {T.~B.}\ \bibnamefont
  {{Littenberg}}}\ and\ \bibinfo {author} {\bibfnamefont {N.~J.}\ \bibnamefont
  {{Cornish}}},\ }\href {\doibase 10.1103/PhysRevD.80.063007} {\bibfield
  {journal} {\bibinfo  {journal} {Phys. Rev. D}\ }\textbf {\bibinfo {volume}
  {80}},\ \bibinfo {eid} {063007} (\bibinfo {year} {2009})},\ \Eprint
  {http://arxiv.org/abs/0902.0368} {arXiv:0902.0368 [gr-qc]} \BibitemShut
  {NoStop}%
\bibitem [{\citenamefont {Braak}(2006)}]{Braak06}%
  \BibitemOpen
  \bibfield  {author} {\bibinfo {author} {\bibfnamefont {C.~J. F.~T.}\
  \bibnamefont {Braak}},\ }\href {\doibase 10.1007/s11222-006-8769-1}
  {\bibfield  {journal} {\bibinfo  {journal} {Statistics and Computing}\
  }\textbf {\bibinfo {volume} {16}},\ \bibinfo {pages} {239} (\bibinfo {year}
  {2006})}\BibitemShut {NoStop}%
\bibitem [{\citenamefont {ter Braak}\ and\ \citenamefont
  {Vrugt}(2008)}]{BraakVrugt08}%
  \BibitemOpen
  \bibfield  {author} {\bibinfo {author} {\bibfnamefont {C.~J.~F.}\
  \bibnamefont {ter Braak}}\ and\ \bibinfo {author} {\bibfnamefont {J.~A.}\
  \bibnamefont {Vrugt}},\ }\href {\doibase 10.1007/s11222-008-9104-9}
  {\bibfield  {journal} {\bibinfo  {journal} {Statistics and Computing}\
  }\textbf {\bibinfo {volume} {18}},\ \bibinfo {pages} {435} (\bibinfo {year}
  {2008})}\BibitemShut {NoStop}%
\bibitem [{\citenamefont {{Graff}}\ \emph {et~al.}(2012)\citenamefont
  {{Graff}}, \citenamefont {{Feroz}}, \citenamefont {{Hobson}},\ and\
  \citenamefont {{Lasenby}}}]{Graff+11}%
  \BibitemOpen
  \bibfield  {author} {\bibinfo {author} {\bibfnamefont {P.}~\bibnamefont
  {{Graff}}}, \bibinfo {author} {\bibfnamefont {F.}~\bibnamefont {{Feroz}}},
  \bibinfo {author} {\bibfnamefont {M.~P.}\ \bibnamefont {{Hobson}}}, \ and\
  \bibinfo {author} {\bibfnamefont {A.}~\bibnamefont {{Lasenby}}},\ }\href
  {\doibase 10.1111/j.1365-2966.2011.20288.x} {\bibfield  {journal} {\bibinfo
  {journal} {Mon. Not. Roy. Astron. Soc.}\ }\textbf {\bibinfo {volume} {421}},\
  \bibinfo {pages} {169} (\bibinfo {year} {2012})},\ \Eprint
  {http://arxiv.org/abs/1110.2997} {arXiv:1110.2997 [astro-ph.IM]} \BibitemShut
  {NoStop}%
\bibitem [{\citenamefont {Skilling}(2006)}]{Skilling06}%
  \BibitemOpen
  \bibfield  {author} {\bibinfo {author} {\bibfnamefont {J.}~\bibnamefont
  {Skilling}},\ }\href {\doibase 10.1214/06-BA127} {\bibfield  {journal}
  {\bibinfo  {journal} {Bayesian Anal.}\ }\textbf {\bibinfo {volume} {1}},\
  \bibinfo {pages} {833} (\bibinfo {year} {2006})}\BibitemShut {NoStop}%
\bibitem [{\citenamefont {{Vallisneri}}(2008)}]{Vallisneri08}%
  \BibitemOpen
  \bibfield  {author} {\bibinfo {author} {\bibfnamefont {M.}~\bibnamefont
  {{Vallisneri}}},\ }\href {\doibase 10.1103/PhysRevD.77.042001} {\bibfield
  {journal} {\bibinfo  {journal} {Phys. Rev. D}\ }\textbf {\bibinfo {volume}
  {77}},\ \bibinfo {eid} {042001} (\bibinfo {year} {2008})},\ \Eprint
  {http://arxiv.org/abs/gr-qc/0703086} {gr-qc/0703086} \BibitemShut {NoStop}%
\bibitem [{\citenamefont {{Barausse}}(2012)}]{Barausse12}%
  \BibitemOpen
  \bibfield  {author} {\bibinfo {author} {\bibfnamefont {E.}~\bibnamefont
  {{Barausse}}},\ }\href {\doibase 10.1111/j.1365-2966.2012.21057.x} {\bibfield
   {journal} {\bibinfo  {journal} {Mon. Not. Roy. Astron. Soc.}\ }\textbf
  {\bibinfo {volume} {423}},\ \bibinfo {pages} {2533} (\bibinfo {year}
  {2012})},\ \Eprint {http://arxiv.org/abs/1201.5888} {arXiv:1201.5888
  [astro-ph.CO]} \BibitemShut {NoStop}%
\bibitem [{\citenamefont {{Planck Collaboration}}(2016)}]{Planck15param}%
  \BibitemOpen
  \bibfield  {author} {\bibinfo {author} {\bibnamefont {{Planck
  Collaboration}}},\ }\href {\doibase 10.1051/0004-6361/201525830} {\bibfield
  {journal} {\bibinfo  {journal} {Astron. Astrophys.}\ }\textbf {\bibinfo
  {volume} {594}},\ \bibinfo {eid} {A13} (\bibinfo {year} {2016})},\ \Eprint
  {http://arxiv.org/abs/1502.01589} {arXiv:1502.01589 [astro-ph.CO]}
  \BibitemShut {NoStop}%
\bibitem [{\citenamefont {Marsat}\ and\ \citenamefont
  {Babak}(2020)}]{MarsatBabak20}%
  \BibitemOpen
  \bibfield  {author} {\bibinfo {author} {\bibfnamefont {S.}~\bibnamefont
  {Marsat}}\ and\ \bibinfo {author} {\bibfnamefont {S.}~\bibnamefont {Babak}},\
  }\href@noop {} {\bibfield  {journal} {\bibinfo  {journal} {In preparation}\ }
  (\bibinfo {year} {2020})}\BibitemShut {NoStop}%
\bibitem [{\citenamefont {{Katz}}\ and\ \citenamefont
  {{Larson}}(2019)}]{KatzLarson18}%
  \BibitemOpen
  \bibfield  {author} {\bibinfo {author} {\bibfnamefont {M.~L.}\ \bibnamefont
  {{Katz}}}\ and\ \bibinfo {author} {\bibfnamefont {S.~L.}\ \bibnamefont
  {{Larson}}},\ }\href {\doibase 10.1093/mnras/sty3321} {\bibfield  {journal}
  {\bibinfo  {journal} {Mon. Not. Roy. Astron. Soc.}\ }\textbf {\bibinfo
  {volume} {483}},\ \bibinfo {pages} {3108} (\bibinfo {year} {2019})},\ \Eprint
  {http://arxiv.org/abs/1807.02511} {arXiv:1807.02511 [gr-qc]} \BibitemShut
  {NoStop}%
\bibitem [{\citenamefont {{Singer}}\ and\ \citenamefont
  {{Price}}(2016)}]{SingerPrice15}%
  \BibitemOpen
  \bibfield  {author} {\bibinfo {author} {\bibfnamefont {L.~P.}\ \bibnamefont
  {{Singer}}}\ and\ \bibinfo {author} {\bibfnamefont {L.~R.}\ \bibnamefont
  {{Price}}},\ }\href {\doibase 10.1103/PhysRevD.93.024013} {\bibfield
  {journal} {\bibinfo  {journal} {Phys. Rev. D}\ }\textbf {\bibinfo {volume}
  {93}},\ \bibinfo {eid} {024013} (\bibinfo {year} {2016})},\ \Eprint
  {http://arxiv.org/abs/1508.03634} {arXiv:1508.03634 [gr-qc]} \BibitemShut
  {NoStop}%
\bibitem [{\citenamefont {{Gerosa}}\ \emph {et~al.}(2019)\citenamefont
  {{Gerosa}}, \citenamefont {{Ma}}, \citenamefont {{Wong}}, \citenamefont
  {{Berti}}, \citenamefont {{O'Shaughnessy}}, \citenamefont {{Chen}},\ and\
  \citenamefont {{Belczynski}}}]{Gerosa+19}%
  \BibitemOpen
  \bibfield  {author} {\bibinfo {author} {\bibfnamefont {D.}~\bibnamefont
  {{Gerosa}}}, \bibinfo {author} {\bibfnamefont {S.}~\bibnamefont {{Ma}}},
  \bibinfo {author} {\bibfnamefont {K.~W.~K.}\ \bibnamefont {{Wong}}}, \bibinfo
  {author} {\bibfnamefont {E.}~\bibnamefont {{Berti}}}, \bibinfo {author}
  {\bibfnamefont {R.}~\bibnamefont {{O'Shaughnessy}}}, \bibinfo {author}
  {\bibfnamefont {Y.}~\bibnamefont {{Chen}}}, \ and\ \bibinfo {author}
  {\bibfnamefont {K.}~\bibnamefont {{Belczynski}}},\ }\href {\doibase
  10.1103/PhysRevD.99.103004} {\bibfield  {journal} {\bibinfo  {journal} {Phys.
  Rev. D}\ }\textbf {\bibinfo {volume} {99}},\ \bibinfo {eid} {103004}
  (\bibinfo {year} {2019})},\ \Eprint {http://arxiv.org/abs/1902.00021}
  {arXiv:1902.00021 [astro-ph.HE]} \BibitemShut {NoStop}%
\bibitem [{\citenamefont {{Moore}}\ \emph {et~al.}(2019)\citenamefont
  {{Moore}}, \citenamefont {{Gerosa}},\ and\ \citenamefont
  {{Klein}}}]{Moore+19}%
  \BibitemOpen
  \bibfield  {author} {\bibinfo {author} {\bibfnamefont {C.~J.}\ \bibnamefont
  {{Moore}}}, \bibinfo {author} {\bibfnamefont {D.}~\bibnamefont {{Gerosa}}}, \
  and\ \bibinfo {author} {\bibfnamefont {A.}~\bibnamefont {{Klein}}},\ }\href
  {\doibase 10.1093/mnrasl/slz104} {\bibfield  {journal} {\bibinfo  {journal}
  {Mon. Not. Roy. Astron. Soc.}\ }\textbf {\bibinfo {volume} {488}},\ \bibinfo
  {pages} {L94} (\bibinfo {year} {2019})},\ \Eprint
  {http://arxiv.org/abs/1905.11998} {arXiv:1905.11998 [astro-ph.HE]}
  \BibitemShut {NoStop}%
\bibitem [{\citenamefont {Toubiana}\ \emph {et~al.}(2018)\citenamefont
  {Toubiana}, \citenamefont {Marsat}, \citenamefont {Babak}, \citenamefont
  {Baker},\ and\ \citenamefont {Dal~Canton}}]{Toubiana+20}%
  \BibitemOpen
  \bibfield  {author} {\bibinfo {author} {\bibfnamefont {A.}~\bibnamefont
  {Toubiana}}, \bibinfo {author} {\bibfnamefont {S.}~\bibnamefont {Marsat}},
  \bibinfo {author} {\bibfnamefont {S.}~\bibnamefont {Babak}}, \bibinfo
  {author} {\bibfnamefont {J.~G.}\ \bibnamefont {Baker}}, \ and\ \bibinfo
  {author} {\bibfnamefont {T.}~\bibnamefont {Dal~Canton}},\ }\href@noop {}
  {\bibfield  {journal} {\bibinfo  {journal} {In preparation}\ } (\bibinfo
  {year} {2018})}\BibitemShut {NoStop}%
\bibitem [{\citenamefont {{Mangiagli}}\ \emph {et~al.}(2019)\citenamefont
  {{Mangiagli}}, \citenamefont {{Klein}}, \citenamefont {{Sesana}},
  \citenamefont {{Barausse}},\ and\ \citenamefont {{Colpi}}}]{Mangiagli+18}%
  \BibitemOpen
  \bibfield  {author} {\bibinfo {author} {\bibfnamefont {A.}~\bibnamefont
  {{Mangiagli}}}, \bibinfo {author} {\bibfnamefont {A.}~\bibnamefont
  {{Klein}}}, \bibinfo {author} {\bibfnamefont {A.}~\bibnamefont {{Sesana}}},
  \bibinfo {author} {\bibfnamefont {E.}~\bibnamefont {{Barausse}}}, \ and\
  \bibinfo {author} {\bibfnamefont {M.}~\bibnamefont {{Colpi}}},\ }\href
  {\doibase 10.1103/PhysRevD.99.064056} {\bibfield  {journal} {\bibinfo
  {journal} {Phys. Rev. D}\ }\textbf {\bibinfo {volume} {99}},\ \bibinfo {eid}
  {064056} (\bibinfo {year} {2019})},\ \Eprint
  {http://arxiv.org/abs/1811.01805} {arXiv:1811.01805 [gr-qc]} \BibitemShut
  {NoStop}%
\bibitem [{\citenamefont {Katz}\ \emph {et~al.}(2020)\citenamefont {Katz},
  \citenamefont {Larson}, \citenamefont {Marsat}, \citenamefont {Babak},\ and\
  \citenamefont {K.}}]{Katz+20}%
  \BibitemOpen
  \bibfield  {author} {\bibinfo {author} {\bibfnamefont {M.~L.}\ \bibnamefont
  {Katz}}, \bibinfo {author} {\bibfnamefont {S.~L.}\ \bibnamefont {Larson}},
  \bibinfo {author} {\bibfnamefont {S.}~\bibnamefont {Marsat}}, \bibinfo
  {author} {\bibfnamefont {S.}~\bibnamefont {Babak}}, \ and\ \bibinfo {author}
  {\bibfnamefont {C.~A.~J.}\ \bibnamefont {K.}},\ }\href@noop {} {\bibfield
  {journal} {\bibinfo  {journal} {In preparation}\ } (\bibinfo {year}
  {2020})}\BibitemShut {NoStop}%
\bibitem [{\citenamefont {{Veitch}}\ \emph {et~al.}(2015)\citenamefont
  {{Veitch}}, \citenamefont {{Raymond}}, \citenamefont {{Farr}}, \citenamefont
  {{Farr}}, \citenamefont {{Graff}}, \citenamefont {{Vitale}}, \citenamefont
  {{Aylott}}, \citenamefont {{Blackburn}}, \citenamefont {{Christensen}},
  \citenamefont {{Coughlin}}, \citenamefont {{Del Pozzo}}, \citenamefont
  {{Feroz}}, \citenamefont {{Gair}}, \citenamefont {{Haster}}, \citenamefont
  {{Kalogera}}, \citenamefont {{Littenberg}}, \citenamefont {{Mandel}},
  \citenamefont {{O'Shaughnessy}}, \citenamefont {{Pitkin}}, \citenamefont
  {{Rodriguez}}, \citenamefont {{R{\"o}ver}}, \citenamefont {{Sidery}},
  \citenamefont {{Smith}}, \citenamefont {{Van Der Sluys}}, \citenamefont
  {{Vecchio}}, \citenamefont {{Vousden}},\ and\ \citenamefont
  {{Wade}}}]{Veitch+14}%
  \BibitemOpen
  \bibfield  {author} {\bibinfo {author} {\bibfnamefont {J.}~\bibnamefont
  {{Veitch}}}, \bibinfo {author} {\bibfnamefont {V.}~\bibnamefont {{Raymond}}},
  \bibinfo {author} {\bibfnamefont {B.}~\bibnamefont {{Farr}}}, \bibinfo
  {author} {\bibfnamefont {W.}~\bibnamefont {{Farr}}}, \bibinfo {author}
  {\bibfnamefont {P.}~\bibnamefont {{Graff}}}, \bibinfo {author} {\bibfnamefont
  {S.}~\bibnamefont {{Vitale}}}, \bibinfo {author} {\bibfnamefont
  {B.}~\bibnamefont {{Aylott}}}, \bibinfo {author} {\bibfnamefont
  {K.}~\bibnamefont {{Blackburn}}}, \bibinfo {author} {\bibfnamefont
  {N.}~\bibnamefont {{Christensen}}}, \bibinfo {author} {\bibfnamefont
  {M.}~\bibnamefont {{Coughlin}}}, \bibinfo {author} {\bibfnamefont
  {W.}~\bibnamefont {{Del Pozzo}}}, \bibinfo {author} {\bibfnamefont
  {F.}~\bibnamefont {{Feroz}}}, \bibinfo {author} {\bibfnamefont
  {J.}~\bibnamefont {{Gair}}}, \bibinfo {author} {\bibfnamefont {C.~J.}\
  \bibnamefont {{Haster}}}, \bibinfo {author} {\bibfnamefont {V.}~\bibnamefont
  {{Kalogera}}}, \bibinfo {author} {\bibfnamefont {T.}~\bibnamefont
  {{Littenberg}}}, \bibinfo {author} {\bibfnamefont {I.}~\bibnamefont
  {{Mandel}}}, \bibinfo {author} {\bibfnamefont {R.}~\bibnamefont
  {{O'Shaughnessy}}}, \bibinfo {author} {\bibfnamefont {M.}~\bibnamefont
  {{Pitkin}}}, \bibinfo {author} {\bibfnamefont {C.}~\bibnamefont
  {{Rodriguez}}}, \bibinfo {author} {\bibfnamefont {C.}~\bibnamefont
  {{R{\"o}ver}}}, \bibinfo {author} {\bibfnamefont {T.}~\bibnamefont
  {{Sidery}}}, \bibinfo {author} {\bibfnamefont {R.}~\bibnamefont {{Smith}}},
  \bibinfo {author} {\bibfnamefont {M.}~\bibnamefont {{Van Der Sluys}}},
  \bibinfo {author} {\bibfnamefont {A.}~\bibnamefont {{Vecchio}}}, \bibinfo
  {author} {\bibfnamefont {W.}~\bibnamefont {{Vousden}}}, \ and\ \bibinfo
  {author} {\bibfnamefont {L.}~\bibnamefont {{Wade}}},\ }\href {\doibase
  10.1103/PhysRevD.91.042003} {\bibfield  {journal} {\bibinfo  {journal} {Phys.
  Rev. D}\ }\textbf {\bibinfo {volume} {91}},\ \bibinfo {eid} {042003}
  (\bibinfo {year} {2015})},\ \Eprint {http://arxiv.org/abs/1409.7215}
  {arXiv:1409.7215 [gr-qc]} \BibitemShut {NoStop}%
\end{thebibliography}%

\end{document}